%% file: component_sigma_paper_v4.tex
\documentclass[11pt, a4paper]{article}
\usepackage{jheppub}

\usepackage{amsmath, amsfonts, amssymb}
\usepackage{hyperref}
\usepackage{booktabs}
\usepackage{mathrsfs}

\include{sugracom_v2}
\renewcommand{\bm}[1]{#1}
\newcommand{\AntT}{(\textrm{antarctic term})}
\newcommand{\antt}{\textrm{a.t.}}
\newcommand{\bbD}{\mathbb D}
\newcommand{\bbA}{\mathbb A}
\newcommand{\bnabla}{{\bar\nabla}}
\newcommand{\trD}{{\rm \scriptscriptstyle D}}
\newcommand{\trA}{{\rm \scriptscriptstyle A}}
\newcommand{\trQ}{{\rm \scriptscriptstyle Q}}
\newcommand{\trSU}{{\rm \scriptscriptstyle SU(2)}}

\newcommand{\Sp}[1]{\ensuremath{{\rm Sp}(#1)}}
\newcommand{\SU}[1]{\ensuremath{{\rm SU}(#1)}}
\newcommand{\gU}[1]{\ensuremath{{\rm U}(#1)}}

\numberwithin{equation}{section}

\title{Projective multiplets and hyperk\"ahler cones in conformal supergravity}
\author{Daniel Butter}
\affiliation{Nikhef Theory Group, \\
Science Park 105, 1098 XG Amsterdam, The Netherlands}
\emailAdd{dbutter@nikhef.nl}
\preprint{NIKHEF-2014-040}

\abstract{Projective superspace provides a natural framework for the
construction of actions coupling hypermultiplets to conformal
supergravity. We review how the off-shell actions are formulated in superspace
and then discuss how to eliminate the infinite number of auxiliary
fields to produce an on-shell $\cN=2$ supersymmetric
sigma model, with the target space corresponding to a generic
$4n$-dimensional hyperk\"ahler cone.
We show how the component action coupling the hypermultiplets
to conformal supergravity may be constructed starting from curved superspace.
The superspace origin of the geometric data -- the hyperk\"ahler potential,
complex structures, and any gauged isometries -- is also addressed.
}

\dedicated{Dedicated to the memory of Bruno Zumino}

\begin{document}
\maketitle

\section{Introduction}
The hypermultiplet is the most general on-shell matter multiplet
with rigid $\cN=2$ supersymmetry in four dimensions. It consists of
a complex doublet of scalar fields and a pair of Weyl fermions
(equivalently, a single Dirac fermion), whose supersymmetry transformations
close only on the equations of motion.
For this reason, specifying the hypermultiplet is equivalent to
specifying the action,
or equivalently, specifying the target space of the sigma model
parametrized by the hypermultiplet scalars.
It has long been known that such target
spaces must be hyperk\"ahler manifolds \cite{A-GF:HK, A-GF:Potentials}.
For locally supersymmetric theories -- by which one usually means
those coupled to supergravity with a canonically normalized
Einstein-Hilbert Lagrangian -- the target space must instead be
quaternion-K\"ahler \cite{BaWi:QK}.
This distinction recalls that between rigid $\cN=1$ theories,
where the target space is K\"ahler \cite{Zumino:Kahler},
and their locally supersymmetric counterparts, which must be
Hodge-K\"ahler \cite{WiBa:Quant}.\footnote{See also
the work of Komargodski and Seiberg for a new perspective
on these results \cite{KS:Supercurrents, Seiberg:2010qd}.}

It is known that off-shell formulations of the hypermultiplet,
employing $\cN=2$ superspace and involving (an infinite number of)
auxiliary fields,
provide a means to generate the required target space geometry from some unconstrained
generating function. To understand this better, let us elaborate first
on the rigid $\cN=1$ case. Recall that
the chiral multiplet $\phi$, the natural $\cN=1$ cousin of the hypermultiplet,
admits a simple off-shell representation $\{\phi, \z_\alpha, F\}$
with a single complex auxiliary field.
(We employ a convention where a single symbol $\phi$ may stand for
both a superfield and its lowest component.)
Zumino showed that the most general two-derivative function of off-shell
chiral multiplets $\phi^\ra$ is described by a superspace integral \cite{Zumino:Kahler}
\begin{align}\label{eq:N1K}
\int \rd^4x \, \rd^2 \q\, \rd^2\bar \q\, K(\phi,\bar\phi)
\end{align}
for an arbitrary real function $K$.\footnote{For now we ignore
the possibility of a superpotential or gauged isometries.}
When the auxiliary fields $F^\ra$ are eliminated, the component
Lagrangian takes a sigma model form\footnote{The fermion
$\zeta_\alpha^\ra$ is normalized here in an unconventional way.
We use the same normalization for $\cN=2$ models, following \cite{GIOS}.}
\begin{align}
\cL &= -g_{\ra \bar \rb} \pa_m \phi^\ra \pa^m \bar\phi^{\bar \rb}
	- \frac{i}{4} g_{\ra \bar \rb} \,\Big(
		\zeta^\ra \sigma^m
		{\widehat\cD}_{m} \bar\zeta^{\bar \rb}
		+ \bar\zeta^{\bar \rb} \bsigma^m
		{\widehat\cD}_{m} \zeta^\ra \Big)
	+ \frac{1}{16} R_{\ra \bar \ra \rb \bar\rb} (\zeta^\ra \zeta^\rb)
		(\bar\zeta^{\bar\ra} \bar\zeta^{\bar \rb})~, \eol
\zeta_\alpha^\ra &:= D_\alpha \phi^\ra ~, \qquad
\widehat \cD_m \zeta^\ra := \pa_m \zeta^\ra + \Gamma_{\rb \rc}{}^\ra \,\pa_m \phi^\rb \zeta^\rc~.
\end{align}
The metric is determined in terms of the function $K$,
$g_{\ra \bar\rb} := \pa_\ra \pa_{\bar \rb} K$; of course, this
is the well-known K\"ahler metric -- in using chiral multiplets
we have automatically diagonalized the complex structure --
and this construction gives the explicit proof that $\cN=1$ sigma models
in four dimensions must possess a K\"ahler target space
geometry \cite{Zumino:Kahler}. The key point is that the general off-shell supersymmetric
action \eqref{eq:N1K}
provides both the means to construct the on-shell action and
the generating function for the target space geometry.

This narrative can be repeated for $\cN=2$ theories. There one has
two complementary formulations of a general off-shell hypermultiplet,
depending on whether one uses harmonic superspace \cite{GIOS, GIKOS}
or projective superspace \cite{KLR, LR88, LR90}.
In harmonic superspace, which augments $\cN=2$ superspace with an
auxiliary $S^2$ manifold with harmonic coordinates $u^{i\pm}$,
one has $n$ pseudoreal superfields $q^{\ra+}$,
$\widetilde{q^{\ra +}} = q^{\rb+} \Omega_{\rb\ra}$ where
$\Omega$ is the canonical symplectic form of $\rm Sp(n)$,
obeying the analyticity condition
\begin{align}
u_i^+ D_{\ul\alpha}^i q^{\ra +} \equiv D_{\ul\alpha}^+ q^{\ra+} = 0~, \qquad
D_{\ul\alpha}^i \equiv (D_\alpha^i, \bar D_\dalpha^i)~.
\end{align}
This $q^+$ hypermultiplet is the natural off-shell matter multiplet
of harmonic superspace and is defined globally on the $S^2$.
The most general two-derivative action is
\begin{align}
\int \rd u\, \int \rd^4x\, \rd^4\q^+ \Big(\frac{1}{2} \Omega_{\ra \rb} \,q^{\ra+} D^{++} q^{\rb +} + H^{+4}(q^+, u^\pm)\Big)
\end{align}
with $H^{+4}(q^+, u^\pm)$ an arbitrary real function \cite{GIOS:HK1, GIOS:HK2}. The striking
resemblance to a Hamiltonian system was explained in \cite{GIOS:HK2} (see also \cite{GO:Hamiltonian}),
where it was shown that any hyperk\"ahler manifold locally possesses a
Hamiltonian structure.

For projective superspace,
one uses the same auxiliary manifold $\mathbb CP^1 \cong S^2$ as in harmonic superspace,
but superfields are taken to be holomorphic on an open domain of
$\mathbb CP^1$ rather than globally
defined on $S^2$, which is the main distinction between the two
approaches.\footnote{We follow the projective superspace conventions of
\cite{Butter:CSG4d.Proj} (similar to those of
\cite{KLRT-M1, KLRT-M2, KT-M:DiffReps, Kuzenko:DualForm}), to which we refer for definitions,
notations, and further references.}
We denote the auxiliary coordinates by $v^{i\pm}$ to distinguish
them from harmonic superspace. Sigma models are described by $n$ complex arctic superfields
$\U^{I+}$ and their antarctic conjugates $\breve \U^{\bar I+} = \widetilde {\U^{I+}}$.
These are Grassmann analytic, $D_{\ul\alpha}^+ \U^{I+} = 0$, and
holomorphic on an open domain of $\mathbb CP^1$, $D^{++} \U^{I+} = 0$.
Arctic multiplets are taken to be holomorphic near the north pole
and antarctics are holomorphic near the south pole.
The combination of $\U^{I+}$ and $\breve \U^{\bar I+}$, collectively
known as a polar multiplet, serves as the general off-shell matter multiplet
in projective superspace, and the most general two-derivative action is
given by \cite{LR88} (see also \cite{Kuzenko:CommentsSigma} for a recent
discussion)
\begin{align}\label{eq:ProjSigmaFlat}
-\frac{1}{2\pi} \oint_\cC v_i^+ \rd v^{i+} \int \rd^4x\, \rd^4\q^+ \,
	\cF^{++}(\U^+, \breve \U^+, v^{i+})~,
\end{align}
involving an arbitrary real function $\cF^{++}$ that is homogeneous
of weight two in its parameters,
$\cF^{++}(\l \U^+, \l \breve \U^+, \l v^{i+}) = \l^2 \cF^{++}(\U^+, \breve \U^+, v^{i+})$.
This function can be
interpreted as the generating function for symplectic transformations
on the hyperk\"ahler manifold \cite{LR:Prop} (see also the pioneering
work of \cite{HiKLR} and the recent approach of \cite{APSV:LinearHK})
and turns out to possess a simple relationship to the harmonic
Hamiltonian $H^{+4}$ \cite{Butter:HarmProj}.

In both the projective and harmonic cases, the data necessary
to define the hyperk\"ahler geometry are encoded in their respective
superspace Lagrangians. In evaluating the component actions,
one must find a way to eliminate the infinite number of
auxiliary fields. In doing so, one generates a
hyperk\"ahler metric and associated complex structures that
describe the sigma model parametrized by the physical scalars.
In principle
this can be done explicitly for a specific generating function, although
only certain classes of hyperk\"ahler metrics have been explicitly constructed
in this way.
Once the auxiliaries are eliminated and the geometric data
constructed, the component action takes its final form.
These are distinct steps because the component
reduction is just as easily done for a general geometry as a specific one,
since one can treat the conversion from generating function to hyperk\"ahler
geometry completely formally. Our concern in this paper will be with
the general action and with the formal elimination of the auxiliaries; we will
return to the subject of specific solvable classes in the conclusion.

What about locally supersymmetric theories? In this case, there are two
options, depending on whether one couples minimally to conformal supergravity
or to Poincar\'e supergravity with a canonical Einstein-Hilbert term.
In the $\cN=1$ setting, a sigma model coupled to conformal supergravity is given by
\begin{align}\label{eq:N1ActionCurved}
\int \rd^4x \, \rd^2 \q\, \rd^2\bar \q\, E\, K(\phi,\bar\phi)~, \qquad
E = \textrm{sdet}\, E_M{}^A~, 
\end{align}
except now the superconformal
algebra includes both dilatations $(\bbD)$ and
chiral $\gU{1}_R$ rotations $(\bbA)$
under which $\phi^\ra$ and $K$ must transform as
\begin{align}
\delta \phi^\ra = \L_{\trD} \chi^\ra + \frac{2i}{3} \L_{\trA} \chi^\ra~, \qquad
\delta K = \delta \phi^\ra K_\ra + \HC = 2 \L_\trD K~.
\end{align}
The chiral function $\chi^\ra(\phi)$ describes a homothetic conformal
Killing vector,
\begin{align}\label{eq:HomoCKV}
\nabla_\rb \chi^\ra = \delta_\rb{}^\ra~, \qquad \nabla_{\bar \rb} \chi^\ra = 0~,
\end{align}
where $\nabla_\ra$ and $\nabla_{\bar \ra}$ are the target space covariant derivatives.
The K\"ahler potential is $K \equiv \chi^\ra \chi^{\bar \rb} g_{\ra \bar \rb}$
and describes a \emph{K\"ahler cone} \cite{GR}.
At the component level, the Lagrangian includes a
contribution from the Ricci scalar of the form $\frac{1}{6} \cR K$.
Usually one prefers instead a canonically-normalized Einstein-Hilbert term,
which can be achieved by imposing the dilatation gauge $K = -3$ and
simultaneously fixing the $\gU{1}_R$ symmetry. This eliminates two scalar
fields from the sigma model and converts the K\"ahler cone of dimension $2n$
into a Hodge-K\"ahler manifold of dimension $2(n-1)$.

In the $\cN=2$ setting, a similar picture emerges. A sigma model coupled
to $\cN=2$ conformal supergravity must be a hyperk\"ahler cone \cite{dWKV}
(see \cite{deWKV:Rigid, GR} for the rigid superconformal case and
\cite{SezginTanii} for a discussion in general dimensions),
which is a hyperk\"ahler manifold possessing a homothetic conformal
Killing vector and $\SU{2}_R$ isometries that rotate the complex structures;
such spaces are also known as Swann bundles \cite{Swann}. There
exists a one-to-one correspondence between $4n$-dimensional hyperk\"ahler cones and
$4(n-1)$-dimensional quaternion-K\"ahler manifolds \cite{Swann} (see 
also \cite{Galicki} as well as \cite{dWRV:QK, APSV:LinearQK} for
recent discussions and references). 
The component action for a hyperk\"ahler
cone coupled to conformal supergravity was given in \cite{dWKV},
where its relation to the Poincar\'e supergravity-coupled quaternion-K\"ahler
action of \cite{BaWi:QK} was also discussed:
the elimination of four scalars in the target space comes
from fixing the dilatation and ${\rm SU}(2)_R$ gauges.

Our goal in this paper is to reproduce in a systematic way
the component action of \cite{dWKV}
for a hyperk\"ahler cone coupled to conformal supergravity directly from
(curved) projective superspace.\footnote{The corresponding actions
in harmonic superspace were discussed in \cite{BaGIO}.
The relation between the unconstrained harmonic potentials and
general quaternion-K\"ahler geometry was established in \cite{GIO:QK}.
The derivation of the bosonic Lagrangian from harmonic superspace appeared
in \cite{IV:QuatMetrics}.}
For this reason, aside from a brief discussion in the conclusion,
we will always be referring to conformal supergravity when we mention
locally supersymmetric sigma models.
To do this, we require a covariant approach to supergravity-matter systems
employing projective superspace. Such an approach was developed in 4D
by Kuzenko, Lindstr\"om, Ro\v{c}ek, and Tartaglino-Mazzucchelli
\cite{KLRT-M1, KLRT-M2, KT-M:DiffReps, Kuzenko:DualForm} using
conventional $N=2$ superspace (in turn based on the work of Kuzenko and
Tartaglino-Mazzucchelli in 5D \cite{KT-M:5DSugra1, KT-M:5DSugra2, KT-M:5DSugra3}).
This approach was later extended in
\cite{Butter:CSG4d.Proj}, which clarified a number of issues and
changed the superspace geometry to conformal superspace \cite{Butter:CSG4d_2},
which has a close relationship with the superconformal tensor calculus.
For this paper, we will employ the conventions and projective superspace
geometry of \cite{Butter:CSG4d.Proj}, but one could also employ those of
\cite{KLRT-M1, KLRT-M2, KT-M:DiffReps, Kuzenko:DualForm}.
The hypermultiplet action is given in curved superspace by 
\begin{align}\label{eq:ProjSigmaCurved}
-\frac{1}{2\pi} \oint_\cC \rd\tau \int \rd^4x\, \rd^4\q^+ \, \cE^{--}
	\cF^{++}(\U^+, \breve \U^+)~.
\end{align}
We explicitly parametrize the $\SU{2}$ contour $\cC$ by
the coordinate $\tau$, and the measure $\cE^{--}$ is a
superdeterminant of the relevant superspace
vielbein.\footnote{In the flat space limit,
$\cE^{--} = v_i^+ \rd v^{i+} / \rd \tau$ so that
$\rd \tau \,\cE^{--}$ reduces to the flat measure
$v_i^+ \rd v^{i+}$. Further details can be found in
\cite{Butter:CSG4d.Proj}.}
The generating function $\cF^{++}$ possesses no explicit
dependence on $v^{i+}$: this generalizes to curved superspace
the superconformal version of \eqref{eq:ProjSigmaFlat},
constructed originally in flat space \cite{Kuzenko:5Dallthat}
and ensures that the target space describes a hyperk\"ahler
cone \cite{Kuzenko:SPH}.
The major barrier to this calculation is that the usual method of eliminating
the infinite number of auxiliary fields in the flat superspace action \eqref{eq:ProjSigmaFlat}
depends on introducing an intermediate $\cN=1$ superspace.
This is easy to do in flat superspace and has been accomplished
recently in AdS superspace
\cite{KT-M:CFlat5D, KT-M:CFlat4D, BKLT-M:AdSPro, BuKuTM:SigmaAdS3},
but it is quite daunting in a general curved geometry.\footnote{The
approach of reducing curved $\cN=2$ superfields to $\cN=1$ superfields
has been discussed in \cite{GKLR:N2_N1, Labastida}, but its application in
this case would seem to be very difficult.} Instead, we will take
inspiration from rigid harmonic superspace
and proceed directly from $\cN=2$ superspace to the component action.
The elimination of the auxiliary fields will seem rather
different at first glance from the $\cN=1$ approach,
but will actually involve solving
the same set of equations; this implies that
the coupling to conformal supergravity will in no way affect the
elimination of the hypermultiplet auxiliaries.\footnote{In the
quaternion-K\"ahler case, this is more subtle.
As is evident from the corresponding harmonic \cite{GIO:QK} and
projective \cite{APSV:LinearQK} descriptions, a hypermultiplet
compensator plays the role of the effective auxiliary variable
for the on-shell hypermultiplet superfields.}
Along the way, we will derive explicit formulae in projective superspace
for all of the geometric quantities necessary for describing the
hyperk\"ahler cone and its sigma model.

This paper is laid out as follows. Section \ref{sec:HKN1Red}
provides a review of how hyperk\"ahler geometry can be derived
from flat projective superspace via the $\cN=1$ superspace method.
Although this method 
seems to be useful mainly for rigid supersymmetric spaces
such as Minkowski or AdS,
many of the same formulae and notation will reoccur in later sections,
so some familiarity will be necessary.
In section \ref{sec:N2super}, we describe how the full $\cN=2$ superfield
equations of motion lead to on-shell $\cN=2$ hypermultiplets whose
target space is a hyperk\"ahler cone, and we derive all the geometric
data we will need from projective superspace. We then describe in section
\ref{sec:Blocks} how to restrict the $\cN=2$ equations of motion to only
the auxiliary sector, eliminating the infinite tail of auxiliary fields
while keeping the physical fields off-shell.
In section \ref{sec:CompActonRigid} we test this approach by deriving
the component action in the rigid supersymmetric case, before addressing
the curved case in section \ref{sec:CompActonCurved}.
We comment on our results and speculate about some open questions in
the conclusion.

Two appendices are included. The first gives our conventions for vector
multiplets and their component fields (these conventions were absent
in \cite{Butter:CSG4d.Proj}). The second provides some technical
details necessary to calculate the final component action.

\section{A review of hyperk\"ahler geometry from flat projective superspace}\label{sec:HKN1Red}
Let us begin by reviewing how projective superspace permits
the construction of hyperk\"ahler sigma models in flat space.
This material is well-known and we refer to
the lecture notes \cite{Kuzenko:Lectures} as well as
\cite{LR:Prop} and \cite{Kuzenko:CommentsSigma, Kuzenko:SigmaDuality}
for further details and the relevant references.

We begin with the flat projective superspace Lagrangian $\cF^{++}$ 
\begin{align}
\cF^{++} = \cF^{++}(\U^+, \breve \U^+, v^{i+})
\end{align}
depending on arctic multiplets $\U^{I+}$, antarctic multiplets
$\breve \U^{\bar I+}$ and the coordinate $v^{i+}$. It
is analytic and holomorphic, $D_{\ul\alpha}^+ \cF^{++} = D^{++} \cF^{++} = 0$,
by construction. The action is 
\begin{align}
S &= -\frac{1}{2\pi} \oint_\cC v_i^+ \rd v^{i+} \int \rd^4x \, \rd^4\q^+\, \cF^{++}
	= -\frac{1}{2\pi} \oint_\cC v_i^+ \rd v^{i+} \int \rd^4x \, (D^-)^4 \cF^{++}~.
\end{align}
This can be evaluated as an integral in $\cN=1$ superspace.
To do this, recast all superfields so that they
depend solely on the complex coordinate $\z = v^{\2+} / v^{\1+}$
rather than $v^{\1+}$ and $v^{\2+}$ separately.
For example, one introduces a new arctic superfield $\U^I(\z)$,
\begin{align}
\U^I := \frac{1}{v^{\1+}} \U^{I+} = \Phi^I + \z \Sigma^I + \sum_{n=2}^\infty \z^n \U_n^I~.
\end{align}
If we interpret the components in this expansion as $\cN=1$ superfields,
then $\Phi^I$ is chiral and $\Sigma^I$ is complex linear, while the
infinite tail of superfields $\U_n^I$ are unconstrained $\cN=1$ superfields.
The antarctic superfield possesses a similar expansion
\begin{align}
\breve \U^{\bar I} = \frac{1}{v^{\2+}} \breve \U^{I+}
	= \bar \Phi^{\bar I} - \frac{1}{\z} \bar \Sigma^{\bar I}
	+ \sum_{n=2}^\infty (-1)^n \z^{-n} \bar \U_n^{\bar I}~.
\end{align}
Rewriting the projective superspace Lagrangian as
$\cF^{++} = i v^{\1+} v^{\2+} \cF(\z)$, we find
\begin{align}\label{eq:ActionCNM}
S & = \int \rd^4x \,\rd^2\q_\1 \,\rd^2\bar\q^\1 \mathscr{L}~, \qquad
\mathscr{L} = \oint_\cC \frac{\rd \z}{2\pi i \z} \cF(\U, \breve \U, \z)~,
\end{align}
in terms of the $\cN=1$ superspace Lagrangian $\mathscr{L}$. Note
that while $\cF^{++}$ had to be
homogeneous of weight two in its parameters, no such constraint
is imposed on $\cF$.

Because the superfields $\U_n^I$ with $n \geq 2$ are unconstrained $\cN=1$
superfields, their equations of motion are purely
algebraic\footnote{These equations of motion were described in \cite{LR88}.
They were given explicitly in \cite{GK:CNM, GK:Cotangent}, for a class
of $\z$-independent functions $\cF$ whose resulting
hyperk\"ahler manifolds were shown to be cotangent bundles of K\"ahler
manifolds, building off a related observation in \cite{Kuzenko:DP}.
The full explicit form, discussed here, appeared later in \cite{LR:Prop}.}
\begin{align}
\oint_\cC \frac{\rd \z}{2\pi i \z} \frac{\pa \cF}{\pa \U^I} \z^{n} = 0 ~, \qquad
\oint_\cC \frac{\rd \z}{2\pi i \z} \frac{\pa \cF}{\pa \breve \U^{\bar I}} \,\z^{-n} = 0 ~,
	\qquad n \geq 2~.
\end{align}
Imposing these, the Lagrangian
becomes solely a function of $\Phi^I$, $\Sigma^I$ and their complex
conjugates.
Now perform a duality transformation, exchanging the complex linear superfield
$\Sigma^I$ for a chiral superfield $\Psi_I$,
\begin{align}
S & = \int \rd^4x \,\rd^2\q_\1 \,\rd^2\bar\q^\1 \Big(\mathscr{L}
	- \Sigma^I \Psi_I - \bar \Sigma^I \bar \Psi_I\Big)~.
\end{align}
The equation of motion for $\Psi_I$ enforces the complex linearity of
$\Sigma^I$, recovering \eqref{eq:ActionCNM}.
Alternatively, we can eliminate $\Sigma^I$ using its own equation of motion,
effecting a Legendre transformation
\begin{align}
K(\Phi, \bar\Phi, \Psi, \bar \Psi) = \mathscr{L}(\Phi, \bar \Phi, \Sigma, \bar \Sigma)
	- \Sigma^I \Psi_I - \bar \Sigma^I \bar \Psi_I~.
\end{align}
The resulting function $K$ is a K\"ahler potential
with complex coordinates $\phi^\ra = (\Phi^I, \Psi_I)$.

This K\"ahler potential describes a hyperk\"ahler manifold.
In addition to the manifest $\cN=1$ supersymmetry,
there is a hidden second supersymmetry on-shell,
which manifests in $\cN=1$ language as \cite{HuKLR}
\begin{align}
\delta \phi^\ra = \omega^{\ra \rb} \,\bar\rho_\dalpha \,\bar D^\dalpha K_\rb
	= \omega^{\ra}{}_{\bar \rb} \,\bar\rho_\dalpha \,\bar D^\dalpha \bar\phi^{\bar \rb}
\end{align}
for constant $\bar \rho_\dalpha \equiv \bar\xi_\dalpha{}^\2$, the second
supersymmetry parameter. The tensor $\omega^{\ra \rb}$ is antisymmetric,
covariantly constant, and obeys
$\omega^\ra{}_{\bar \rb} \omega^{\bar \rb}{}_\rc= \omega^{\ra \rb} \omega_{\rb \rc} = - \delta^\ra{}_\rc$.
The special coordinates $\phi^\ra = (\Phi^I, \Psi_I)$ are Darboux
coordinates for which \cite{LR:Prop, Kuzenko:SigmaDuality}
\begin{align}
\omega^{\ra\rb} =
\begin{pmatrix}
0 & \delta^I{}_J \\
-\delta_I{}^J & 0
\end{pmatrix}~, \qquad \omega_{\ra\rb} =
\begin{pmatrix}
0 & \delta_I{}^J \\
-\delta^I{}_J & 0
\end{pmatrix}~.
\end{align}
The presence of such an antisymmetric covariantly constant tensor ensures that the K\"ahler
manifold is actually hyperk\"ahler, with a triplet of closed hyperk\"ahler two-forms
$\Omega_{ij}$,
\begin{subequations}
\begin{align}
\Omega_{\1\1} &= \frac{1}{2} \omega_{\ra \rb} \,\rd \phi^\ra \wedge \rd \phi^{\rb}
	= \rd \Phi^I \wedge \rd \Psi_I~, \\
\Omega_{\1\2} &= \Omega_{\2\1} = \frac{1}{2} g_{\ra \bar \rb}\, \rd \phi^\ra \wedge \rd \bar\phi^{\bar\rb}~, \\
\Omega_{\2\2} &= \frac{1}{2} \bar\omega_{\bar\ra \bar\rb} \,
	\rd \bar\phi^{\bar\ra} \wedge \rd \bar\phi^{\bar\rb}
	= \rd \bar\Phi^{\bar I} \wedge \rd \bar \Psi_I~.
\end{align}
\end{subequations}
The three hyperk\"ahler two-forms are related to three covariantly constant
complex structures
$(\cJ_{ij})^\mu{}_\nu = g^{\mu\rho} (\Omega_{ij})_{\rho \nu}$,
\begin{align}\label{eq:Jij}
(\cJ_{\1\1})^\mu{}_\nu = 
\begin{pmatrix}
0 & 0 \\
\omega^{\bar \ra}{}_{\rb} & 0
\end{pmatrix}~, \qquad
(\cJ_{\1\2})^\mu{}_\nu = 
\begin{pmatrix}
-\frac{1}{2} \delta^{\ra}{}_{\rb} & 0 \\
0 & \frac{1}{2} \delta^{\bar \ra}{}_{\bar \rb}
\end{pmatrix}~, \qquad
(\cJ_{\2\2})^\mu{}_\nu = 
\begin{pmatrix}
0 & \omega^{\ra}{}_{\bar\rb} \\
0 & 0 
\end{pmatrix}~,
\end{align}
which obey the multiplication rule
\begin{align}\label{eq:Jmult1}
\cJ_{ij} \cJ_{kl}
	&= \frac{1}{2} \eps_{i(k} \eps_{l)j}
	+ \frac{1}{2} \Big(\eps_{i(k} \cJ_{l)j} + \eps_{j(k} \cJ_{l)i}\Big)~.
\end{align}
Introducing $\cJ_A := -i (\tau_A)^i{}_j \cJ^i{}_j$, $A=1,2,3$ with the Pauli
matrices $(\tau_A)^i{}_j$, the multiplication rule becomes that of the quaternions,
$\cJ_A \cJ_B = -\delta_{AB} + \eps_{ABC} \cJ_C$.
The complex structures in this form are given by
\begin{align}
(\cJ_1)^\mu{}_\nu =
\begin{pmatrix}
0 & -i \omega^\ra{}_{\bar \rb} \\
i \omega^{\bar \ra}{}_{\rb} & 0
\end{pmatrix}~, \quad
(\cJ_2)^\mu{}_\nu =
\begin{pmatrix}
0 & \omega^\ra{}_{\bar \rb} \\
\omega^{\bar \ra}{}_{\rb} & 0
\end{pmatrix}~, \quad
(\cJ_3)^\mu{}_\nu =
\begin{pmatrix}
i \delta^\ra{}_\rb & 0 \\
0 & -i \delta^{\bar \ra}{}_{\bar \rb}
\end{pmatrix}~.
\end{align}
$\cJ_3 = -2i \cJ_{\1\2}$ is the complex structure associated
with the manifest $\cN=1$ supersymmetry.

We will eventually be interested in the case where the model
is superconformal \cite{Kuzenko:SPH}. This amounts to the condition
that $\cF^{++}$ is homogeneous of degree two,
$2 \cF^{++} = \cF_I^+ \U^{I+} + \cF_{\bar I}^+ \breve \U^{\bar I+}$,
which is equivalent to requiring the projective Lagrangian $\cF^{++}$
to be independent of $v^{i+}$.
In contrast to the $\cN=1$ situation, there is no requirement that $\cF^{++}$ be separately
homogeneous in $\U^{I+}$.  This is because arctic and antarctic multiplets
are both inert under $\gU{1}_R$, so there is no superconformal symmetry
that distinguishes between them.\footnote{Moreover, imposing a separate homogeneity condition
for the arctics and antarctics is equivalent to assigning an additional
global $\gU{1}$ isometry to the projective Lagrangian, which descends
to the hyperk\"ahler manifold as a new triholomorphic isometry. Such an
isometry is not generically present in hyperk\"ahler cones. This issue was
already noted in the context of $3D$ sigma models with $(3,0)$ AdS
supersymmetry \cite{BuKuTM:SigmaAdS3}.}

Now the K\"ahler potential $K$ turns out to possess a chiral
homothetic conformal Killing vector (CKV) $\chi^\ra$ obeying \eqref{eq:HomoCKV},
implying that the K\"ahler potential can be chosen (up to a K\"ahler
transformation) as $K = \chi^\ra \chi_\ra$.
For the Darboux coordinate system, the homothetic conformal
Killing vector takes the simple form $\chi^\ra = (\Phi^I, \Psi_I)$.
The presence of $\chi^\ra$ ensures that the
hyperk\"ahler manifold is actually a hyperk\"ahler cone.
In addition to the two supersymmetry transformations, it admits
a full set of $\cN=2$ superconformal transformations,
including dilatation and $\SU{2}_R$ transformations. These
manifest as \cite{Kuzenko:SigmaDuality}
\begin{align}
\delta \phi^\ra =
	\L_\trD \chi^\ra
	+ \l^{\1\2} \chi^\ra
	- \l^{\2\2} \omega^{\ra \rb} \chi_\rb~,
\end{align}
where $\L_\trD$ is the scale parameter and
$\l^{i}{}_j$ is the $\SU{2}_R$ transformation parameter.
The fields $\phi^\ra$ are inert under $\gU{1}_R$.

\section{Hyperk\"ahler geometry and on-shell $\cN=2$ superfields}\label{sec:N2super}
Our goal in this section is to establish the geometric properties of
the target space geometry of \eqref{eq:ProjSigmaCurved}
(including the results of the previous section)
without explicitly reducing to $\cN=1$ superspace.
This is necessary in order to derive the component action in
the presence of supergravity, where an $\cN=1$ superspace is not readily
available. Our goal will be to reduce the arctic superfields to
on-shell $\cN=2$ superfields: we will define these as $\phi^\ra = (\Phi^I, \Psi_I)$
in analogy to the $\cN=1$ superfields of the previous section.

Our starting point is equivalent to that discussed in \cite{GK2} and \cite{LR:Prop}:
we will analyze the full $\cN=2$ superfield equations of motion.
If the action \eqref{eq:ProjSigmaCurved}
is stationary under arbitrary variations of $\U^{I+}$,
\begin{align}\label{eq:N2Eoms} 
-\frac{1}{2\pi} \oint_\cC \rd \tau \int \rd^4x \, \rd^4\q^+ \cE^{--} \,\delta \U^{I+} \frac{\pa \cF^{++}}{\pa \U^{I+}} = 0~, 
\end{align}
then $\pa \cF^{++} / \pa \U^{I+}$ must itself be an arctic
superfield. (The integral vanishes in this case since the contour $\cC$
can be retracted to the north pole without encountering singularities.)
This result holds both in the rigid and locally supersymmetric situations.
This leads one to introduce superfields
$\Gamma_I^+$ and $\breve\G_{\bar I}^+$, defined by
\begin{align}\label{eq:defGamma}
\Gamma_I^+ := -i \frac{\pa \cF^{++}}{\pa \U^{I+}}~, \qquad
\breve\Gamma_{\bar I}^+ := i \frac{\pa \cF^{++}}{\pa \breve\U^{\bar I+}}~.
\end{align}
The equations of motion require $\Gamma_I^+$ and $\breve \Gamma_{\bar I}^+$
to be, respectively, arctic and antarctic.\footnote{These can be
interpreted as dual superfields; see \cite{KuLivU} for
a discussion of polar-polar duality.}
The superfields
$\Phi^I$ and $\Psi_I$ correspond to the leading terms in the
$\z$ expansions of $\U^{I+}$ and $\Gamma_I^+$,
\begin{align}\label{eq:ArcticExp}
\U^{I+} = v^{\1+} \Big(\Phi^I + \cO(\z)\Big)~, \qquad
\Gamma_I^+ = v^{\1+} \Big(\Psi_I + \cO(\z)\Big)
\end{align}
and can be defined equivalently via contour
integration,\footnote{In defining the $\cN=2$ superfields
$\Phi^I$ and $\Psi_I$, we implicitly choose to work in the central gauge
of projective superspace as discussed in \cite{Butter:CSG4d.Proj}.}
\begin{align}\label{eq:cdefPhiPsi}
\Phi^I = \oint_\cC \frac{\rd \z}{2 \pi i \z} \frac{\U^{I+}}{v^{\1+}}~, \qquad
\Psi_I = \oint_\cC \frac{\rd \z}{2 \pi i \z} \frac{\G^{+}_I}{v^{\1+}}~.
\end{align}
We will assume that the contour $\cC$ winds around the north pole
(and thus the south pole as well) exactly once and that the arctic
(antarctic) multiplets possess no singularities in the northern
(southern) chart bounded by $\cC$.
Consistency with the flat space $\cN=1$ analysis implies that
the on-shell $\cN=2$ superfields $\U^{I+}$ and $\breve \U^{\bar I+}$ must be
given by power series in the $\cN=2$ superfields $\phi^\ra = (\Phi^I, \Psi_I)$ and
their complex conjugates. Because no fields of the conformal supergravity
multiplet appear in the solution of the power series, the sigma model
for local supersymmetry will be identical as for the rigid superconformal case.

The $\cN=2$ superconformal transformations of $\Phi^I$ and $\Psi_I$ can be derived from their definitions
\eqref{eq:cdefPhiPsi},
\begin{align}
\delta \Phi^I = \oint_\cC \frac{\rd \z}{2 \pi i \z} \frac{\delta \U^{I+}}{v^{\1+}}~, \qquad
\delta \Psi_I = \oint_\cC \frac{\rd \z}{2 \pi i \z} \frac{\delta \G^{+}_I}{v^{\1+}}~,
\end{align}
where $\delta$ consists of any local (super)symmetry transformation.
Consistency dictates that
$\delta \U^{I+}$ can equivalently be calculated by
\begin{align}
\delta \U^{I+} = \delta \phi^\ra \pa_{\ra} \U^{I+} + \delta \bar\phi^{\bar \ra} \pa_{\bar \ra} \U^{I+}~.
\end{align}
Using only these results, let us briefly discuss how the geometry of the target space follows.

\subsection{Hyperk\"ahler geometry}
Here we follow closely the approach of \cite{LR:Prop}.
The two-form\footnote{For target space
quantities such as $\Omega^{++}$, we follow the standard conventions for
differential forms rather than the superspace conventions of e.g. \cite{WessBagger}.}
\begin{align}
\Omega^{++} &= \rd \U^{I+} \wedge \rd\G_I^+
	= \rd \breve\U^{\bar I+} \wedge \rd\breve\G_{\bar I}^{+}
\end{align}
is both arctic and antarctic when the full $\cN=2$ equations of motion
are imposed, so it must be globally-defined on the
auxiliary manifold. This means it is given by
$\Omega^{++} = \Omega_{ij} \,v^{i+} v^{j+}$ for a triplet of two-forms
$\Omega_{ij}$. It is obvious from its definition that
$\Omega_{\1\1} = \rd \Phi^I \wedge \rd \Psi_I$
and
$\Omega_{\2\2} = \rd \bar\Phi^{\bar I} \wedge \rd \Psi_{\bar I}$,
while $\Omega_{\1\2}$ can only be a $(1,1)$ form,
$\Omega_{\1\2} = \frac{1}{2} \rd \phi^\ra \wedge \rd \bar\phi^{\bar \rb} g_{\ra \bar \rb}$,
for some tensor $g_{\ra \bar \rb}$.
The expression for $\Omega^{++}$ can then be written as
\begin{align}
\Omega^{++} &= v^{\1+} v^{\2+} \left(
	\frac{1}{2\z} \rd \phi^\ra \wedge \rd \phi^\rb \, \omega_{\ra \rb}
	+ \rd \phi^\ra \wedge \rd\bar\phi^{\bar \rb} g_{\ra \bar \rb}
	+ \frac{\z}{2} \rd \bar\phi^{\bar \ra} \wedge \rd \bar\phi^{\bar \rb} \, \omega_{\bar \ra \bar \rb} \right)~,
\end{align}
identifying
\begin{subequations}\label{eq:OmegaComps}
\begin{align}
\omega_{\ra\rb} &= \frac{1}{v^{\1+} v^{\1+}} (\pa_\ra \U^{I+} \pa_\rb \G_I^+
	- \pa_\ra \G_I^{+} \pa_\rb \U^{I+} )~, \label{eq:defomega}\\
g_{\ra\bar\rb} &= \frac{1}{v^{\1+} v^{\2+}} (\pa_\ra \U^{I+} \pa_{\bar\rb} \G_I^+
	- \pa_\ra \G_I^{+} \pa_{\bar \rb} \U^{I+} )~, \label{eq:defg} \\
\omega_{\bar\ra\bar\rb} &= \frac{1}{v^{\2+} v^{\2+}} (\pa_{\bar\ra} \U^{I+} \pa_{\bar\rb} \G_I^+
	- \pa_{\bar\ra} \G_I^{+} \pa_{\bar\rb} \U^{I+} )~.
\end{align}
\end{subequations}
These relations also hold upon replacing $\U^{I+} \rightarrow \breve \U^{\bar I+}$ and
$\G_I^+ \rightarrow \breve \G_{\bar I}^+$.

Because $\Omega^{++}$ is closed, it follows that both
$\omega_{\ra\rb}$ and $g_{\ra \bar \rb}$ must be closed
when viewed respectively as $(2,0)$ and $(1,1)$ forms.
(The closure of $\omega_{\ra\rb}$ is obvious in the Darboux coordinates.)
The closure of $g_{\ra\bar\rb}$ implies that it
must be the second derivative of a function $K$.
This function can be chosen as in the explicit $\cN=1$ reduction as
\begin{align}\label{eq:defK}
K = \mathscr{L} - \Sigma^I \Psi_I - \bar \Sigma^{\bar I} \bar\Psi_{\bar I}~, \qquad
\mathscr{L} := \oint_\cC \frac{\rd \z}{2\pi i \z} \frac{\cF^{++}}{i v^{\1+} v^{\2+}}~.
\end{align}
Identifying $\Sigma^I$ and $\Sigma_I$ as the second terms in the expansions
of $\U^{I+}$ and $\G_I^+$, one can show
\begin{align}\label{eq:paK}
\pa_{\ra} K &= \Sigma_I \pa_\ra \Phi^I - \Sigma^I \pa_\ra \Psi_I~, \eol
\pa_{\bar \rb} \pa_\ra K &= \pa_{\bar \rb} \Sigma_I \pa_\ra \Phi^I - \pa_{\bar \rb} \Sigma^I \pa_\ra \Psi_I
	= \oint_\cC \frac{\rd \z}{2\pi i \z} \frac{1}{\z} (\pa_\ra \U^I \pa_{\bar \rb} \G_I
		- \pa_\ra \G_I \pa_{\bar \rb} \U^I)
	\equiv g_{\ra \bar \rb}~.
\end{align}

Let us next establish that $\omega_{\ra \rb} = g_{\ra \bar \rc} \omega^{\bar \rc \bar \rd} g_{\bar \rd \rb}$
where $\omega^{\bar\ra\bar\rb}$ is the inverse of $\omega_{\bar\ra \bar\rb}$, given by
\begin{align}
\omega^{\bar\ra \bar\rb} =
\begin{pmatrix}
0 & \delta^{\bar I}{}_{\bar J} \\
-\delta_{\bar I}{}^{\bar J} & 0
\end{pmatrix}~, \qquad \omega^{\bar\ra \bar\rb} \omega_{\bar\rb \bar\rc} = -\delta^{\bar\ra}{}_{\bar\rc}~.
\end{align}
A proof of this follows by using the explicit antarctic
expression for $g_{\ra \bar\rb}$ and writing
\begin{align}
g_{\ra \bar \rc} \omega^{\bar \rc \bar \rd} g_{\bar \rd \rb} &= -\frac{1}{(v^{\1+} v^{\2+})^2}
(\pa_\ra \breve \U^{\bar I+} \pa_{\bar K} \breve\G_{\bar I}^+ -
 \pa_\ra \breve \G_{\bar I}^+ \pa_{\bar K} \breve\U^{\bar I+})
(\pa^{\bar K} \breve \U^{\bar J+} \pa_\rb \breve\G_{\bar J}^+ -
 \pa^{\bar K} \breve \G_{\bar J}^+ \pa_\rb \breve\U^{\bar J+})
 \eol & \quad
+ \frac{1}{(v^{\1+} v^{\2+})^2}
(\pa_\ra \breve \U^{\bar I+} \pa^{\bar K} \breve\G_{\bar I}^+ -
 \pa_\ra \breve \G_{\bar I}^+ \pa^{\bar K} \breve\U^{\bar I+})
(\pa_{\bar K} \breve \U^{\bar J+} \pa_\rb \breve\G_{\bar J}^+ -
 \pa_{\bar K} \breve \G_{\bar J}^+ \pa_\rb \breve\U^{\bar J+})~, \eol
\pa_{\bar I} &:= \frac{\pa}{\pa \bar\Phi^{\bar I}}~, \qquad \pa^{\bar I} = \frac{\pa}{\pa \bar \Psi_{\bar I}}~.
\end{align}
This expression must be independent of $v^{i+}$, so we can discard all
terms that go as negative powers of $\z$.
Using
\begin{align}
\pa_{\bar J} \breve \U^{\bar I+} &= v^{\2+} (\delta^{\bar I}_{\bar J} + \cO(1/\z))~, \qquad
\pa^{\bar J} \breve \G_{\bar I}^+ = v^{\2+} (\delta_{\bar I}^{\bar J} + \cO(1/\z))~, \eol
\pa_\ra \breve \U^{\bar I+} &\sim \pa_\ra \breve \G^{\bar I+} \sim
\pa^{\bar J} \breve \U^{\bar I+} \sim
\pa^{\bar J} \breve \G_{\bar I}^+ \sim v^{\2+} \cO(1/\z)~,
\end{align}
we see that the only terms that contribute are
\begin{align}\label{eq:GoG}
g_{\ra \bar \rc} \omega^{\bar \rc \bar \rd} g_{\bar \rd \rb}
= \frac{1}{(v^{\1+})^2} \Big(
\pa_\ra \breve \U^{\bar I+} \pa_\rb \breve\G_{\bar I}^+
-\pa_\ra \breve \G_{\bar I}^+  \pa_\rb \breve\U^{\bar I+}
\Big)
= \omega_{\ra\rb}~,
\end{align}
which is what we wished to establish.
This result actually guarantees that $g_{\ra\bar\rb}$ is invertible
because both sides of \eqref{eq:GoG} must have non-vanishing determinant.
In other words, the non-degeneracy of the metric $g_{\ra \bar \rb}$ is
implied if we can solve the equations \eqref{eq:N2Eoms} completely in
terms of the coordinates $\Phi^I$ and $\Psi_I$.
This equality also
guarantees that $\omega_{\ra\rb}$ is covariantly constant.
It follows that the manifold is hyperk\"ahler.

\subsection{Gauged isometries from projective superspace}
Suppose that the projective superspace Lagrangian possesses some gauge
invariance -- that is, the arctic multiplets possess gauged holomorphic
isometries of the form
\begin{align}
\delta_g \U^{I+} = \l^r \cJ_r^{I+}~, \qquad \cJ_r^{I+} = \cJ_r^{I+}(\U^+, v^{i+})~,
\end{align}
for an arctic function $\cJ_r^{I+}$, and similarly for the
antarctic multiplets, with $r$ labelling the adjoint
representation of the gauge group.
Let us show how these descend to triholomorphic isometries in
the hyperk\"ahler manifold, rederiving the results
of \cite{HuKLR}.\footnote{A treatment based on $\cN=1$
superspace methods can be found in
\cite{GonzalezRey:FR2} and \cite{Kuzenko:Superpotentials} for
flat 4D and 5D cases, and
\cite{BKLT-M:AdSPro, BuKuTM:SigmaAdS3} for AdS geometries.}

Because the projective Lagrangian is gauge-invariant,\footnote{It is
actually not necessary for the Lagrangian to be fully gauge invariant,
provided that one can consistently introduce a naked prepotential to
counter its gauge transformation property. As discussed in \cite{HuKLR},
the prepotential can be absorbed in a covariant framework
by introducing a fictitious multiplet that drops out of the action
(its metric vanishes),
except for its modification of the moment map. In this way,
we retain the description above with only covariant hypermultiplets
and no prepotentials. We thank Martin Ro\v{c}ek for pointing out this important subtlety.}
\begin{align}
\delta_g \cF^{++} = i \l^r (\Gamma_I^+ \cJ_r^{I+} - \breve \Gamma_{\bar I}^+ \breve \cJ_r^{\bar I+}) = 0~,
\end{align}
we may introduce a real quantity
\begin{align}\label{eq:N2KPproj}
D_r^{++} := \Gamma_I^+ \cJ_r^{I+} = \breve \Gamma_{\bar I}^+ \breve \cJ_r^{\bar I+}~.
\end{align}
This is the $\cN=2$ moment map (or Killing potential) in projective
superspace.\footnote{The $\cN=1$ formulation of the $\cN=2$
Killing potential appeared in \cite{HuKLR}.
Harmonic and projective superspace formulations appeared 
explicitly in \cite{BaGIO} and \cite{Kuzenko:Superpotentials}.}
By construction, the gauge transformation of $\G_I^+$ is
\begin{align}
\delta_g \G_I^+ \equiv \l^r \cJ_{rI}^+ =  -\l^r \pa_{I+} \cJ_r^{J+}\, \G_J^{+} ~, \qquad \pa_{I+} := \frac{\pa}{\pa \U^{I+}}
\end{align}
and so it follows that $D_r^{++}$ transforms in the co-adjoint of the gauge group,
\begin{align}
\delta_g D_r^{++}
	= \l^s \G_I^+ (\cJ_s^{J+} \pa_{J+} \cJ_r^{I+} - \cJ_r^{J+} \pa_{J+} \cJ_s^{I+})
	= -\l^s f_{sr}{}^t D_t^{++}~.
\end{align}

When the equations of motion are imposed, the gauge transformations of
the arctic fields must be manifested on the target space of
complex fields. As a consequence of the transformation properties of $\U^{I+}$
and $\G_I^+$, the two-form $\Omega^{++}$ 
is gauge-invariant. Interpreting the gauge transformation as a
target space transformation, it follows that $\cL_J \Omega^{++}$ vanishes,
ensuring that gauged isometries in projective superspace descend
to triholomorphic isometries in the target space.
To derive this explicitly, observe that for 
$\cJ_r^{I+}(\U^+, v^{i+}) \equiv v^{\1+} \cJ_r^I(\U, \z)$,
\begin{align}
\delta_g \phi^\ra = \l^r J_r^\ra~, \qquad 
J_r^I = \cJ_r^I(\Phi, 0)~, \qquad J_{r\, I} = -\pa_I J_r^J(\Phi) \,\Psi_J~.
\end{align}
Now $D_r^{++}$ must be both arctic and antarctic
and so must be given by $D_{r \,ij} v^{i+} v^{j+}$, where
\begin{align}\label{eq:N2KPcomp}
D_r^{++} &= v^{\1+} v^{\2+}
	\Big(
	\frac{1}{\z} \Lambda_r + i D_r + \z \bar\Lambda_r
	\Big)~, \eol
\Lambda_r &= \Psi_I J_r^I(\Phi)~, \qquad
D_r = -i J_r^\ra K_\ra = i J_r^{\bar \ra} K_{\bar \ra}~.
\end{align}
The quantity $D_r$ is the $\cN=1$ Killing potential and is related to the
holomorphic quantity $\Lambda_r$ and the Killing vector $J_r^\ra$ via \cite{HuKLR}
\begin{align}\label{eq:KVecFromPot1}
J_r^\ra = i g^{\ra \bar \rb} \pa_{\bar \rb} D_r = \omega^{\ra \rb} \pa_\rb \L_r~, \qquad
J_r^{\bar\ra} = -i g^{\bar\ra \rb} \pa_{\rb} D_r = \omega^{\bar\ra \bar\rb} \pa_{\bar\rb} \bar\L_r~.
\end{align}
The relations \eqref{eq:KVecFromPot1} can equivalently be written
\begin{align}
\nabla_\mu D_r{}^{ij} = - (\Omega^{ij})_{\mu\nu} J_r^\nu~.
\end{align}
From this equation, one can prove that the Killing vector $J_r^\mu$
is triholomorphic.

It is worth mentioning that if $\cF^{++}$ is independent of $v^{i+}$, so that
the target space is a hyperk\"ahler cone, $\cJ_r^{I+}$ must also be independent of $v^{i+}$
and homogeneous in $\U^{I+}$ of degree one, $\U^{J+} \pa_{J+} \cJ_r^{I+} = \cJ_r^{I+}$.
The $\cN=2$ moment map can then equivalently be defined as
\begin{align}
D_r^{++} = \frac{1}{2} \Gamma_I^+ X_r \U^{I+} - \frac{1}{2} \U^{I+} X_r \Gamma_I^+
	= \frac{1}{2} \Gamma_I^+ \cJ_r^{I+} - \frac{1}{2} \U^{I+} \cJ_{rI}^+~.
\end{align}
This can be derived from $\Omega^{++}$, replacing one $\rd$ with
$D^0$ and the other with the gauge generator $X_r$, leading to
an explicit expression for its components,
\begin{align}
D_{r\, ij} = -\frac{1}{2} (\Omega_{ij})_{\mu\nu} \chi^\mu J_r^\nu \quad \implies \quad
\L_r = -\frac{1}{2} \omega_{\ra \rb} \chi^\ra J_r{}^\rb~, \quad
D_r = -\frac{i}{2} (J_r^\ra \chi_\ra - J_r^{\bar\ra} \chi_{\bar \ra})~,
\end{align}
in terms of the homothetic conformal Killing vector $\chi^\mu$.

\subsection{Superconformal isometries}
Now let us analyze the superconformal properties of the target space,
following roughly the same approach as \cite{Kuzenko:SPH, Kuzenko:SigmaDuality},
to which we refer for further details.
The arctic multiplet $\U^{I+}$ transforms locally under dilatations
and $\SU{2}_R$ transformations as
\begin{align}
\delta \U^{I+} &= \L_\trD \bbD \U^{I+} + \l^i{}_j I^j{}_i \U^{I+} =
(\L_\trD + \l^{-+}) \U^{I+} - \l^{++} D^{--} \U^{I+}~, \eol
\l^{++} &= \l^{ij} v_i^+ v_j^+~, \quad \l^{+-} = \l^{ij} v_i^+ v_j^-~,
\end{align}
and similarly for $\breve\U^{\bar I+}$. Both are inert under $\gU{1}_R$,
and so are their dual fields $\Gamma_I^+$ and $\breve \Gamma_{\bar I}^+$.
The above transformation should map to the target space as
\begin{align}
\delta \phi^\ra &= \L_{\trD} k_{\trD}^\ra + \l^i{}_j k^j{}_i{}^\ra
	= \L_\trD k_\trD^\ra - \l^{ij} k_{ij}{}^\ra
\end{align}
for some choice of vectors $k_\trD^\ra := \bbD \phi^\ra$ and
$k_{ij}{}^\ra := I_{ij} \phi^\ra$. 
Let us recover their properties 
using projective superspace.

Superconformal invariance dictates that the projective Lagrangian
transforms as
\begin{align}
\delta \cF^{++} &= (2 \L_\trD + 2 \l^{+-}) \cF^{++} - \l^{++} D^{--} \cF^{++}~,
\end{align}
implying that $\cF^{++}$ is homogeneous of degree two in
the projective multiplets
and possesses no explicit dependence on $v^{i+}$. It follows that
the fields $\Gamma_I^+$ and $\breve \Gamma_{\bar I}^+$ transform
in the same way as $\U^{I+}$ and $\breve \U^{\bar I+}$ under
superconformal transformations.
From the definitions \eqref{eq:cdefPhiPsi} of $\Phi^I$ and $\Psi_I$, it is clear that they possess
unit dilatation weight, so we establish
$k_\trD^\ra := \bbD \phi^\ra = (\Phi^I, \Psi_I)$.
A similar calculation with $\Sigma^I$, the second component in the
$\z$-expansion of $\U^{I+}$, establishes that it also has unit dilatation weight.
This leads to
\begin{align}\label{eq:DilK}
2 K = k_\trD^\ra \pa_\ra K + k_{\trD}^{\bar \ra} \pa_{\bar \ra} K
\end{align}
using the definition \eqref{eq:defK} of the K\"ahler potential.

Next, we establish the $\SU{2}_R$ transformation properties of the fields.
Consider first the diagonal $\gU{1}$ subgroup of
$\SU{2}_R$ generated by $I^{\1}{}_\1 = -I^\2{}_\2$.\footnote{This transformation
was called the \emph{shadow chiral rotation} in \cite{Kuzenko:SPH, Kuzenko:SigmaDuality}.}
It acts as
\begin{align}
I^\1{}_\1 \phi^\ra = -\frac{1}{2} k_\trD^\ra~, \qquad
I^\1{}_\1 \bar\phi^{\bar\ra} = +\frac{1}{2} k_\trD^{\bar\ra}~.
\end{align}
Using $I^\1{}_\1 \Sigma^I = \frac{1}{2} \Sigma^I$,
it is easy to show that the K\"ahler potential \eqref{eq:defK} is inert.
This implies that $k_\trD^\ra \equiv \chi^\ra$ is a homothetic conformal Killing vector.
Now the off-diagonal $\SU{2}_R$ component $I^\1{}_\2 = I_{\2\2}$
annihilates the antichiral fields and acts on the chiral ones as
\begin{align}
I_{\2\2} \Phi^I = \frac{\pa K}{\pa \Psi_I}~, \qquad
I_{\2\2} \Psi_I = -\frac{\pa K}{\pa \Phi^I} \quad \implies \quad
I_{\2\2} \phi^\ra = \omega^{\ra \rb} K_\rb = \omega^\ra{}_{\bar \rb} \chi^{\bar \rb}~.
\end{align}
Putting these results together, we deduce that
\begin{align}
I^i{}_j \phi^\mu = (\cJ^i{}_j)^\mu{}_\nu \chi^\nu~.
\end{align}
This implies that the K\"ahler potential is
invariant under all of the $\SU{2}$ generators.

At this stage, we should point out how these isometries act on the
two-form $\Omega^{++}$. Under dilatations,
\begin{align}
\delta_{\trD} \Omega^{++} = \L_\trD \cL_\chi \Omega^{++} = 2 \L_\trD \Omega^{++} \quad \implies\quad
\cL_\chi \omega_{\ra\rb} = 2 \omega_{\ra\rb}~, \quad
\cL_\chi g_{\ra \bar \rb} = 2 g_{\ra \bar \rb}~,
\end{align}
while $\SU{2}$ transformations rotate the complex structures,
\begin{align}
\delta_\trSU \Omega^{++} &= -\l^{++} D^{--} \Omega^{++} + 2 \l^0 \Omega^{++}
	= -2 \l^k{}_i \Omega_{jk} \,v^{i+} v^{j+} \quad \implies\quad \eol
\delta_\trSU \Omega_{ij} &= -2 \l^k{}_{(i} \Omega_{j)k}~.
\end{align}
This is consistent with the target space transformations
\begin{align}\label{eq:DSU2trans}
\delta \phi^\mu = \L_\trD \chi^\mu + \l^i{}_j (\cJ^j{}_i)^\mu{}_\nu \chi^\nu~.
\end{align}
These comprise the isometries required of a hyperk\"ahler cone.

\subsection{Supersymmetry and fermion transformations}
In addition to the $2n$ complex bosons $\phi^\ra$ parametrizing the
target space, there must be $2n$ Weyl fermions. It will be convenient
for us to define the fermions to be consistent with the $\cN=1$
reduction -- that is, we will associate one left-handed Weyl fermion
with each of the $\phi^\ra$ and one right-handed Weyl fermion
with each of the $\bar\phi^{\bar \ra}$. Using the on-shell superfields
$\phi^\ra$ and $\bar\phi^{\bar\ra}$, we define (using the curved superspace
spinor derivatives $\nabla_{\alpha}^i$ and $\bar\nabla_{\dalpha i}$)
\begin{align}
\z_\alpha^\ra := \nabla_\alpha^\1 \,\phi^\ra~, \qquad
\bar\z_\dalpha^{\bar\ra} := \bar\nabla_\dalpha{}_\1 \,\bar\phi^{\bar\ra}~.
\end{align}
The set of fields
$\{\phi^\ra, \bar\phi^{\bar\ra}, \z_\alpha^\ra, \bar\z_\dalpha^{\bar \ra}\}$
constitute the on-shell field content of the supersymmetric sigma model.
Our goal in this section is to derive their supersymmetry transformations.

We first establish the action of the spinor derivatives
on the scalars $\phi^\mu$:\footnote{Such superfields $\phi^\mu$ 
were called deformed Fayet-Sohnius multiplets in \cite{BuKu:AdS4}.}
\begin{alignat}{4}\label{eq:SpinorDervPhi}
\nabla_\alpha^\1 \phi^\ra &\equiv \z_\alpha^\ra~, &\qquad
\nabla_\alpha^\2 \phi^\ra &= 0~, &\qquad
\nabla_\alpha^\1 \bar\phi^{\bar\ra} &= 0~, &\qquad
\nabla_\alpha^\2 \bar\phi^{\bar\ra} &= \omega^{\bar\ra}{}_{\rb} \z_\alpha^{\rb}~,
\end{alignat}
and similarly for their complex conjugates.
Take the partial pullback of $\Omega^{++}$ to the supermanifold,
replacing one of the exterior target space derivatives with a spinor derivative:
\begin{align}
&\rd \phi^\mu \pa_\mu \U^{I+} \,\nabla_\alpha^i \G_I^+ - 
\rd \phi^\mu \pa_\mu \G_I^+ \,\nabla_\alpha^i \U^{I+}\eol
&= v^{\1+} v^{\2+} \Big(
\frac{1}{\z} \rd \phi^\ra \,\nabla_\alpha^i \phi^b \omega_{\ra\rb}
+ \rd \phi^\ra \,\nabla_\alpha^i \bar\phi^{\bar b} g_{\ra \bar \rb}
+ \rd \bar\phi^{\bar \rb} \,\nabla_\alpha^i \phi^\ra g_{\ra \bar \rb}
+ \z \rd \bar\phi^{\bar\ra} \,\nabla_\alpha^i \bar\phi^{\bar \rb} \omega_{\ra\rb}
\Big)~.
\end{align}
Now contract with $v_i^+$ and the desired results follow.

The spinor derivatives of $\z_\alpha^\ra$ and $\bar\z_\dalpha^{\bar \ra}$
can be derived directly. Noting, for example, that
$\z_\alpha^\ra = \nabla_\alpha^\1 \phi^\ra =
	-\omega^\ra{}_{\bar \rb} \nabla_\alpha^\2 \bar\phi^{\bar \rb}$
one can evaluate any spinor derivative of $\z_\alpha^\ra$ by
exploiting the fact that each spinor derivative annihilates either
$\phi^\ra$ or $\bar\phi^{\bar \rb}$. This leads to
\begin{subequations}
\begin{alignat}{2}
\nabla_\beta^\1 \z_\alpha^\ra
	&= 2 \,\eps_{\beta\alpha} \bar W^r \omega^\ra{}_{\bar \rb} J_r^{\bar \rb}
		- \Gamma_{\rc \rb}{}^\ra \,\z_\beta^\rc  \,\z_\alpha^\rb~, &\qquad
\nabla_\beta^\2 \z_\alpha^\ra
	&= 2\, \eps_{\beta\alpha} \bar W^r J_r^\ra~, \\
\bar\nabla_{\dbeta \2} \z_\alpha^\ra &=
	2 i \,\omega^\ra{}_{\bar \rb} \nabla_{\alpha\dbeta} \bar\phi^{\bar \rb}
	- \Gamma_{\rc \rd}{}^\ra \,\omega^\rc{}_{\bar \rc} \,\bar\z_\dbeta^{\bar \rc} \,\z_\alpha^\rd~, &\qquad
\bar\nabla_{\dbeta \1} \z_\alpha^\ra &= -2i \,\nabla_{\alpha\dbeta} \phi^\ra~,
\end{alignat}
\end{subequations}
and similarly for their complex conjugates.
The superfield $W^r$ is the chiral field strength of
the $\cN=2$ vector multiplet, involving components
$\{W^r, A_m^r, \l^r_{\alpha i}, Y^{r\,ij}\}$.
We collect their definitions and supersymmetry transformations in
appendix \ref{app:VectorMult}.

We will also need the dilatation, $\gU{1}$, and $\SU{2}_R$
transformations of the fermions,
\begin{alignat}{3}
\bbD \z_\alpha^\ra &= \frac{3}{2} \z_\alpha^\ra - \Gamma_{\rb \rc}{}^\ra  (\bbD \phi^\rb) \, \z_\alpha^\rc~, &\quad
\bbA \z_\alpha^\ra &= -i \z_\alpha^\ra~, &\quad
I^i{}_j \z_\alpha^\ra &= -\Gamma_{\rb \rc}{}^\ra  (I^i{}_j \phi^\rb) \, \z_\alpha^\rc~,
\end{alignat}
and their $S$-supersymmetry transformations,
\begin{align}
S_{\beta \1} \z_\alpha^\ra = 4 \eps_{\beta \alpha} \chi^\ra~, \qquad
S_{\beta \2} \z_\alpha^\ra = -4 \eps_{\beta \alpha} \omega^\ra{}_{\bar \rb} \chi^{\bar \rb}~, \qquad
\bar S^{\dbeta i} \z_\alpha^\ra = 0~.
\end{align}
All of the matter fields are invariant under the special conformal
generator $K_a$.

\subsection{\Sp{n}-covariant formulation and summary}\label{sec:N2super.Spn}
We have not commented yet on one important feature of hyperk\"ahler manifolds:
the tangent space group is actually $\Sp{n} \times \Sp{1}$ (due to
the existence of the covariantly constant holomorphic tensor $\omega_{\ra\rb}$)
and the $\Sp{1}$ part of the target space connection vanishes.
Following \cite{SierraTownsend} (see also \cite{BaWi:QK}), we can introduce
a tensor $f_\mu{}_i{}^{\ra}$ and its inverse $f_{\ra}{}^{i\,\mu}$,
with an \Sp{n} index $\ra = 1, \cdots, 2n$. These obey the conditions
\begin{align}
f_\mu{}_i{}^{\ra} f_{\ra}{}^{i\,\nu} = \delta_\mu{}^\nu~, \qquad
f_{\ra}{}^{i \,\mu} f_\mu{}_j{}^{\rb} = \delta_\ra{}^\rb \delta^i{}_j~, \qquad
f_\mu{}_i{}^\ra = -\eps_{ij} \,\omega^{\ra\rb} g_{\mu \nu}\, f_\rb{}^{j\,\nu }
\end{align}
and allow one to convert any vector $V^\mu$ into an $\Sp{n} \times \Sp{1}$ vector,
$V_i{}^{\ra} = V^\mu f_\mu{}_i{}^{\ra}$.
They are related to the metric, the hyperk\"ahler two-forms and the complex structures via
\begin{align}
g_{\mu\nu} = \eps^{ij} \omega_{\ra\rb}\, f_\mu{}_i{}^{\ra} f_{\nu}{}_j{}^{\rb} ~, \quad
(\Omega^{ij})_{\mu\nu} = f_\mu{}^{\ra (i} f_\nu{}^{\rb j)} \,\omega_{\ra\rb}~, \quad
(\cJ_A)^\mu{}_\nu = i  f_{\nu}{}_i{}^{\ra} (\tau_A)^i{}_j f_{\ra}{}^{j \,\mu}~.
\end{align}
Requiring  $f_\mu{}_i{}^{\ra}$ to be covariantly constant,
\begin{align}
\nabla_\nu f_\mu{}_i{}^{\ra} := \pa_\nu f_\mu{}_i{}^{\ra} - \Gamma_{\nu \mu}{}^\rho f_\rho{}_i{}^{\ra}
	+ \Gamma_{\nu\rb}{}^\ra f_\mu{}_i{}^{\rb} = 0~,
\end{align}
defines the \Sp{n} connection $\Gamma_{\mu\rb}{}^\ra$.

We are actually interested in the situation where the indices $\ra, \rb, \cdots$ are not quite
flat tangent space $\Sp{n}$ indices, but rather complex world indices in the coordinate system
that diagonalizes the complex structure $\cJ_3$. We impose the pseudoreality condition
\begin{align}
\Big(\rd \phi^\mu f_\mu{}_i{}^{\ra}\Big)^* = \rd \phi^\mu f_\mu{}^{i \bar \ra}
	= \eps^{ij} g^{\bar \ra \ra} \omega_{\ra \rb} \, \rd \phi^\mu f_\mu{}_j{}^{\rb} ~,
\end{align}
where $g_{\ra \bar \rb}$ is the K\"ahler metric associated with $\cJ_3$.
In our conventions, the tensors $f_\mu{}_i{}^{\ra}$, $f_\mu{}^{i \bar \ra}$ and their
inverses are given by
\begin{alignat}{4}
f_\mu{}_\1{}^{\ra} &= \delta_\mu{}^\ra~, &\qquad
f_\mu{}_\2{}^{\ra} &= g_{\mu \rb} \,\omega^{\rb \ra}~, &\qquad
f_\mu{}^{\1 \bar\ra} &= \delta_\mu{}^{\bar\ra}~, &\qquad
f_\mu{}^{\2 \bar\ra} &= g_{\mu \bar\rb} \,\omega^{\bar\rb \bar\ra}~, \eol
f_{\ra}{}^{\1 \,\mu} &= \delta_\ra{}^\mu~, &\qquad
f_{\ra}{}^{\2 \,\mu} &= -\omega_{\ra \rb} \,g^{\rb \mu}~, &\qquad
f_{\bar \ra \1}{}^\mu &= \delta_{\bar \ra}{}^\mu~, &\qquad
f_{\bar \ra \2}{}^\mu &= -\omega_{\bar \ra \bar \rb} \,g^{\bar \rb \mu}~,
\end{alignat}
and the \Sp{n} connection is identical to the K\"ahler connection.

The advantage of introducing the tensors $f_\mu{}_i{}^{\ra}$ is that they
simplify the equations given in the preceding sections.
For example, the spinor derivatives of $\phi^\mu$ in \eqref{eq:SpinorDervPhi} become
\begin{align}\label{eq:SpinorDervPhi2}
\nabla_\alpha^i \phi^\mu = \z_\alpha^\rb \,f_{\rb}{}^{i\,\mu}~, \qquad
\bar\nabla^{\dalpha}_i \phi^\mu = \bar \z^{\dalpha \bar \rb} \,f_{\bar \rb i}{}^{\mu}~,
\end{align}
equivalent to the supersymmetry transformations
\begin{align}
\delta_{\trQ} \phi^\mu = \xi_i \z^\rb \, f_\rb{}^i{}^\mu + \bar \xi^{i} \bar \z^{\bar \rb}\, f_{\bar \rb i}{}^\mu~,
\end{align}
where $\xi_i^\alpha$ and $\bar\xi^i_\dalpha$ are the supersymmetry parameters.

Similarly, if we introduce the pseudoreal $\rm Sp(n) \times Sp(1)$ sections $A_i{}^{\ra}$
associated with the conformal Killing vectors $\chi^\mu$ \cite{dWKV}
\begin{align}
A_i{}^{\ra} := \chi^\mu f_\mu{}_i{}^{\ra}~, \qquad
A^{i \bar\ra} := \chi^\mu f_\mu{}^{i \bar \ra}~, \qquad
(A_i{}^{\ra})^* = A^{i \bar \ra}= \eps^{ij} \omega^{\bar \ra}{}_\rb A_j{}^\rb\, ~,
\end{align}
then the supersymmetry and $S$-supersymmetry transformations of the fermions can be written compactly as
\begin{align}
\delta \z_\alpha^\ra &=
	2 i \,(\widehat\nabla_{\alpha\dbeta} A_i{}^\ra) \,\bar\xi^{\dbeta i} 
	- 2 \bar W^r J_r{}_i{}^\ra\, \eps^{ij} \z_{\alpha j}
	- 4 \eta_\alpha^i A_i{}^\ra
	- \Gamma_{\rb \rc}{}^\ra \delta \phi^\rb \, \z_\alpha^\rc~, \eol
\delta \bar\z^{\dalpha \bar\ra} &=
	2 i \,(\widehat\nabla^{\dalpha\beta} A^{i \bar \ra})\, \xi_{\beta i}
	+ 2 W^r J_r{}^{i \bar \ra}\, \eps_{ij} \bar\z^{\dalpha j} 
	- 4 \bar\eta^\dalpha_i A^{i \bar \ra}
	- \Gamma_{\bar\rb \bar\rc}{}^{\bar\ra} \delta \bar\phi^{\bar\rb}
		\, \bar\z^{\dalpha \bar\rc}~,
\end{align}
where $\eta^{i\alpha}$ and $\bar \eta_{i \dalpha}$ are the $S$-supersymmetry parameters,
$J_r{}_i{}^\ra := J_r{}^\mu f_\mu{}_i{}^\ra$ is the Killing vector associated
with the gauged isometries, and $\widehat\nabla_a$ includes the \Sp{n} connection.
For reference, we also give the transformations of the fermions under gauged isometries,
\begin{align}
\delta_g \z_\alpha^\ra = \l^r \z_\alpha^\rb \nabla_\rb J_r^\ra - \Gamma_{\rb \rc}{}^\ra \delta_g \phi^\rb \z_\alpha^\rc~, \qquad
\delta_g \bar\z^{\dalpha \bar\ra} = 
	\l^r \bar\z^{\dalpha \bar\rb} \nabla_{\bar\rb} J_r^{\bar\ra}
	- \Gamma_{\bar\rb \bar\rc}{}^{\bar\ra} \delta_g \bphi^{\bar\rb} \bar\z^{\dalpha \bar \rc}~.
\end{align}

Note that the scalar fields $\phi^\mu$ and the fermions $\z_\alpha^\rb$
transform into each other (and into the components of the vector multiplet)
under supersymmetry.
The conditions \eqref{eq:SpinorDervPhi} (equivalently
\eqref{eq:SpinorDervPhi2}) have eliminated the hypermultiplet auxiliary fields and
placed the entire multiplet on-shell: in particular, one can check that the supersymmetry algebra
closes only up to the equations of motion.
These results match those of \cite{dWKV}, up to differences in
conventions discussed at the end of section \ref{sec:CompActonCurved}.

\section{Building blocks of a component reduction}\label{sec:Blocks}
In the previous section, we addressed the on-shell structure of the
target space multiplets $\phi^\mu$, where the auxiliary fields were
completely eliminated. In a manifestly supersymmetric setting --
which we have implicitly been using -- this actually is a stronger
condition than what we want. It corresponds, in the
$\cN=1$ situation, to specifying not only the algebraic equations of motion for
the auxiliaries $F^{\ra}$ but also the dynamical equations of motion
for the physical fields $\phi^\ra$ and $\z_\alpha^\ra$. Our goal in this
section now is to describe how to analyze the superfield equations of motion
for the arctic multiplet so that only the auxiliary fields are
placed on-shell.

\subsection{Component expansions of the projective multiplets}\label{sec:Blocks.Comps}
We begin by analyzing the component structure of the projective
multiplets. The conventional approach is to reduce to $\cN=1$
superfields and superspace, but as we have already discussed in the introduction,
this is far more difficult in the curved case.
Instead, it will be more convenient to reduce
directly to component fields. Much insight can be gleaned from the
harmonic superspace approach to sigma models (see section 11.4 of \cite{GIOS}),
where no $\cN=1$ subspace is readily available.
The first step is to choose an appropriate system of coordinates. In
projective superspace, the following coordinates are well-defined in the
north chart and suitable for arctic superfields \cite{KT-M:CFlat4D, BKLT-M:AdSPro}:
\begin{gather}
x_{N}^m := x^m - \frac{i}{v^{\1+}} (\q^+ \sigma^m \bar \q^\1 + i \q^\1 \sigma^m \bar \q^+)~, \qquad
\q_{\ul\alpha}^\pm := v_i^\pm \q_{\ul\alpha}{}^i~, \eol
\z := \frac{v^{\2+}}{v^{\1+}}~, \qquad
v^{\1+}~, \qquad
z_1^{--} := \frac{v^{\1-}}{v^{\1+}} = \frac{v_\2^-}{v^{\1+}} = e^{-2i \psi} \bar\z~.
\end{gather}
These lead to the simple expressions
\begin{align}
D_{\ul\alpha}^+ = \frac{\pa}{\pa \q^{\ul\alpha-}}~, \qquad
D^{++} = \frac{\pa}{\pa z_1^{--}}~,
\end{align}
implying that holomorphic analytic multiplets are independent of
$z_1^{--}$ and $\q^{\ul\alpha-}$.
An arctic multiplet $\U^{I+}$, well-defined in the north chart,
is then simply specified as a function
$\U^{I+}(x_N,  v^{\1+}, \z, \q^+)$
possessing an arctic expansion in $\z$.
Schematically, such a superfield admits a decomposition (following
closely the approach of \cite{GIOS})
\begin{align}
\U^{I+}&= \U^{I+}\vert_{\q^+=0}
	+ \q^{\alpha+} \,\Psi_\alpha^I
	+ \bar \q_\dalpha^+ \,\bar\Psi^{I\dalpha}
	+ (\q^+)^2 \,M^{I-}
	+ (\bar\q^+)^2 \,N^{I-}
	\eol & \quad
	+ i \q^+ \sigma^a \bar\q^+ \, A_a^{I-}
	- 2 (\bar\q^+)^2 \q^{\alpha+} \,\Xi_\alpha^{I--}
	- 2 (\q^+)^2 \bar\q_\dalpha^+ \,\bar\Xi^{I\dalpha--}
	+ (\q^+)^2 (\bar \q^+)^2 \,P^{I(-3)}~,
\end{align}
where the component fields $\Psi_{\ul\alpha}^I, \cdots, P^{I(-3)}$ depend on
$x_N$, $v^{\1+}$ and $\z = v^{\2+} / v^{\1+}$. Their dependence
on $v^{\1+}$ is indicated by their charge, that is,
\begin{align}
\Psi_{\ul\alpha}^I = \Psi_{\ul\alpha}^I(x_N, \z)~, \quad \cdots ~, \quad
P^{I(-3)} = \frac{1}{(v^{\1+})^3} \, P^I(x_N, \z)
\end{align}
and all expressions are arctic in $\z$.

In a curved background, it is more convenient to define the components of
$\U^{I+}$ in a covariant way. The closest analogue
to the expansion given above is to define
\begin{subequations}\label{eq:defComps}
\begin{alignat}{2}
\Psi_\alpha^I &:= \frac{1}{v^{\1+}} \nabla_\alpha^{\1} \U^{I+}~, &\qquad
\Psi_\dalpha^I &:= \frac{1}{v^{\1+}} \bar\nabla_\dalpha^{\1} \U^{I+}~, \\
M^{I-} &:= -\frac{1}{4} \frac{1}{(v^{\1+})^2} (\nabla^{\1})^2 \U^{I+}~, &\qquad
N^{I-} &:= -\frac{1}{4} \frac{1}{(v^{\1+})^2} (\bar\nabla^{\1})^2 \U^{I+}~, \\
A_{\alpha\dbeta}^{I-} &:= -i \frac{1}{(v^{\1+})^2} \nabla_\alpha^\1 \nabla_\dbeta^\1 \U^{I+}~, \\
\Xi_\alpha^{I--} &:= \frac{1}{8} \frac{1}{(v^{\1+})^3} \nabla_\alpha^\1 (\bar\nabla^{\1})^2 \U^{I+}~, &\qquad
\Xi_\dalpha^{I--} &:= \frac{1}{8} \frac{1}{(v^{\1+})^3} \bar\nabla_\dalpha^\1 (\nabla^{\1})^2 \U^{I+}~, \\
P^{I(-3)} &:= \frac{1}{16} \frac{1}{(v^{\1+})^4} (\nabla^{\1})^2 (\bar\nabla^{\1})^2 \U^{I+}~.
\end{alignat}
\end{subequations}
Each of these components is manifestly holomorphic and well-defined in
the north chart. The corresponding formulae for the components of an
antarctic multiplet $\breve\U^{\bar I+}$ are found by replacing
$v^{\1+} \rightarrow v^{\2+}$
and $\nabla_{\ul\alpha}^\1 \rightarrow \nabla_{\ul\alpha}^\2$.

An inconvenient feature of these component fields is that the arctic and
antarctic fields are naturally defined in terms of different sets of
spinor derivatives. Moreover, when we actually analyze the
component action, we will encounter the spinor derivatives
$\nabla_{\ul\alpha}^- := v_i^- \nabla_{\ul\alpha}^i$,
which involve a linear combination of $\nabla_{\ul\alpha}^\1$
and $\nabla_{\ul\alpha}^\2$.
Using $\nabla_{\ul\alpha}^- = \frac{1}{v^{\1+}} \nabla_{\ul\alpha}^\1 + z_1^{--} \nabla_{\ul\alpha}^+$,
it is straightforward to determine the relation between
expressions like $\nabla_{\ul\alpha}^- \U^{I+}$ and $\nabla_{\ul\alpha}^\1 \U^{I+}$.
The expressions for the lowest few components are rather simple:
\begin{subequations}\label{eq:DUp}
\begin{alignat}{3}
\nabla_{\ul\alpha}^- \U^{I+} &= \Psi_{\ul\alpha}^I~, &\quad
\nabla_\alpha^- \bar\nabla_\dbeta^- \U^{I+}
	&= i A_{\alpha \dbeta}^{I-} - 2i z_1^{--} \nabla_{\alpha\dbeta} \U^{I+}~, \\
-\frac{1}{4} (\bar\nabla^-)^2 \U^{I+}
	&= N^{I-} - z_1^{--} W^r \cJ_r^{I+}  ~, &\quad
-\frac{1}{4} (\nabla^-)^2 \U^{I+}
	&= M^{I-} - z_1^{--} \bar W^r \cJ_r^{I+}~.
\end{alignat}
Recall that the arctic function $\cJ_r^{I+}$ arises from the action of the gauge
generator on $\U^{I+}$, so some of the terms in the second line
above are present only when isometries are gauged.
The expressions with three spinor derivatives are
\begin{align}
\frac{1}{8} \nabla_\alpha^- (\bar\nabla^-)^2 \U^{I+}
	&= \Xi_\alpha^{I--}
		+ z_1^{--} \Big(
		- \frac{i}{2} \nabla_{\alpha \dbeta} \bar\nabla^{\dbeta -} \U^{I+}
		+ 2 \bm\lambda_\alpha^{r-} \cJ_r^{I+}
		+ \frac{1}{2} W^{r} \nabla_\alpha^- \cJ_r^{I+}
		\eol & \qquad\qquad\qquad\quad
		- \frac{1}{2} W_\alpha{}^\beta \nabla_\beta^- \U^{I+}
		- \frac{3}{2} \chi_\alpha^- \U^{I+}
		+ \frac{3}{2} \chi_\alpha^+ D^{--} \U^{I+}
		\Big)
		\eol & \quad
		- (z_1^{--})^2 \bm\lambda_\alpha^{r+} \cJ_r^{I+}~, \\
\frac{1}{8} \bar\nabla^\dalpha{}^- (\nabla^-)^2 \U^{I+}
	&= \bar\Xi_\dalpha^{I--}
		+ z_1^{--} \Big(
		\frac{i}{2} \nabla^{\dalpha \beta} \nabla_\beta^- \U^{I+}
		- 2 \bm{\bar\lambda}^{\dalpha r -} \cJ_r^{I+}
		+ \frac{1}{2} \bar W^{r} \bar\nabla^\dalpha{}^- \cJ_r^{I+}
		\eol & \qquad\qquad\qquad\quad
		- \frac{1}{2} \bar W^\dalpha{}_\dbeta \bar\nabla^\dbeta{}^- \U^{I+}
		+ \frac{3}{2} \bar\chi^\dalpha{}^- \U^{I+}
		- \frac{3}{2} \bar\chi^\dalpha{}^+ D^{--} \U^{I+}
		\Big)
		\eol & \quad
		+ (z_1^{--})^2 \bm{\bar\lambda}^{\dalpha r+} \cJ_r^{I+}~,
\end{align}
and involve the covariant conformal supergravity fields
$W_{\alpha\beta}$ and $\chi_{\alpha i}$ as well as the gaugino $\l_{\alpha i}^{r}$.
The term with four spinor derivatives is the most complicated:
\begin{align}
\frac{1}{16} (\nabla^-)^2 (\bar\nabla^-)^2 \U^{I+}
	&= P^{I(-3)}
	+ z_1^{--} \Big\{
	- \frac{i}{2} \nabla^{\dalpha \alpha} \nabla_{\alpha}^- \nabla_\dalpha^- \U^{I+}
	- 3 D \,D^{--} \U^{I+}
	\eol & \qquad \qquad
	+ \frac{3}{2} \chi^{\alpha +} D^{--} \nabla_\alpha^- \U^{I+}
	- \frac{3}{2} \bar\chi_\dalpha^+ D^{--} \bar\nabla^{\dalpha -} \U^{I+}
	\eol & \qquad \qquad	
	+ 2 \bm\lambda^{\alpha r -}\nabla_\alpha^- \cJ_r^{I+}
	- 2 \bm{\bar \lambda}_\dalpha^{r-} \bar\nabla^\dalpha{}^- \cJ_r^{I+}
	\eol & \qquad \qquad
	+ \frac{1}{4} \bar W^{r} (\bar\nabla^-)^2 \cJ_r^{I+}
	+ \frac{1}{4} W^{r} (\nabla^-)^2 \cJ_r^{I+}
	+ 3 Y^{r--} \cJ_r^{I+}
	\Big\}
	\eol & \quad
	+ (z_1^{--})^2 \Big\{
	\frac{3}{2} D \U^{I+}
	- \bm\lambda^\alpha{}^{r+} \nabla_\alpha^- \cJ_r^{I+}
	+ \bm{\bar\lambda}_\dalpha^{r+} \bar\nabla^{\dalpha -} \cJ_r^{I+}
	\eol & \qquad \qquad		
	- \frac{1}{2} W^{r} \bar W^{s} \,\{X_{r}, X_{s}\} \U^{I+}
	- 3 Y^{r -+} \cJ_r^{I+}
	+ \frac{1}{2} \nabla_{\alpha \dalpha} \nabla^{\dalpha \alpha} \U^{I+}
	\Big\}
	\eol & \quad
	+ (z_1^{--})^3  Y^{r++}\cJ_r^{I+}~.
\end{align}
\end{subequations}
Similar equations to \eqref{eq:DUp} arise if we replace the arctic
$\U^{I+}$ with the antarctic $\breve \U^{\bar I+}$ -- the only change is
the replacement of $z_1^{--}$ with $z_2^{--} := v^{\2-} / v^{\2+}$.

\subsection{Auxiliary field equations of motion}
Now we need to understand what conditions
arise from placing \emph{only} the auxiliary fields on-shell.
It helps to recall again the rigid $\cN=1$ situation where, in the absence
of a superpotential, the on-shell equation of motion is given by
\begin{align}
\int \rd^4x \, \rd^2 \q\, (\bar D^2 K_\ra) \delta \phi^\ra = 0 \quad \implies \quad \bar D^2 K_\ra = 0~.
\end{align}
The chiral superfield $\bar D^2 K_\ra$ is constrained to vanish.
At the component level, this corresponds to three
distinct equations, corresponding to the equations of motion of the three
components of $\phi^\ra$. Setting to zero the lowest component of
$\bar D^2 K_\ra$ amounts to constraining the auxiliary field, while constraining
the higher two components leads to dynamical equations of motion.
The situation can be rendered schematically
as in table \ref{tab:N1Comp}.
\begin{table*}[h]
\centering
\renewcommand{\arraystretch}{1.4}
\begin{tabular}{@{}ccc@{}} \toprule
component e.o.m. & $\implies$ & constrained field \\ \midrule
$F^\ra$ & $\implies$& $\bar D^2 K_\ra \sim g_{\ra \bar \rb} \bar F^{\bar \rb}$ \\
$\z_\alpha^\ra$ & $\implies$& $D_\alpha \bar D^2 K_\ra \sim g_{\ra \bar \rb} \,\pa_{\alpha \dalpha} \bar \z^{\bar \rb \dalpha}$ \\
$\phi^\ra$ & $\implies$& $D^2 \bar D^2 K_\ra \sim g_{\ra \bar \rb} \Box \phi^{\bar \rb}$ \\
\bottomrule
\end{tabular}
\caption{$\cN=1$ component equations of motion}\label{tab:N1Comp}
\end{table*}

These statements apply equally well in the $\cN=2$ setting.
There we have the equations \eqref{eq:N2Eoms}, equivalently written as
\begin{align}\label{eq:ArcticEOM}
-\frac{1}{2\pi} \oint_\cC \rd \tau \int \rd^4x \, \rd^4\q^+ \cE^{--} \,\G_I^+ \delta \U^{I+} = 0 \quad \implies \quad
\text{$\Gamma_I^+$ arctic}~,
\end{align}
which place the arctic multiplet on-shell and imply
that the composite $\G_I^+$ is an arctic multiplet.
If we introduce the component fields
$\Psi_{\ul\alpha I}, \cdots, P_I^{(-3)}$ for $\G_I^+$ in analogy to \eqref{eq:defComps},
then the equations of motion for each of the $\U^{I+}$ component
fields implies the corresponding arctic nature of these components of $\G_I^+$.
For example, by considering the component reduction of \eqref{eq:ArcticEOM},
the equation of motion for the \emph{highest} component $P^{I(-3)}$ of $\U^{I+}$
must set the \emph{lowest} component of $\Gamma_I^+$ to be arctic,
and similarly throughout the multiplet. The precise relations
are given in table \ref{tab:N2Comp}.

\begin{table*}[h]
\centering
\renewcommand{\arraystretch}{1.4}
\begin{tabular}{@{}ccc@{}} \toprule
component e.o.m. & $\implies$ & constrained arctic field\\ \midrule
$P^{I(-3)}$ & $\implies$& $\Gamma_I^+$ \\
$\Xi_{\ul\alpha}^{I--}$ & $\implies$& $\Psi_{I \ul\alpha}$ \\
$N^{I-}$ & $\implies$& $M_I^-$ \\
$A_{\alpha\dalpha}^{I-}$ & $\implies$& $A_{I \,\alpha\dalpha}^-$ \\
$M^{I-}$ & $\implies$& $N_I^-$ \\
$\Psi_{\ul\alpha}^I$ & $\implies$& $\Xi_{I \ul\alpha}^{--}$ \\
$\U^{I+}$ & $\implies$& $P_I^{(-3)}$ \\
\bottomrule
\end{tabular}
\caption{$\cN=2$ component equations of motion}\label{tab:N2Comp}
\end{table*}

The key issue here is that the auxiliary field equations
consist of all but the final two lines of table \ref{tab:N2Comp},
which for dimensional reasons must contain the field equations for
the physical fermions and scalars.
Keeping in mind that $\Gamma_I^+$ is a composite quantity,
it follows that setting the components $\G_I^+$ through $N_I^-$ to be arctic
must correspond to fixing them and the original quantities
$\U^{I+}$ through $N^{I-}$ to be given by their on-shell expressions.
For the on-shell superfields discussed in the previous section,
these expressions are easy to work out. We already know that
the lowest components $\phi^\ra = (\Phi^I, \Psi_I)$ are defined by \eqref{eq:ArcticExp}.	
The components at next order in $\q$ are
\begin{subequations}\label{eq:onshDU}
\begin{align}
\Psi_\alpha^{I} &=
	\frac{1}{v^{\1+}} \z_\alpha^\rb \pa_\rb \U^{I+}
		= - \frac{1}{v^{\2+}} \z_\alpha^\rb \,\omega_\rb{}^{\bar \rb} \pa_{\bar \rb} \U^{I+}~, \\
\Psi_\dalpha^{I} &=
	-\frac{1}{v^{\2+}} \bar\z_\dalpha^{\bar \rb} \pa_{\bar \rb} \U^{I+}
		= - \frac{1}{v^{\1+}} \bar\z_\dalpha^{\bar \rb} \,\omega_{\bar \rb}{}^{\rb} \pa_{\rb} \U^{I+}~.
\end{align}
\end{subequations}
Analogous expressions follow for $\Psi_{I \ul\alpha}$ by replacing $\U^{I+}$ with $\G_I^+$.
The $\q^2$ components are
\begin{subequations}\label{eq:onshA}
\begin{align}
A_{\alpha \dalpha}^{I-}
	= \frac{2}{v^{\1+} v^{\2+}} \nabla_{\alpha\dalpha} \bar\phi^{\bar \rb} \,\pa_{\bar \rb} \U^{I+}
	+ \frac{i}{v^{\1+} v^{\2+}} \zeta_\alpha^b \bar \zeta_{\dalpha}^{\bar \rb} \,\pa_\rb \pa_{\bar \rb} \U^{I+}~,\\
A_{\alpha \dalpha}^{\bar I-}
	= -\frac{2}{v^{\1+} v^{\2+}} \nabla_{\alpha\dalpha} \phi^{\rb} \,\pa_{\rb} \breve\U^{I+}
	+ \frac{i}{v^{\1+} v^{\2+}} \zeta_\alpha^b \bar \zeta_{\dalpha}^{\bar \rb} \,\pa_\rb \pa_{\bar \rb} \breve\U^{I+}~,
\end{align}
\end{subequations}
and
\begin{subequations}\label{eq:onshMN}
\begin{align}
M^{I-}
	&= \frac{1}{v^{\1+} v^{\2+}} \Big(
	\bar W^{r} \,\bar J_{r}^{\bar \rb} \,\pa_{\bar \rb} \U^{I+}
	+ \frac{1}{4} \z^{\alpha \ra} \z_\alpha^{\rb} \,\omega_\rb{}^{\bar \rb} \,\pa_\ra \pa_{\bar \rb} \U^{I+}
	\Big)~, \\
N^{I-}
	&= \frac{1}{v^{\1+} v^{\2+}} \Big(
	\bar W^{r} \,\bar J_{r}^{\bar \rb} \,\pa_{\bar \rb} \U^{I+}
	- \frac{1}{4} \bar\z_\dalpha^{\bar \ra} \bar\z^{\dalpha \bar \rb} \,\omega_{\bar \rb}{}^\rb \,\pa_{\bar \ra} \pa_{\rb} \U^{I+}
	\Big)~, \\
M^{\bar I -}
	&= \frac{1}{v^{\1+} v^{\2+}} \Big(
	- \bar W^{r} \,J_{r}^{\rb} \,\pa_{\rb} \breve \U^{\bar I +}
	+ \frac{1}{4} \z^{\alpha \ra} \z_\alpha^{\rb} \,\omega_\rb{}^{\bar \rb} \,\pa_\ra \pa_{\bar \rb} \breve \U^{\bar I +}
	\Big)~, \\	
N^{\bar I -}
	&= \frac{1}{v^{\1+} v^{\2+}} \Big(
	- W^{r} \,J_{r}^{\rb} \,\pa_{\rb} \breve\U^{I+}
	- \frac{1}{4} \bar\z_\dalpha^{\bar \ra} \bar\z^{\dalpha \bar \rb} \,\omega_{\bar \rb}{}^\rb \,\pa_{\bar \ra} \pa_{\rb} \breve\U^{I+}
	\Big)~.
\end{align}
\end{subequations}
These conditions, as well as the corresponding ones for
the components of $\G_I^+$ and $\bar \G_{\bar I}^+$,
constitute the set of auxiliary equations
of motion. We cannot specify the components $\Xi_{\ul\alpha}^{I --}$
and $P^{I(-3)}$
without placing the physical fields on-shell.

\newcommand{\homega}{\Omega}

\subsection{Some relevant quantities}
In performing the component reduction, there are a number
of geometric quantities that we will encounter that arise from simple
expressions in terms of the arctic superfields. Rather than discuss
them piecemeal, one at a time as we come across them, we present them here
together to emphasize their common features.
They can be grouped loosely into three classes: quantities corresponding
to pullbacks of the hyperk\"ahler two-forms, which exist even in the
flat limit; quantities associated with gauged isometries; and
quantities that arise from the cone structure.

\subsubsection{Hyperk\"ahler two-forms and their pullbacks}
Recall that the hyperk\"ahler two-forms can be represented equivalently as
\begin{align}\label{eq:OmegaSec4.3}
\Omega^{++} &= \rd \U^{I+} \wedge \rd\G_I^+
	= \rd \breve\U^{\bar J+} \wedge \rd\breve\G_{\bar J}^+ \eol
&= v^{\1+} v^{\2+} \left(
	\frac{1}{2\z} \rd \phi^\ra \wedge \rd \phi^\rb \, \omega_{\ra \rb}
	+ \rd \phi^\ra \wedge \rd\bar\phi^{\bar \rb} g_{\ra \bar \rb}
	+ \frac{\z}{2} \rd \bar\phi^{\bar \ra} \wedge \rd \bar\phi^{\bar \rb} \, \omega_{\bar \ra \bar \rb} \right)~.
\end{align}
We can interpret the second line as a superfield expression with on-shell
superfields $\phi^\ra$ provided we only work at most to order $\q^2$.
Taking the pullback to the supermanifold and using two spinor derivatives
of opposite chirality, we find
\begin{align}
\homega_{\alpha \dbeta} &:= \nabla_\alpha^- \U^{I+}\, \bar\nabla_\dbeta^- \Gamma_I^+
	- \nabla_\alpha^- \Gamma_I^+\, \bar\nabla_\dbeta^- \U^{I+}
		= \nabla_\alpha^- \breve \U^{\bar I+}\, \bar\nabla_\dbeta^- \breve \Gamma_{\bar I}^+
	- \nabla_\alpha^- \breve \Gamma_{\bar I}^+\, \bar\nabla_\dbeta^- \breve \U^{\bar I+}~, \eol
	&= - \frac{1}{v^{\1+} v^{\2+}} \z_\alpha^\ra \bar \z_\dbeta^{\bar \rb}
		\Big(\pa_\ra \U^{I+} \, \pa_{\bar \rb} \Gamma_{I}^+ - \pa_\ra \Gamma_{I}^+ \, \pa_{\bar \rb} \U^{I+})
	= - \z_\alpha^\ra \, \bar \z_\dbeta^{\bar \rb} \, g_{\ra \bar \rb}~,
\end{align}
where the second line follows from \eqref{eq:onshDU} and \eqref{eq:defg}.
Similar expressions for other quantities can be derived:
\begin{align}
\homega_{\alpha \beta} &:= \nabla_\alpha^- \U^{I+}\, \nabla_\beta^- \Gamma_I^+
	- \nabla_\alpha^- \Gamma_I^+\, \nabla_\beta^- \U^{I+} = \z_\alpha^\ra \, \z_\beta^\rb \, \omega_{\ra \rb}~, \eol
\homega_{\dalpha \dbeta} &:= \nabla_\dalpha^- \U^{I+}\, \nabla_\dbeta^- \Gamma_I^+
	- \nabla_\dalpha^- \Gamma_I^+\, \nabla_\dbeta^- \U^{I+}
	= \z_\dalpha^{\bar \ra} \, \z_\dbeta^{\bar \rb} \, \omega_{\bar \ra \bar \rb}~.
\end{align}
Note that each of the quantities $\Omega_{\alpha \beta}$, $\Omega_{\alpha \dbeta}$
and $\Omega_{\dalpha \dbeta}$ are harmonic-independent at lowest order in their
$\q$ expansion. Similar pullbacks can be defined using the vector derivatives,
\begin{align}\label{eq:homegaSV}
\homega_{\alpha\, b}^+ &:= \nabla_\alpha^- \U^{I+}\, \nabla_{b} \Gamma_I^+
	- \nabla_\alpha^- \Gamma_I^+\, \nabla_{b} \U^{I+}
	= v^{\1+} \z_\alpha^\ra \,\omega_{\ra \rb} \nabla_{b} \phi^\rb
	+ v^{\2+} \z_\alpha^\ra \,g_{\ra \bar \rb} \nabla_{b} \phi^{\bar \rb}~, \eol
\homega_{\dalpha\, b}^+ &:= \nabla_\dalpha^- \U^{I+}\, \nabla_{b} \Gamma_I^+
	- \nabla_\dalpha^- \Gamma_I^+\, \nabla_{b} \U^{I+}
	= - v^{\2+} \bar\z_\dalpha^{\bar \ra} \,\omega_{\bar \ra \bar \rb} \nabla_{b} \phi^{\bar \rb}
	+ v^{\1+} \bar\z_\dalpha^{\bar \ra} \,g_{\bar \ra \rb} \nabla_{b} \phi^{\rb}~,
\end{align}
or with the gauge generator,
\begin{align}\label{eq:homegaGS}
\homega_{r \beta}^+ &:= X_r \U^{I+}\, \nabla_\beta^- \Gamma_I^+
	- X_r \Gamma_I^+\, \nabla_\beta^- \U^{I+}
	=  v^{\1+} J_r^\ra \omega_{\ra \rb} \z_\beta^\rb - v^{\2+} J_r^{\bar \ra} g_{\bar \ra \rb} \z_\beta^\rb~, \eol
\homega_{r \dbeta}^+ &:= X_r \U^{I+}\, \bar\nabla_\dbeta^- \Gamma_I^+
	- X_r \Gamma_I^+\, \bar\nabla_\dbeta^- \U^{I+}
	=  -v^{\2+} J_r^{\bar \ra} \omega_{\bar \ra \bar \rb} \bar\z_\dbeta^{\bar \rb}
		- v^{\1+} J_r^{\ra} g_{\ra \bar \rb} \bar\z_\dbeta^{\bar \rb}~.
\end{align}
It is also possible to define other pullbacks such as
$\Omega_{r b}^{++}$ or $\Omega_{ab}^{++}$, but these will not play
a major role in our discussion so we do not give their explicit forms.

It can be useful to employ a condensed notation to simplify
the right-hand sides of \eqref{eq:homegaSV} and \eqref{eq:homegaGS}.
Using the $\rm Sp(n)$ vielbeins $f_{\mu}{}_i{}^\ra$ introduced in
section \ref{sec:N2super.Spn}, we define
\begin{align}
\z_\beta{}^{i\mu} := \nabla_\beta^i \phi^\mu = \z_\beta^\ra f_\ra{}^{i\,\mu}~, \qquad
\bar\z_{\dbeta i}{}^{\mu} := \bar\nabla_{\dbeta i} \phi^\mu = \bar\z_\dbeta^{\bar \ra} f_{\bar \ra i}{}^\mu~.
\end{align}
Then one can alternatively write (using also the fields $A_i{}^\ra$ and $J_{r}{}_i{}^\ra$),
\begin{alignat}{2}
\homega_{\alpha\, b}^+ &= -\z_\alpha^{+ \mu} g_{\mu \nu} \nabla_b \phi^\nu = -\z_\alpha^{\ra} \nabla_b A^{+\bar \rb} g_{\ra \bar \rb}~, &\qquad
\homega_{\dalpha\, b}^+ &= -\bar\z_\dalpha^{+ \mu} g_{\mu \nu} \nabla_b \phi^\nu = -\bar\z_\dalpha^{\bar \ra} \nabla_b A^{+ \rb} g_{\bar \ra \rb}~, \eol
\homega_{r\, \beta}^+ &= \z_\alpha^{+ \mu} J_{r\mu} = \z_\alpha^\ra J_{r}^{+\bar \rb} g_{\ra \bar \rb}~, &\qquad
\homega_{r\, \dbeta}^+ &= \bar\z_\dalpha^{+ \mu} J_{r\mu} = \bar \z_\dalpha^{\bar \ra} J_{r}^{+\rb} g_{\bar \ra \rb}~.
\end{alignat}
It is also useful to note some relations between these various quantities.
For example, $\nabla_\beta^j \homega_{\alpha \dalpha}$
and $\bar\nabla_\dbeta^j \homega_{\alpha \dalpha}$ are found by taking
\begin{align}
\nabla_\beta^+ \homega_{\alpha \dalpha}
	&= \phantom{+}2\eps_{\beta \alpha} \bar W^s \homega_{s \dalpha}^+ + 2i \homega_{\alpha\, \beta \dalpha}^+
\quad \implies \quad
\nabla_\beta^j \homega_{\alpha \dalpha}
	= \phantom{+}2\eps_{\beta \alpha} \bar W^s \homega_{s \dalpha}^j + 2i \homega_{\alpha\, \beta \dalpha}^j~, \eol
\bar\nabla_\dbeta^+ \homega_{\alpha \dalpha}
	&= -2\eps_{\dbeta \dalpha} W^r \homega_{r \alpha}^+ - 2i \homega_{\dalpha\, \alpha \dbeta}^+
\quad \implies \quad
\bar\nabla_\dbeta^j \homega_{\alpha \dalpha}
	= -2\eps_{\dbeta \dalpha} W^r \homega_{r \alpha}^j - 2i \homega_{\dalpha\, \alpha \dbeta}^j~.
\end{align}
Relations such as these will be extremely useful in our analysis.

\subsubsection{Components of the $\cN=2$ moment map}
Recall that the $\cN=2$ moment map $D_r^{++}$ is given in projective superspace
by \eqref{eq:N2KPproj}.
With the auxiliary field equations imposed, $D_r^{++}$ is 
a globally defined superfield to order $\q^2$.
Its lowest component was given already in \eqref{eq:N2KPcomp}.
Its higher components can be analyzed by applying spinor derivatives to \eqref{eq:N2KPproj}.
A single spinor derivative gives
\begin{align}
\nabla_{\ul\alpha}^+ D_r^{++} = 0~, \qquad
\nabla_{\ul\alpha}^- D_r^{++} = \homega_{r \,\ul\alpha}^+~.
\end{align}
For two spinor derivatives, one finds
\begin{align}
(\nabla^-)^2 D_r^{++}
	&= - 4 \bar W^s f_{sr}{}^t D_t^{+-}
	- 2 \bar W^s J_s^\mu J_{r\mu}
	- \z^\ra \z^\rb \nabla_\ra (\omega_{\rb \rc} J_r^\rc)~,
\end{align}
which implies
\begin{align}\label{eq:DDDij}
\frac{1}{3} \nabla_{ij} D_r^{ij} &=
	- 2 \bar W^s J_s^\mu J_{r\mu}
	- \z^\ra \z^\rb \nabla_\ra (\omega_{\rb \rc} J_r^\rc)~.
\end{align}

\subsubsection{Hyperk\"ahler cone potential and hyperk\"ahler one-forms}
Recall that the hyperk\"ahler cone possesses a globally defined function,
the hyperk\"ahler potential $K$.
In terms of superfield quantities, $K$ is given by the pullback of
$\Omega^{++}$ onto the auxiliary $\SU{2}$ manifold,
\begin{align}\label{eq:hyperKdef}
K &:= D^{--} \U^{I+}\, D^0 \Gamma_I^+ - D^0 \U^{I+}\, D^{--} \Gamma_I^+ 
 = \Gamma_I^+ D^{--} \U^{I+} - \U^{I+} D^{--} \Gamma_I^+~.
\end{align}
Let us prove this result.
Because $\cF^{++}$ is independent of $v^{i+}$ in the superconformal case,
this expression is equivalent to its conjugate so $K$ is real.
It is easy to check that $D^{++} K = 0$.
Because $K$ is both arctic and antarctic, it must actually be harmonic-independent.
To prove that it is the hyperk\"ahler potential, observe that
an $\SU{2}_R$ transformation acts on projective multiplets and
on the target space respectively as
\begin{align}
\l^i{}_j I^j{}_i = -\l^{++} D^{--} + \l^0 D^0 + \l^{--} D^{++}
	= \l^i{}_j (\cJ^j{}_i)^\mu{}_\nu \chi^\nu \pa_\mu~.
\end{align}
For projective multiplets $\U^{I+}$ and $\G_I^+$, this implies
\begin{align}
D^{--} = (\cJ^{--})^\mu{}_\nu \chi^\nu \pa_\mu~, \qquad
D^0 = 2 (\cJ^{-+})^\mu{}_\nu \chi^\nu \pa_\mu~, \qquad
D^{++} = -(\cJ^{++})^\mu{}_\nu \chi^\nu \pa_\mu~.
\end{align}
It follows that \eqref{eq:hyperKdef} can be written
\begin{align}
K &= 2 (\cJ^{--})^\mu{}_\nu \chi^\nu 
	(\cJ^{-+})^\rho{}_\sigma \chi^\sigma
	(\pa_\mu \U^{I+}\pa_\rho \G_I^+ - \pa_\mu \G_I^+\pa_\rho \U^{I+}) \eol
	&= -2 \chi_\nu (\cJ^{--} \cJ^{++} \cJ^{-+})^\nu{}_\sigma \chi^\sigma~.
\end{align}
From \eqref{eq:Jmult1}, one finds
$(\cJ^{++} \cJ^{-+})^\mu{}_\nu = \frac{1}{2} (\cJ^{++})^\mu{}_\nu$
and
$(\cJ^{--} \cJ^{++})^\mu{}_\nu = -\frac{1}{2} \delta^\mu{}_\nu - (\cJ^{+-})^\mu{}_\nu$.
This implies that $K = \frac{1}{2} \chi_\mu \chi^\mu$, so $K$ is indeed
the hyperk\"ahler potential.
It should be emphasized that \eqref{eq:hyperKdef} defines the
hyperk\"ahler potential to order $\q^2$ if we impose only the auxiliary equations
of motion. This is because at higher order, $\Gamma_I^+$ ceases to be arctic
(without imposing all the equations of motion)
and so $K$ ceases to be harmonic-independent.

It will be useful to have explicit expressions for various derivatives
of $K$, such as
\begin{align}
\nabla_{\ul \alpha}^\pm K = K_\mu \,\z_\alpha^{\pm\mu}~,\qquad
\nabla_b \nabla_\alpha^i K
	= K_\mu \widehat \nabla_b \z_\alpha{}^{i \mu} - \homega_{\alpha\, b}^i~,
\end{align}
where $\widehat\nabla_b$ carries the target space connection.

Although we will not explicitly make use of them,
it is also interesting to note that the hyperk\"ahler one-forms $k^{ij}$
can be written in superspace as
\begin{align}
k^{++} &= \Gamma_I^{+} \rd \U^{I+} - \U^{I+} \rd \Gamma_I^+~, \qquad
\rd k^{++} = -2 \Omega^{++}~.
\end{align}
These are given by a partial pullback of $\Omega^{++}$, replacing one
$\rd$ with the derivative $D^0$ of the auxiliary manifold. Because
they are globally defined, $\Omega^{++}$ is exact.
Because $\Gamma_I^+$ and $\U^{I+}$ are both Weyl weight one,
we can rewrite this on the target space in the familiar way,
\begin{align}
k^{++} &= \rd \phi^\mu\, \chi^\nu \Big(
	\pa_\mu \U^{I+} \pa_\nu \Gamma_I^+ - \pa_\mu \Gamma_I^+ \pa_\nu \U^{I+} 
	\Big) = \rd \phi^\mu \,\Omega_{\mu\nu}^{++} \chi^\nu \eol
	&=  v^{\1+} v^{\2+} \Big(
		\frac{1}{\z} \rd \phi^\ra \omega_{\ra \rb} \chi^\rb 
		+ (\rd\phi^\ra \,\chi_\ra - \rd \phi^{\bar \rb} \, \chi_{\bar \rb} )
		+ \z \rd \phi^{\bar\ra} \omega_{\bar\ra \bar\rb} \chi^{\bar\rb}
	\Big)~.
\end{align}

\section{The component action in rigid projective superspace}\label{sec:CompActonRigid}
Now that we have established a great deal of preliminary material,
we now can turn to deriving the component action from superspace.
This is fairly involved, so we have chosen to separate the task into
two distinct stages. In this section, we will derive the component
action from \emph{rigid} projective superspace.
This calculation will yield a subset of the terms we actually need.
Of course, the result can already be derived via a reduction to $\cN=1$
superspace as discussed in section \ref{sec:HKN1Red}, and we will be able to compare our result to the
component version of that action. The point of this exercise is
to introduce the techniques we will need in the curved case.
In fact, it will turn out
that reconstructing the rigid terms is actually more involved than
finding the additional supergravity contributions! For this reason, we
will be rather explicit in the calculation.
To emphasize the applicability to the rigid case, we will avoid
assuming in this section that the target space is a cone.

The action we seek to evaluate is
\begin{align}
S = -\frac{1}{2\pi} \oint v_{i}^+ \rd v^{i+}\, \int \rd^4x\, e\, \cL^{--}~, \qquad
\cL^{--} = \frac{1}{16} (\cD^-)^2 (\bar\cD^-)^2 \cF^{++}
\end{align}
with $\cD_A$ the gauge covariant derivative associated with rigid
projective superspace. It will be convenient for later reference in the curved
case to refer to $\cL^{--}$ above as $T_0:= \frac{1}{16} (\cD^-)^2 (\bar\cD^-)^2 \cF^{++}$.
In the curved case, there will be additional terms.

The function $\cF^{++}$ depends on the collective set of projective multiplets
$\cQ^{+} = (\U^{I+}, \breve \U^{\bar I +})$. We will suppress any index on $\cQ^+$
to keep the notation compact, denoting derivatives of the function
$\cF^{++}$ as $\cF_\cQ^+$, $\cF_{\cQ\cQ}$, etc.
The Lagrangian $T_0$ can now be written
\begin{align}\label{eq:T0}
T_0 &= \frac{1}{16} (\cD^-)^2 (\bar \cD^-)^2 \cQ^{+} \,\cF_\cQ^+
	+ \frac{1}{8} \cD^{\alpha -} \bar\cD^{\dalpha -} \cQ^{+}\,
		\cD_\alpha^- \bar\cD_\dalpha^- \cF_\cQ^+
	\eol & \quad
	+ \frac{1}{8} (\cD^-)^2 \bar\cD_\dalpha^- \cQ^{+} \,\bar\cD^{\dalpha -} \cF_\cQ^+
	+ \frac{1}{8} (\bar\cD^-)^2 \cD^\alpha{}^- \cQ^{+} \,\cD_\alpha^- \cF_\cQ^+
	\eol & \quad
	+ \frac{1}{32} (\cD^-)^2 \cQ^{+} \,(\bar\cD^-)^2 \cF_\cQ^+
	+ \frac{1}{32} (\bar\cD^-)^2 \cQ^{+} \,(\cD^-)^2 \cF_\cQ^+
	\eol & \quad
	+ \frac{1}{32} (\cD^-)^2 \cQ^+ \,\bar\cD_\dalpha^- \cQ^+ \,\bar\cD^{\dalpha-} \cQ^+\,
		\cF_{\cQ\cQ\cQ}^-
	+ \frac{1}{32} (\bar\cD^-)^2 \cQ^+ \,\cD^\alpha{}^- \cQ^+ \,\cD_\alpha^- \cQ^+\,
		\cF_{\cQ\cQ\cQ}^-	
	\eol & \quad
	+ \frac{1}{8} \cD^{\alpha -} \bar\cD^{\dalpha -} \cQ^+ \,
		\cD_\alpha^- \cQ^+ \,\bar\cD_\dalpha^- \cQ^+\,
		\cF_{\cQ\cQ\cQ}^{-}
	\eol & \quad
	+ \frac{1}{16} \cD^{\alpha -} \cQ^{+} \,\cD_\alpha^- \cQ^{+}\,
		\bar\cD_\dalpha^- \cQ^{+} \,\bar\cD^{\dalpha -} \cQ^{+} \,
		\cF_{\cQ\cQ\cQ\cQ}^{--}~.
\end{align}
In the first three lines,
we have chosen to write explicit spinor derivatives of $\cF_\cQ^+$
rather than expanding them out. The reason is that
$\cF_Q^+ = (i \Gamma_I^+, -i \breve \Gamma_{\bar I}^+)$ is arctic or
antarctic to order $\q^2$, and applying the auxiliary field equations tells
us quite a bit about these quantities. In contrast, we cannot say
anything about $\cD_\alpha^+ (\bar \cD^+)^2 \cF_\cQ^+$ without applying
dynamical equations of motion, so we have written $T_0$ in a particular
way to avoid such terms.

Even in the rigid case, the expressions for the various terms we
will encounter can be involved. To simplify the analysis, we will
first consider only those terms that contribute in the rigid
ungauged limit. Afterwards, we will include the covariant terms
associated with the gauged isometries.

\subsection{Rigid ungauged terms}
Denote the first line of \eqref{eq:T0} by $T_{0.1}$. It can be rewritten as
\begin{align}
T_{0.1} &= \frac{i}{16} (\cD^-)^2 (\bar\cD^-)^2 \U^{I+}\, \Gamma_I^+
	+ \frac{i}{8} \cD^{\alpha -} \bar\cD^{\dalpha -} \U^{I+}\,
		\cD_\alpha^- \bar\cD_\dalpha^- \Gamma_I^+
	\eol & \quad
	- \frac{i}{16} (\cD^-)^2 (\bar\cD^-)^2 \breve \U^{\bar I+}\, \breve\Gamma_{\bar I}^+
	- \frac{i}{8} \cD^{\alpha -} \bar\cD^{\dalpha -} \breve\U^{\bar I+}\,
		\cD_\alpha^- \bar\cD_\dalpha^- \breve\Gamma_{\bar I}^+~.
\end{align}
Now let us apply the equations \eqref{eq:DUp} for the components of the arctic multiplet.
Taking only those terms that survive in the rigid ungauged limit, we find
\begin{align}\label{eq:T0.1rigid}
T_{0.1} &\sim i \,\Gamma_I^+\, P^{I(-3)}
	+ \frac{1}{2} z^{--}_1 \Gamma_I^+\, \cD^{\dalpha \alpha} \cD_{\alpha}^- \bar\cD_\dalpha^- \U^{I+}
	+ \frac{i}{2} (z^{--}_1)^2 \Gamma_I^+ \cD_{\alpha \dalpha} \cD^{\dalpha \alpha} \U^{I+}
	\eol & \quad
	- \frac{i}{8} A^{\dalpha \alpha}{}^{I-}\, A_{\alpha\dalpha}{}_I^{-}
	+ \frac{i}{4} z^{--}_1 A^{\dalpha \alpha}{}^{I-} \,\cD_{\alpha \dalpha} \Gamma_I^+
	+ \frac{i}{4} z^{--}_1 \cD^{\dalpha \alpha} \U^{I+}\, A_{\alpha\dalpha}{}_I^{-}
	\eol & \quad
	- \frac{i}{2} (z^{--}_1)^2 \cD^{\dalpha \alpha} \U^{I+}\, \cD_{\alpha\dalpha} \Gamma_I^+
	- (\textrm{antarctic term})~.
\end{align}
The antarctic term is found by replacing $\U^{I+} \rightarrow \breve\U^{\bar I +}$,
$\Gamma_I^+ \rightarrow \breve\Gamma_{\bar I}^+$ and $z^{--}_1 \rightarrow z^{--}_2$.
We will use the symbol $\sim$ to denote the terms we are examining at
each stage of the calculation.

The above expression is valid off-shell, that is, without assuming that
$\Gamma_I^+$ is an arctic superfield. If we did not make this assumption,
we could begin to derive it now. The above expression is the only place where
the arctic component field $P^{I(-3)}$ appears, and so it acts as a Lagrange
multiplier enforcing that the lowest component of $\Gamma_I^+$ is arctic.
We could proceed in this way, rederiving all of the auxiliary equations of
motion, but it would be quite involved. Because we already know their content
-- the lower components of $\Gamma_I^+$ must be arctic --
it is easier to simply assume them without comment. Proceeding in this way,
we see that in the expression \eqref{eq:T0.1rigid},
the terms $\Gamma_I^+ P^{I(-3)}$ and $A^{\dalpha \alpha}{}^{I-} A_{\alpha\dalpha}{}_I^{-}$
drop out under the contour integral as they are purely arctic
expressions -- that is, they are of the form
$\frac{1}{(v^{\1+})^2} \sum_{n=0}^\infty C_n \z^n$
for some field-dependent coefficients $C_n$, and these vanish under the contour integral.
Similarly, their antarctic conjugates, which take the form
$\frac{1}{(v^{\2+})^2} \sum_{n=0}^\infty (-1)^n \bar C_n \z^{-n}$
also vanish. The remaining terms can be rearranged to
\begin{align}
T_{0.1} &\sim
	\frac{i}{4} z_1^{--} \cD^{\dalpha \alpha} \U^{I+}\, A_{\alpha\dalpha}{}_I^{-}
	- \frac{i}{4} z_1^{--} A^{\dalpha \alpha}{}^{I-} \,\cD_{\alpha \dalpha} \Gamma_I^+
	\eol & \quad	
	+ \frac{1}{2} \cD^{\dalpha \alpha} \Big(
		z_1^{--} \Gamma_I^+\, \cD_{\alpha}^- \bar\cD_\dalpha^- \U^{I+}
		+ i (z_1^{--})^2 \Gamma_I^+ \cD^{\dalpha \alpha} \U^{I+}
	\Big)
	- \textrm{(antarctic term)}~.
\end{align}
The total derivative can be discarded in the rigid case.
Now we exploit the on-shell conditions \eqref{eq:onshA} for $A_{\alpha\dalpha}{}^{I-}$
and $A_{\alpha\dalpha}{}_I^{-}$. Using the definition \eqref{eq:defg} of $g_{\ra\bar\rb}$
as well as the relations
\begin{gather}
z_1^{--} - z_2^{--} = \frac{1}{v^{\1+} v^{\2+}}~, \qquad
\cD_{\alpha \dalpha} \U^{I+} = \cD_{\alpha \dalpha} \phi^\mu\, \pa_\mu \U^{I+}~,
\end{gather}
one finds
\begin{align}
T_{0.1} &\sim
	\frac{i}{2} \frac{1}{v^{\1+} v^{\2+}} \,\cD^{\dalpha \alpha} \phi^\ra\,
		\cD_{\alpha\dalpha} \phi^{\bar \rb}\, g_{\ra \bar \rb}
	\eol & \quad
	+ \frac{1}{4} \frac{1}{v^{\1+} v^{\2+}} \,\z_\alpha^\rb \,\bar \z_\dalpha^{\bar \rb}\,
		\cD^{\dalpha \alpha} \phi^\mu\,
		\Big(
			z_1^{--} \pa_\rb \pa_{\bar \rb} \U^{I+} \pa_\mu \Gamma_I^+
			- z_1^{--} \pa_\rb \pa_{\bar \rb} \G_I^+ \pa_\mu \U^{I+}
			- \antt
		\Big)~.
\end{align}
(We will occasionally abbreviate antarctic terms as ``a.t.'')

Next, we consider the second line of $T_0$. We have
\begin{align}
T_{0.2} &=
	\frac{i}{8}  \bar\cD_\dalpha^- \G_I^+\, (\cD^-)^2 \bar\cD^{\dalpha-} \U^{I+}
	- \frac{i}{8} \bar\cD_\dalpha^- \breve \G_I^+\, (\cD^-)^2 \bar\cD^{\dalpha-} \breve\U^{\bar I+}
	+ \HC~, \eol
	&\sim
	i \Psi_{I\dalpha}\, \bar \Xi^{\dalpha I--}
	- \frac{z_1^{--}}{2} \Psi_{I \dalpha}\, \cD^{\dalpha \alpha} \Psi_\alpha^I
	- \textrm{(antarctic term)} + \HC
\end{align}
The leading term is purely arctic and vanishes under the contour integral, leaving
\begin{align}
T_{0.2} &\sim
	\frac{1}{4} z_1^{--} \Psi_I^{\alpha} \overleftrightarrow\cD_{\alpha \dalpha} \Psi^{I\dalpha}
	- \frac{1}{4} z_1^{--} \Psi_{I\dalpha}\, \overleftrightarrow\cD^{\dalpha \alpha} \Psi_\alpha^I
	\eol & \quad
	+ \frac{1}{4} \cD^{\dalpha \alpha} \Big(
		z_1^{--} \cD_\alpha^- \Gamma_I^+ \bar\cD_\dalpha^- \U^{I+}
		- z_1^{--} \bar\cD_\dalpha^- \Gamma_I^+ \cD_\alpha^- \U^{I+}
	\Big)
	- \AntT~.
\end{align}
In the rigid case, we can discard the total derivative.
Using the relations \eqref{eq:onshDU},
\begin{align}
\Psi_I^{\alpha} \overleftrightarrow\cD_{\alpha \dalpha} \Psi^{I \dalpha}
	= \frac{1}{v^{\1+} v^{\2+}} \,
		(\z^{\alpha \rb} \pa_\rb \Gamma_I^+) \overleftrightarrow \cD_{\alpha\dalpha}
		(\bar \z^{\dalpha \bar \rb} \pa_{\bar \rb} \U^{I+})
\end{align}
which leads to
\begin{align}\label{eq:Temp5.14}
T_{0.2} &\sim
	\frac{1}{4} \frac{1}{v^{\1+} v^{\2+}}\, g_{\rb \bar \rb}\,
		(\z^{\alpha \rb} \overleftrightarrow{\widehat\cD}_{\alpha\dalpha} \bar\zeta^{\dalpha \bar \rb})
	+ \frac{1}{4} \frac{1}{v^{\1+} v^{\2+}} 
		\z^{\alpha \rb} \bar \z^{\dalpha \bar \rb} \cD_{\alpha\dalpha} \phi^\mu
		\times
		\eol & \qquad \qquad \Big(
			z_1^{--} \nabla_\mu \pa_\rb \Gamma_I^+ \,\pa_{\bar \rb} \U^{I+}
			- z_1^{--} \nabla_\mu \pa_{\bar \rb} \U^{I+} \,\pa_{\rb} \Gamma_I^+
			- z_1^{--} \nabla_\mu \pa_\rb \U^{I+} \,\pa_{\bar \rb} \Gamma_I^+
			\eol & \qquad \qquad\qquad
			+ z_1^{--} \nabla_\mu \pa_{\bar \rb} \Gamma_I^+\, \pa_\rb \U^{I+}
			- \antt
		\Big)~.
\end{align}

At this stage we have recovered the kinetic terms for both scalars
and fermions, along with some extra terms involving two fermions
and a spacetime derivative of $\phi^\mu$ that should be absent
in the final action. The only other contribution involving such terms is
in the fifth line of $T_0$:
\begin{align}
T_{0.5} &= \frac{1}{8} \cD^{\alpha-} \bar\cD^{\dalpha-} \U^{I+}\,
	\cD_\alpha^- \cQ^+\, \bar\cD_\dalpha^- \cQ^+\, \cF_{I\cQ\cQ}^-
	+ \antt \eol
	&\sim \frac{i}{4} \Big(\frac{1}{v^{\1+} v^{\2+}} \cD^{\dalpha\alpha} \bar\phi^{\bar \rb} \pa_{\bar \rb} \U^{I+}
		- z_1^{--} \cD^{\alpha\dalpha} \U^{I+}\Big)
		\cD_\alpha^- \cQ^+\, \bar\cD_\dalpha^- \cQ^+\, \cF_{I\cQ\cQ}^-
	\eol & \quad
	- \frac{i}{4} \Big(\frac{1}{v^{\1+} v^{\2+}} \cD^{\dalpha\alpha} \phi^{\rb} \pa_{\rb} \breve\U^{\bar I+}
		+ z_2^{--} \cD^{\alpha\dalpha} \breve\U^{\bar I+}\Big)
		\cD_\alpha^- \cQ^+\, \bar\cD_\dalpha^- \cQ^+\, \cF_{\bar I\cQ\cQ}^-
	\eol & \quad
	- \frac{1}{8} \frac{1}{v^{\1+} v^{\2+}} \z^{\alpha \rb} \bar\z^{\dalpha \bar \rb}
		\pa_\rb \pa_{\bar \rb} \cQ^+ \,\cD_\alpha^- \cQ^+\, \bar\cD_\dalpha^- \cQ^+\, \cF_{\cQ \cQ\cQ}^-~.
\end{align}
The terms in question can be written as
\begin{align}
T_{0.1} + T_{0.2} + T_{0.5}\sim \frac{1}{4 v^{\1+} v^{\2+}} \z^{\alpha \rb} \,\bar \z^{\dalpha \bar \rb} \,\cD_{\alpha\dalpha} \phi^\rc \cS_{\rc \rb \bar \rb}
	+ \HC
\end{align}
where $\cS_{\rc \rb \bar \rb}$ is given by
\begin{align}
\cS_{\rc \rb \bar \rb} &=
	z_1^{--} (\nabla_\rc \pa_\rb \Gamma_I^+ \, \pa_{\bar \rb} \U^{I+}
	- \nabla_\rc \pa_\rb \U^{I+}\, \pa_{\bar \rb} \Gamma_I^+)
	+ 2 z_1^{--} \nabla_{[\rc}  \Big(
		\pa_{\rb]} \U^{I+}\, \pa_{\bar \rb} \Gamma_I^+
		- \pa_{\rb]} \Gamma_I^+\, \pa_{\bar \rb} \U^{I+}
	\Big)
	\eol & \quad
	- z_2^{--} (\nabla_\rc \pa_\rb \breve\Gamma_{\bar I}^+ \, \pa_{\bar \rb} \breve\U^{\bar I+}
	- \nabla_\rc \pa_\rb \breve\U^{\bar I+}\, \pa_{\bar \rb} \breve\Gamma_{\bar I}^+)
	- 2 z_2^{--} \nabla_{[\rc}  \Big(
		\pa_{\rb]} \breve\U^{\bar I+}\, \pa_{\bar \rb} \breve\Gamma_{\bar I}^+
		- \pa_{\rb]} \breve\Gamma_{\bar I}^+\, \pa_{\bar \rb} \breve\U^{\bar I+}
	\Big)
	\eol & \quad
	+ i z_1^{--} \pa_\rc \cQ^+\, \pa_\rb\cQ^+\, \pa_{\bar \rb}\cQ^+\, \cF_{\cQ\cQ\cQ}^-~.
\end{align}
We want to show that this expression vanishes.
The second and fourth terms of $\cS_{\rc \rb \bar \rb}$
are proportional to $\nabla_{[\rc} g_{\rb] \bar \rb}$,
which vanishes. With the useful identity
\begin{align}\label{eq:dddF1}
\pa_\mu \cQ^+\, \pa_\nu \cQ^+\, \pa_\rho \cQ^+\, \cF_{\cQ\cQ\cQ}^-
	&= -i \,\pa_\mu \pa_\nu \U^{I+} \, \pa_\rho \Gamma_I^+ + i \,\pa_\rho \U^{I+}\, \pa_\mu \pa_\nu \Gamma_I^+
	\eol & \quad
	+ i \,\pa_\mu \pa_\nu \breve\U^{\bar I+} \, \pa_\rho \breve\Gamma_{\bar I}^+ - i \,\pa_\rho \breve \U^{\bar I+}\, \pa_\mu \pa_\nu \breve \Gamma_{\bar I}^+~,
\end{align}
the remaining terms can be rewritten as
\begin{align}
\cS_{\rc \rb \bar \rb} &=
	\frac{1}{v^{\1+} v^{\2+}} \Big[\nabla_\rc (\pa_\rb \breve\Gamma_{\bar I}^+ \, \pa_{\bar \rb} \breve\U^{\bar I+}
		- \pa_\rb \breve\U^{\bar I+} \, \pa_{\bar \rb} \breve\Gamma_{\bar I}^+)
	- (\pa_\rb \breve\Gamma_{\bar I}^+\, \pa_{\bar \rb} \pa_\rc \breve\U^{\bar I+}
		- \pa_\rb \breve\U^{\bar I +} \, \pa_{\bar \rb} \pa_\rc \breve\Gamma_{\bar I}^+)\Big]~.
\end{align}
The first set of terms involves $\nabla_\rc g_{\rb \bar \rb}$, which vanishes.
The second set is purely antarctic and so vanishes under the contour integral.
Thus $\cS_{\rc \rb \bar \rb}$ does indeed drop out.
This leaves
\begin{align}
T_{0.1} + T_{0.2} + T_{0.5}
	&\sim
	\frac{i}{2} \frac{1}{v^{\1+} v^{\2+}} \,\cD^{\dalpha \alpha} \phi^\ra\,
		\cD_{\alpha\dalpha} \phi^{\bar \rb}\, g_{\ra \bar \rb}
	+ \frac{1}{4} \frac{1}{v^{\1+} v^{\2+}}\, g_{\rb \bar \rb}\,
		(\z^{\alpha \rb} \overleftrightarrow{{\widehat\cD}}_{\alpha\dalpha} \bar\zeta^{\dalpha \bar \rb})
	\eol & \quad
	+ \frac{1}{8} \frac{1}{(v^{\1+} v^{\2+})^2} \z^\ra \z^\rb \bar \z^{\bar \ra} \bar\z^{\bar \rb}
		\,\pa_\ra \pa_{\bar \ra} \cQ^+ \, \pa_\rb \cQ^+ \pa_{\bar\rb} \cQ^+ \cF_{\cQ \cQ\cQ}^-
	~.
\end{align}

The only term we must still reconstruct is the four fermion term
involving the hyperk\"ahler curvature. It should be
found by including the remaining terms in the third, fourth and sixth lines of $T_0$.
The third line is purely arctic or antarctic in the absence of gauged isometries
so we may ignore it. The fourth and sixth lines give
\begin{align}
T_{0.4} &= \frac{1}{32} (\cD^-)^2 \U^{I+}\, \bar\cD_\dalpha^- \cQ^+\, \bar\cD^\dalpha{}^- \cQ^+ \, \cF_{I\cQ\cQ}^-
	+ \frac{1}{32} (\bar\cD^-)^2 \U^{I+}\, \cD^\alpha{}^- \cQ^+\, \cD_\alpha^- \cQ^+ \, \cF_{I\cQ\cQ}^-
	+ \antt \eol
	&\sim \frac{1}{32} \frac{\z^\ra \z^\rb \bar \z^{\bar \ra} \bar \z^{\bar \rb}}{(v^{\1+} v^{\2+})^2}
	\Big(
	\nabla_\ra \pa_{\rb} \cQ^+ \, \pa_{\bar \ra} \cQ^+\, \pa_{\bar \rb} \cQ^+
	+ \nabla_{\bar \ra} \pa_{\bar\rb} \cQ^+ \, \pa_{\ra} \cQ^+\, \pa_{\rb} \cQ^+
	\Big)\, \cF_{\cQ\cQ\cQ}^-~, \\
T_{0.6} &\sim
	\frac{1}{16} \frac{1}{(v^{\1+} v^{\2+})^2} \,\z^\ra \z^\rb \, \bar \z^{\bar \ra} \bar \z^{\bar \rb}
		(\pa_\ra \cQ^+ \, \pa_\rb \cQ^+\, \pa_{\bar \rb} \cQ^+\, \pa_{\bar \rb} \cQ^+\, \cF_{\cQ\cQ\cQ\cQ}^{--})
\end{align}
Combining all terms, we have
\begin{align}
T_0 &\sim
	\frac{i}{v^{\1+} v^{\2+}} \Big[
	\frac{1}{2}  \,\cD^{\dalpha \alpha} \phi^\ra\,
		\cD_{\alpha\dalpha} \phi^{\bar \rb}\, g_{\ra \bar \rb}
	- \frac{i}{4} g_{\rb \bar \rb}\,
		(\z^{\alpha \rb} \overleftrightarrow{{\widehat\cD}}_{\alpha\dalpha} \bar\zeta^{\dalpha \bar \rb})\Big]
	+ \frac{1}{16} \frac{1}{(v^{\1+} v^{\2+})^2}\,
		\z^\ra \z^\rb \bar \z^{\bar \ra} \bar\z^{\bar \rb}\, \cR_{\ra \rb \bar \ra \bar \rb}^{++}
\end{align}
where
\begin{align}
\cR_{\ra \rb \bar \ra \bar \rb}^{++}
	&= \frac{1}{2} \nabla_{\bar\ra} \Delta_{\bar \rb \ra \rb}^{++} + \HC~, \eol
\Delta_{\bar \ra \rb\rc}^{++} &:=
	\pa_{\bar \ra} \cQ^+\, \pa_\rb\cQ^+\, \pa_\rc \cQ^+\, \cF^-_{\cQ\cQ\cQ} \eol
	&= -i \nabla_\rb \pa_\rc \U^{I+}\, \pa_{\bar \ra} \Gamma_I^+
	+ i \nabla_\rb \pa_\rc \Gamma_I^+\, \pa_{\bar \ra} \U^{I+}
	+ i \pa_{\bar \ra} \pa_\rb \breve\U^{\bar I+}\, \pa_\rc \breve\Gamma_{\bar I}^+
	- i \pa_{\bar \ra} \pa_\rb \breve\Gamma_{\bar I}^+\,\pa_\rc \breve\U^{\bar I+}~.
\end{align}
The contribution to $\Delta_{\bar \rb \ra \rb}^{++}$ from the antarctic fields
is purely antarctic when divided by $(v^{\1+} v^{\2+})^2$, so it drops out.
The remaining terms contribute to $R^{++}_{\ra \rb \bar \ra \bar \rb}$ as
\begin{align}
\cR_{\ra\rb \bar \ra \bar \rb}^{++} &\sim
	-i \nabla_{\bar \ra} \nabla_\rb \pa_\ra \U^{I+}\, \pa_{\bar \rb} \Gamma_I^+
	+ i \nabla_{\bar \ra} \nabla_\rb \pa_\ra \Gamma_I^+\, \pa_{\bar \rb} \U^{I+}
	\eol & \quad
	-i \nabla_\rb \pa_\ra \U^{I+}\, \nabla_{\bar \ra} \pa_{\bar \rb} \Gamma_I^+
	+ i \nabla_\rb \pa_\ra \Gamma_I^+\, \nabla_{\bar \ra} \pa_{\bar \rb} \U^{I+}
	+ \HC
\end{align}
The second line vanishes under the contour integral since
\begin{align}
\nabla_{\bar\ra} \pa_{\bar\rb} \U^{I+}
	= \Big(\frac{v^{\2+}}{v^{\1+}}\Big)^2 \omega_{\bar \rb}{}^\rb \omega_{\bar \ra}{}^\ra \nabla_\ra \pa_\rb \U^{I+}
\end{align}
is purely arctic when divided by $(v^{\1+} v^{\2+})^2$.
The first line simplifies in the same way after commuting $\nabla_{\bar \ra}$
to act on the arctic superfields. This leaves
\begin{align}
\cR_{\ra\rb \bar \ra \bar \rb}^{++} &\sim
	-i [\nabla_{\bar \ra}, \nabla_\rb] \pa_\ra \U^{I+}\, \pa_{\bar \rb} \Gamma_I^+
	+ i [\nabla_{\bar \ra}, \nabla_\rb] \pa_\ra \Gamma_I^+\, \pa_{\bar \rb} \U^{I+}
	+ \HC \eol
	&= i R_{\rb \bar \ra \ra}{}^\rc \Big(\pa_\rc \U^{I+}\, \pa_{\bar \rb} \Gamma_I^+ -
	\pa_\rc \G_I^{+}\, \pa_{\bar \rb} \U^{I+} \Big)
	= i v^{\1+} v^{\2+} R_{\ra \bar \ra \rb \bar \rb}~.
\end{align}

The final result for the rigid ungauged contributes to $T_0$, including the previously
discarded total covariant derivatives, is 
\begin{align}
T_0 &\sim \frac{i}{v^{\1+} v^{\2+}} \Big(
	\frac{1}{2} \cD^{\dalpha \alpha} \phi^\ra\,
		\cD_{\alpha\dalpha} \phi^{\bar \rb}\, g_{\ra \bar \rb}
	- \frac{i}{4} \, g_{\rb \bar \rb}\,
		(\z^{\alpha \rb} \overleftrightarrow{{\widehat\cD}}_{\alpha\dalpha} \bar\zeta^{\dalpha \bar \rb})
	+ \frac{1}{16} \z^\ra \z^\rb \bar \z^{\bar \ra} \bar\z^{\bar \rb}\, R_{\ra \bar \ra\, \rb \bar \rb}
	\Big)
	\eol & \quad
	+ \cD^{\dalpha \alpha} \Big(
		\frac{1}{2} z_1^{--} \Gamma_I^+\, \cD_{\alpha}^- \bar\cD_\dalpha^- \U^{I+}
		+ \frac{i}{2} (z_1^{--})^2 \Gamma_I^+ \cD^{\dalpha \alpha} \U^{I+}
		\eol & \qquad\qquad
		+ \frac{1}{4} z_1^{--} \cD_\alpha^- \Gamma_I^+ \bar\cD_\dalpha^- \U^{I+}
		- \frac{1}{4} z_1^{--} \bar\cD_\dalpha^- \Gamma_I^+ \cD_\alpha^- \U^{I+}
		-\antt
	\Big)~.
\end{align}
We will return to this expression later when we address the supergravity contributions.
Now let us remain with the rigid case and simplify.
The contour integral is completely trivial since only the prefactor varies along $\cC$,
\begin{align}
-\frac{1}{2\pi} \oint_\cC v_i^+ \rd v^{i+} \frac{i}{v^{\1+} v^{\2+}}
	= \oint_\cC \frac{\rd \z}{2\pi i \z} = 1~.
\end{align}
This leads (dropping the total covariant derivative) to
\begin{align}\label{eq:CompLagRigidUngauged}
\cL = \frac{1}{2} \cD^{\dalpha \alpha} \phi^\ra\,
		\cD_{\alpha\dalpha} \phi^{\bar \rb}\, g_{\ra \bar \rb}
	- \frac{i}{4} \, g_{\rb \bar \rb}\,
		(\z^{\alpha \rb} \overleftrightarrow{{\widehat\cD}}_{\alpha\dalpha} \bar\zeta^{\dalpha \bar \rb})
	+ \frac{1}{16} \z^\ra \z^\rb \bar \z^{\bar \ra} \bar\z^{\bar \rb}\, R_{\ra \bar \ra\, \rb \bar \rb}
\end{align}
for the component Lagrangian. This is the general form of the hypermultiplet
action in holomorphic coordinates $\phi^\ra$ in a rigid background in the
absence of gauged isometries.

\subsection{Rigid gauged terms}
Our next step is to collect the terms corresponding to gauged isometries.
We return to the expression \eqref{eq:T0} and collect the terms
involving the scalar field $W^r$, the gaugino $\l_{\alpha i}^r$
and the auxiliary field $Y^{r ij}$  of the vector multiplet.

Let us begin with terms involving $Y^{r ij}$.
These come only from the first term of $T_0$, which gives
\begin{align}
T_0 &\sim 3i \, z_1^{--} \,Y^{r --} \cJ_r^{I+}\, \Gamma_I^+
	- 3i (z_1^{--})^2 \, Y^{r -+} \cJ_r^{I+}\, \Gamma_I^+
	+ i (z_1^{--})^3 Y^{r++} \cJ_r^{I+}\, \Gamma_I^+
	- \antt \eol
	&\sim i (z_1^{--}-z_2^{--}) \Big[Y^{r--} D_r^{++}
		- 2 Y^{r-+} D_r^{+-}
		+ Y^{r++} D_r^{--} \Big]
	\eol & \quad
	+ D^{--} \Big[
		i (z_1^{--}-z_2^{--}) \Big(2 Y^{r+-} D_r^{++}
		- Y^{r++} D_r^{+-}
		- \frac{1}{2} (z_1^{--}+ z_2^{--}) Y^{r++} D_r^{++} \Big)
	\Big]
\end{align}
after identifying $\cJ_r^{I+}\, \Gamma_I^+ \equiv D_r^{++}$ as the $\cN=2$
moment map.
The argument of $D^{--}$ is holomorphic away from the poles,
so it can be discarded, leaving
\begin{align}
T_0 \sim \frac{i}{v^{\1+} v^{\2+}} Y^r_{ij} \, D_r^{ij}~.
\end{align}

Now let's collect all terms involving the chiral gaugino $\lambda^r_{\alpha i}$.
These are
\begin{align}
T_{0} &\sim i \Gamma_I^+ (2 z_1^{--} \bm\lambda^{\alpha r-} \cD_{\alpha}^- \cJ_r^{I+}
	- (z_1^{--})^2 \bm\lambda^{\alpha r+} \cD_\alpha^- \cJ_r^{I+})
	\eol & \quad
	+ i \cD^{\alpha -} \Gamma_I^+ (2 z_1^{--} \bm\lambda_\alpha^{r-} \cJ_r^{I+}
		- (z_1^{--})^2 \bm\lambda_\alpha^{r+} \cJ_r^{I+} )
	- \AntT \eol
	&= i (z_1^{--} - z_2^{--}) \bm\lambda^{\alpha r -} \cD_\alpha^- D_r^{++}
	- i (z_1^{--} - z_2^{--}) \bm\lambda^{\alpha r+} D^{--} \cD_\alpha^- D_r^{++}
	\eol & \quad
	+ i D^{--} \Big[(z_1^{--} - z_2^{--} )
		\bm\lambda^{\alpha r +} \cD_\alpha^- D_r^{++}
	\Big]~.
\end{align}
Recalling that
$\cD_\alpha^- D_r^{++} = \homega_{r \,\alpha}^+ = v_i^+ \, \homega_{r \,\alpha}^i$
is holomorphic, the argument of $D^{--}$ is clearly holomorphic so it can be discarded.
The remaining terms can be rewritten as
\begin{align}
T_0 \sim \frac{i}{v^{\1+} v^{\2+}} \bm\lambda^\alpha_i{}^r \homega_{r \,\alpha}^i~.
\end{align}

Now let's address all terms involving $W^r$ or its conjugate.
Those from the first two lines of $T_0$ give (after some algebra)
\begin{align}
T_{0.1} + T_{0.2} &\sim
	\frac{i}{12} z_1^{--} \Big(\bar W^s \,\bar\cD_{ij} D_s^{ij} + \HC\Big)
	- \frac{i}{4} z_1^{--} \bar W^s (\bar\cD^-)^2 \Gamma_I^+\, \cJ_s^{I+}
	- \frac{i}{4} z_1^{--} W^r (\cD^-)^2 \Gamma_I^+\, \cJ_r^{I+}
	\eol & \quad
	- \frac{i}{2} (z_1^{--})^2 W^r \bar W^s \Gamma_I^+ \{X_r, X_s\} \U^{I+}
	- \AntT~.
\end{align}
Now we combine these with those terms from $T_{0.3}$, which can be rewritten as
\begin{align}
T_{0.3} &\sim \frac{i}{2} \Big(M_I^- - z_1^{--} \bar W^s \cJ_{sI}^+\Big)
	\Big(N^{I-} - z_1^{--} W^r \cJ_r^{I+}\Big)
	\eol & \quad
	+ \frac{i}{2} \Big(N_I^- - z_1^{--} W^r \cJ_{rI}^+\Big)
		\Big(M^{I-} - z_1^{--} \bar W^s \cJ_s^{I+}\Big) - \AntT~.
\end{align}
The terms in $T_{0.3}$ independent of $z_1^{--}$ vanish under the contour
integral as they are purely arctic. Combining all the other terms gives
\begin{align}
T_{0.1} + T_{0.2} + T_{0.3} &\sim
	\frac{i}{2} \frac{1}{v^{\1+} v^{\2+}} \, W^r \bar W^s J_r^\mu J_s^\nu g_{\mu\nu}
	+ \frac{i}{12} \frac{1}{v^{\1+} v^{\2+}} \, \Big(\bar W^s \,\bar\cD_{ij} D_s^{ij} + \HC\Big)
	\eol & \quad
	+ \bigg[\frac{i}{8} \frac{z_1^{--}}{v^{\1+} v^{\2+}} W^r J_r^\mu\, \z^\ra \z^\rb \omega_\rb{}^{\bar \rb} \Big(
	\pa_\mu \U^{I+}\, \pa_\ra \pa_{\bar \rb} \Gamma_I^+ -  \pa_\mu \Gamma_I^+ \, \pa_\ra \pa_{\bar \rb} \U^{I+}
	\Big) - \antt \bigg]
	\eol & \quad
	- \bigg[\frac{i}{8} \frac{z_1^{--}}{v^{\1+} v^{\2+}} \bar W^r J_r^\mu\, \bar\z^{\bar \ra} \bar\z^{\bar \rb} \omega_{\bar \rb}{}^{\rb} \Big(
	\pa_\mu \U^{I+}\, \pa_{\bar \ra} \pa_{\rb} \Gamma_I^+ - \pa_\mu \Gamma_I^+ \, \pa_{\bar \ra} \pa_{\rb} \U^{I+}
	\Big)- \antt\bigg]~.
\end{align}
The only other term containing explicit $W^r$ or $\bar W^r$ contributions comes from
$T_{0.4}$:
\begin{align}
T_{0.4} &\sim
	\frac{1}{8} \frac{1}{v^{\1+} v^{\2+}}
	\bar W^s J_s^\mu \bar\z^{\bar \ra} \bar\z^{\bar \rb} \omega_{\bar \rb}{}^\rb \Big(
	z_1^{--} \pa_\mu \U^{I+} \, \pa_{\bar \ra} \cQ^+\, \pa_{\rb} \cQ^+\, \cF^-_{I\cQ\cQ}
	+ \antt
	\Big)
	\eol & \quad
	- \frac{1}{8} \frac{1}{(v^{\1+} v^{\2+})^2} \, \bar W^s J_s^{\bar \rc} \bar \z^{\bar \ra} \bar\z^{\bar \rb}
		\omega_{\bar \rb}{}^{\rb} \Big(
	\pa_{\bar \rc} \U^{I+} \, \pa_{\bar \ra} \cQ^+\, \pa_{\rb} \cQ^+\, \cF^-_{I\cQ \cQ}
	\Big)
	\eol & \quad
	+ \frac{1}{8} \frac{1}{(v^{\1+} v^{\2+})^2} \, \bar W^s J_s^{\rc} \bar \z^{\bar \ra} \bar\z^{\bar \rb}
		\omega_{\bar \rb}{}^{\rb} \Big(
	\pa_\rc \breve \U^{\bar I+} \, \pa_{\bar \ra} \cQ^+\, \pa_{\rb} \cQ^+\, \cF^-_{\bar I\cQ \cQ}
	\Big) + \HC
\end{align}
Combining all terms that go as $W^r \z^\ra \z^\rb$, we find (after some simplifications)
\begin{align}
	&\frac{1}{8} \frac{1}{(v^{\1+} v^{\2+})^2} W^r J_r^\rc \z^\ra \z^{\rb} \omega_\rb{}^{\bar \rb} \Big(
	i \pa_\rc \breve\U^{\bar I+} \, \pa_\ra \pa_{\bar \rb} \breve\Gamma_{\bar I}^+
	- i \pa_\rc \breve\Gamma_{\bar I}^+ \, \pa_\ra \pa_{\bar \rb} \breve\U^{\bar I+}
	\Big) \eol
& + \frac{1}{8} \frac{1}{(v^{\1+} v^{\2+})^2} W^r J_r^{\bar \rc} \z^\ra \z^{\rb} \omega_\rb{}^{\bar \rb} \Big(
	i \pa_{\bar \rc} \U^{I+} \, \pa_\ra \pa_{\bar \rb} \Gamma_I^+
	- i \pa_{\bar \rc} \Gamma_I^+ \, \pa_\ra \pa_{\bar \rb} \U^{I+}
	\Big) ~.
\end{align}
The first line is purely antarctic and the second is arctic, so these drop out.

The full contribution from gauged isometries yields
\begin{align}
T_0 &\sim 
	\frac{i}{v^{\1+} v^{\2+}}
	\Big(Y^r_{ij} \, D_r^{ij}
	+ \frac{1}{2} W^r \bar W^s J_r^\mu J_s^\nu g_{\mu\nu}
	+ (\bm\lambda^\alpha_i{}^r \homega_{r \,\alpha}^i + \frac{1}{12} W^r \,\cD_{ij} D_r^{ij} + \HC)
	\Big)~.
\end{align}
Using the explicit relation \eqref{eq:DDDij}
for $\cD_{ij} D^{ij}_r$,
performing the contour integral, and combining with the result
\eqref{eq:CompLagRigidUngauged} of the previous section,
we find the full rigid Lagrangian
\begin{align}\label{eq:CompLagRigid}
\cL &= \frac{1}{2} \cD^{\dalpha \alpha} \phi^\ra\,
		\cD_{\alpha\dalpha} \phi^{\bar \rb}\, g_{\ra \bar \rb}
	- \frac{i}{4} \, g_{\rb \bar \rb}\,
		(\z^{\alpha \rb} \overleftrightarrow{{\widehat\cD}}_{\alpha\dalpha} \bar\zeta^{\dalpha \bar \rb})
	+ \frac{1}{16} \z^\ra \z^\rb \bar \z^{\bar \ra} \bar\z^{\bar \rb}\, R_{\ra \bar \ra\, \rb \bar \rb}
	\eol & \quad
	+ Y^r_{ij} \, D_r^{ij}
	+ (\bm\lambda^\alpha_i{}^r \homega_{r \,\alpha}^i + \HC)
	- \frac{1}{2} W^r \bar W^s J_r^\mu J_s^\nu g_{\mu\nu}
	\eol & \quad
	- \frac{1}{4} W^r \z^\ra \z^\rb \nabla_\ra (\omega_{\rb \rc} J_r^\rc)
	- \frac{1}{4} \bar W^r \bar\z^{\bar\ra} \bar\z^{\bar\rb} \nabla_\ra (\omega_{\bar\rb \bar\rc} J_r^{\bar\rc})~.
\end{align}
This coincides with the component version of the $\cN=1$ action \cite{HuKLR}
\begin{align}\label{eq:N1version}
S = \int \rd^4x\, \rd^4\q\, K - \int \rd^4x\, \rd^2\q\, W^r \L_r
	- \int \rd^4x\, \rd^2\bar\q\, \bar W^r \bar\L_r
\end{align}
upon eliminating the auxiliary fields associated with the
$\cN=1$ chiral multiplets $\phi^\ra$.\footnote{A proof
using flat projective superspace was given in
\cite{GonzalezRey:FR2}
for the case of a single frozen vector multiplet
(see also \cite{Kuzenko:Superpotentials, dWRV:QK, BKLT-M:AdSPro, BuKuTM:SigmaAdS3}).}
The superpotential contribution couples the chiral superfield $W^r$ of
the $\cN=2$ vector multiplet to the chiral
component $\L_r$ of the $\cN=2$ moment map.
The remaining component of the moment map is contained
implicitly within the K\"ahler potential.

\section{The component action in curved projective superspace}\label{sec:CompActonCurved}
Having derived the correct component action in the rigid limit, we must now
include all of the effects of supergravity. The component action from curved
projective superspace is
\begin{align}\label{eq:ProjCompAction}
S = -\frac{1}{2\pi} \oint_\cC v_{i}^+ \rd v^{i+}\, \int \rd^4x\, e\, \cL^{--}
	+ \frac{1}{2\pi} \oint_\cC v_i^- \rd v^{i-}\, \int \rd^4x\, e\, \cL^{++}~,
\end{align}
where
\begin{align}\label{eq:Lmm}
\cL^{--} &= \frac{1}{16} (\nabla^-)^2 (\bar\nabla^-)^2 \cF^{++}
	- \frac{i}{8} (\bar\psi_m^- \bsigma^m)^\alpha \nabla_\alpha^- (\bar\nabla^-)^2 \cF^{++}
	- \frac{i}{8} (\psi_m^- \sigma^m)_\dalpha \bar\nabla^{\dalpha -} (\nabla^-)^2 \cF^{++}
	\eol & \quad
	+ \frac{1}{4} \bigg(
	(\psi_n^- \sigma^{nm})^\alpha \bar\psi_m{}^{\dalpha-}
	+ \psi_n{}^{\alpha-}  (\bsigma^{nm}\bar\psi_m^-)^\dalpha
	- i \cV_m^{--} (\sigma^m)_{\alpha \dalpha} \bigg) [\nabla_\alpha^{-}, \bar\nabla_\dalpha^{-}]  \cF^{++}
	\eol & \quad
	+ \frac{1}{4} (\psi_m^- \sigma^{mn} \psi_n^-) (\nabla^-)^2 \cF^{++}
	+ \frac{1}{4} (\bar\psi_m^- \bsigma^{mn} \bar\psi_n^-) (\bar \nabla^-)^2 \cF^{++}
	\eol & \quad
	- \bigg(
	\frac{1}{2} \eps^{mnpq} (\psi_m^- \sigma_n \bar\psi_p^-) \psi_q^{\alpha -}
	- 2 (\psi_m^- \sigma^{mn})^\alpha \cV_n^{--} \bigg) \nabla_\alpha^- \cF^{++}
	\eol & \quad
	+ \bigg(
	\frac{1}{2} \eps^{mnpq} (\bar\psi_m^- \bsigma_n \psi_p^-) \bar\psi_{q\dalpha}^{-}
	- 2 (\bar\psi_m^- \bsigma^{mn})_\dalpha \cV_n^{--} \bigg) \bar\nabla^{\dalpha -} \cF^{++}
	\eol & \quad
	+ 3 \eps^{mnpq} (\psi_m^- \sigma_n \bar\psi_p^-) \cV_q^{--} \cF^{++}~, \\
\cL^{++}
	&= -\Big[3 D 
	+ 4 f_a{}^a 
	- 4 (\bar \psi_m^- \bsigma^{mn} \hat{\bar\phi}_n^+) 
	+ 4 (\psi_m^{-} \sigma^{mn} \hat\phi_n^{+}) 
	- 3\, \eps^{mnpq}
		 (\psi_m^{-} \sigma_n \bar\psi_p^{-}) \cV_q^{++}
	\Big] \cF^{++}
	\eol & \quad
	+ \Big[\frac{3}{2} \chi^{\alpha +}
	- i (\bar \phi_m^{+} \bsigma^m)^\alpha 
	+ 2 (\psi_m^- \sigma^{mn})^\alpha \cV_n^{++} \Big] \nabla_\alpha^- \cF^{++}
	\eol & \quad
	- \Big[\frac{3}{2} \chi_\dalpha^{+}
	- i (\phi_m^{+} \sigma^m)_\dalpha 
	+ 2 (\bar\psi_m^- \bsigma^{mn})_\dalpha \cV_n^{++}\Big] \bar\nabla^{\dalpha -} \cF^{++}
	\eol & \quad
	- \frac{i}{4} \cV_m^{++} (\bsigma^m)^{\dalpha \alpha} [\nabla_\alpha^-, \bar\nabla_\dalpha^-] \cF^{++}~.
\end{align}
As discussed in \cite{Butter:CSG4d.Proj}, it is always possible to complexify the
auxiliary $\SU{2}$ manifold to $\rm SL(2, \mathbb C)$ and then choose
a contour where $u_i$ is constant.
The resulting formulation of projective superspace is exactly that given in
\cite{KLRT-M1, KLRT-M2, KT-M:DiffReps, Kuzenko:DualForm} and is clearly advantageous for evaluating
component actions: the integral involving $\cL^{++}$ automatically vanishes.

However, we will take an even larger shortcut which avoids the
need to make any such choice.
Because we are dealing with a hyperk\"ahler cone, we can write, using \eqref{eq:defGamma},
\begin{align}\label{eq:Fhomo}
\cF^{++} = \frac{i}{2} \Gamma_I^+ \U^{I+} - \frac{i}{2} \breve\Gamma_{\bar I}^+ \breve\U^{\bar I+}~.
\end{align}
If we were to take all of the component fields to be on-shell,
then $\Gamma_I^+$ would be an arctic multiplet to all orders in its $\q$ expansion.
But then the superspace Lagrangian would be built out of the sum of an
arctic and an antarctic superfield and \emph{such Lagrangians vanish even in
curved space}. The proof of this statement is quite simple
(see \cite{KT-M:DiffReps} for an equivalent argument in $\SU{2}$ superspace)
and we will review it in appendix \ref{app:ArcticAction}.

Of course, what we seek to do is to put only the auxiliary fields on-shell,
so that the above argument holds
only up through $\q^2$. This means that we can
use \eqref{eq:Fhomo} to evaluate all terms \emph{except} the leading ones
in $\cL^{--}$. If we write the action $S$ as
\begin{align}\label{eq:CleverCompAction}
S &= -\frac{1}{2\pi} \oint_\cC v_{i}^+ \rd v^{i+}\, \int \rd^4x\, e\, (T_0 + T_1) + S_{\rm rest}~, \eol
T_0 &= \frac{1}{16} (\nabla^-)^2 (\bar\nabla^-)^2 \cF^{++}~, \qquad
T_1 = - \frac{i}{8} (\bar\psi_m^- \bsigma^m)^\alpha \nabla_\alpha^- (\bar\nabla^-)^2 \cF^{++}
	+ \HC~,
\end{align}
we will find that all contributions to the component action arise solely from $T_0$
and $T_1$. There will be additional remainder terms within
$T_0$ and $T_1$ that involve
only the combination $\Gamma_I^+ \U^{I+}$ and its conjugate; when combined
with $S_{\rm rest}$, which also depends solely on this combination, all these
terms will turn out to vanish.

One final point: to avoid confusing the CKV $\chi^\mu$ of the target space
with the spinor field $\chi_{\alpha i}$ of the conformal supergravity multiplet,
we will from now on always arrange to lower the target space index of the CKV
so that we deal instead with $K_\mu = (K_\ra, K_{\bar \ra}) \equiv \chi_\mu$
or we will rewrite it as $A_i{}^\ra = \chi^\mu f_\mu{}_i{}^\ra$.

\subsection{Remaining evaluation of $T_0$}
The first term $T_0$ is the most complicated as it must generate all
the interactions present even in the rigid supersymmetric theory.
For this reason, we have already performed the majority of its
evaluation in the previous section. We again can decompose $T_0$
as in \eqref{eq:T0}, replacing gauged covariant derivatives with
supergravity covariant derivatives.
Taking all the terms that contributed in the rigid limit, we found
\begin{align}
T_0 &\sim \frac{i}{v^{\1+} v^{\2+}} \Big[
	\frac{1}{2} \nabla^{\dalpha \alpha} \phi^\ra\,
		\nabla_{\alpha\dalpha} \phi^{\bar \rb}\, g_{\ra \bar \rb}
	- \frac{i}{4} \, g_{\rb \bar \rb}\,
		(\z^{\alpha \rb} \overleftrightarrow{{\widehat\nabla}}_{\alpha\dalpha} \bar\zeta^{\dalpha \bar \rb})
	+ \frac{1}{16} \z^\ra \z^\rb \bar \z^{\bar \ra} \bar\z^{\bar \rb}\, R_{\ra \bar \ra\, \rb \bar \rb}
	\eol & \qquad \qquad
	+ Y^r_{ij} \, D_r^{ij}
	+ (\bm\lambda^\alpha_i{}^r \homega_{r \,\alpha}^i + \HC)
	- \frac{1}{2} W^r \bar W^s J_r^\mu J_s^\nu g_{\mu\nu}
	\eol & \qquad \qquad
	- \frac{1}{4} W^r \z^\ra \z^\rb \nabla_\ra (\omega_{\rb \rc} J_r^\rc)
	- \frac{1}{4} \bar W^s \bar\z^{\bar\ra} \bar\z^{\bar\rb} \nabla_\ra (\omega_{\bar\rb \bar\rc} J_r^{\bar\rc})
	\Big]
	\eol & \quad
	+ \nabla^{\dalpha \alpha} \Big[
		\frac{1}{2} z_1^{--} \Gamma_I^+\, \nabla_{\alpha}^- \bar\nabla_\dalpha^- \U^{I+}
		+ \frac{i}{2} (z_1^{--})^2 \Gamma_I^+ \nabla^{\dalpha \alpha} \U^{I+}
		\eol & \qquad\qquad
		+ \frac{1}{4} z_1^{--} \nabla_\alpha^- \Gamma_I^+ \bar\nabla_\dalpha^- \U^{I+}
		- \frac{1}{4} z_1^{--} \bar\nabla_\dalpha^- \Gamma_I^+ \nabla_\alpha^- \U^{I+}
		-\antt
	\Big]~,
\end{align}
up to terms involving covariant supergravity fields.
The reader may check that all of the steps followed in the rigid case
to derive this expression follow equally well here.
Now we must include the covariant supergravity fields.
The steps required to derive these terms are quite similar to those
used for the covariant fields of the vector multiplet, so we will
omit an explicit discussion. Finally, it will be convenient to rewrite
\begin{align}
\frac{1}{2} \nabla^{\alpha\dalpha} \phi^\ra \nabla_{\alpha\dalpha} \phi^{\bar \rb}\,
	g_{\ra \bar\rb}
	&= \frac{1}{4} \nabla^{\dalpha \alpha} \nabla_{\alpha\dalpha} K 
	+ \frac{1}{2} K_\mu \widehat\nabla_a \nabla^a \phi^\mu
\end{align}
where $\widehat\nabla_a$ carries the target space connection.
(This formula is valid when $K$ describes a hyperk\"ahler cone.)
After this rewriting, we find
\begin{align}\label{eq:T0final}
T_0 &= \frac{i}{v^{\1+} v^{\2+}} \bigg[
	\frac{1}{2} K_\mu \widehat {\Box} \phi^\mu
	- \frac{i}{4} \, g_{\rb \bar \rb}\,
		(\z^{\alpha \rb} \overleftrightarrow{{\widehat\nabla}}_{\alpha\dalpha} \bar\zeta^{\dalpha \bar \rb})
	+ \frac{1}{16} \z^\ra \z^\rb \bar \z^{\bar \ra} \bar\z^{\bar \rb}\, R_{\ra \bar \ra\, \rb \bar \rb}
	\eol & \qquad\qquad
	+ Y^r_{ij} \, D_r^{ij}
	+ (\bm\lambda^\alpha_i{}^r \homega_{r \,\alpha}^i + \HC)
	- \frac{1}{2} W^r \bar W^s J_r^\mu J_s^\nu g_{\mu\nu}
	\eol & \qquad\qquad
	- \frac{1}{4} W^r \z^\ra \z^\rb \nabla_\ra (\omega_{\rb \rc} J_r^\rc)
	- \frac{1}{4} \bar W^r \bar\z^{\bar\ra} \bar\z^{\bar\rb} \nabla_\ra (\omega_{\bar\rb \bar\rc} J_r^{\bar\rc})~.
	\eol & \qquad\qquad
	-\frac{3}{2} D\, K
	- \frac{1}{4} (W^{\alpha\beta} \z_\beta^\rb \z_\alpha^\ra \,\omega_{\rb \ra} + \HC)
	- \frac{3}{4} (\bar \chi_\dalpha^j \,\bar\nabla^{\dalpha}_j K  + \HC)
	\bigg]
	\eol & \quad
	+ \frac{3}{2} \Big[i z_1^{--} \bar \chi_\dalpha^- \bar\nabla^{\dalpha -} (\Gamma_I^+ \U^{I+})
	- \frac{i}{2} (z_1^{--})^2 \bar \chi_\dalpha^+ \bar\nabla^{\dalpha -} (\Gamma_I^+ \U^{I+})
	- \AntT
	+ \HC \Big]
	\eol & \quad
	+ \nabla^{\dalpha \alpha} B_{\alpha\dalpha}^{--}
	+ D^{--} B~.
\end{align}
In the final line, we have collected a total covariant derivative, with
\begin{align}
B_{\alpha\dalpha}^{--} &= 
		z_1^{--} \Big(\frac{1}{2} \Gamma_I^+\, \nabla_{\alpha}^- \bar\nabla_\dalpha^- \U^{I+}
		+ \frac{1}{4} \nabla_\alpha \Gamma_I^+ \,\bar\nabla_\dalpha \U^{I+}
		- \frac{1}{4} \bar\nabla_\dalpha \Gamma_I^+ \,\nabla_\alpha \U^{I+}\Big)
		\eol & \quad
		+ \frac{i}{2} (z_1^{--})^2 \Gamma_I^+ \nabla^{\dalpha \alpha} \U^{I+}
		- \antt
		+ \frac{1}{4} \frac{i}{v^{\1+} v^{\2+}} \,\nabla_{\alpha\dalpha} K 
\end{align}
and a total contour derivative with
\begin{align}\label{eq:DefB}
B &= - \frac{3i}{2} z_1^{--} D \,\U^{I+} \Gamma_I^+ + \frac{3i}{2} z_2^{--} D \,\breve\U^{\bar I+} \breve\Gamma_{\bar I}^+
	\eol & \quad
	- \frac{3}{4} \Big(i z_1^{--} \bar \chi_\dalpha^+ \, \bar\nabla^{\dalpha -} (\Gamma_I^+ \U^{I+})
	- i z_2^{--} \bar \chi_\dalpha^+ \, \bar\nabla^{\dalpha -} (\breve\Gamma_{\bar I}^+ \breve\U^{\bar I+})
	+ \HC\Big)~.
\end{align}
Because $B$ is not holomorphic, it cannot be dropped.

Before moving on to evaluate the first set $T_1$ of gravitino terms, we must address what to do with
$\nabla^{\dalpha \alpha} B_{\alpha \dalpha}^{--}$. Because the vector
derivative contains a number of connections -- including supersymmetry,
$S$-supersymmetry, and $\SU{2}$ which are quite non-trivial --
it does not vanish identically and must be separately analyzed.

\subsection{Simplification of the total covariant derivative}\label{sec:CompActionCurved.Simp}
Let us denote the total covariant derivative by
$T_{0.\rm TD} = \nabla^{\dalpha \alpha} B_{\alpha\dalpha}^{--}$.
This can be rewritten as
\begin{align}\label{eq:T0TD}
T_{0. \rm TD} &=
	- \frac{1}{2} \psi^{\dalpha \alpha\, \beta}{}_j \nabla_\beta^j B_{\alpha\dalpha}^{--}
	- \frac{1}{2} \psi^{\dalpha \alpha\,}{}_\dbeta{}^j \bar\nabla^\dbeta_j B_{\alpha\dalpha}^{--}
	\eol & \quad
	- \cV^{\dalpha \alpha \, --} D^{++} B_{\alpha\dalpha}^{--}
	+ D^{--} (\cV^{\dalpha \alpha\, ++} B_{\alpha \dalpha}^{--})
	\eol & \quad
	- \frac{1}{2} \phi^{\dalpha \alpha\, \beta j} S_{\beta j} B_{\alpha\dalpha}^{--}
	- \frac{1}{2} \phi^{\dalpha \alpha}{}_{\dbeta j} \bar S^{\dbeta j} B_{\alpha\dalpha}^{--}
	- f^{\dalpha \alpha}{}^b K_b B_{\alpha\dalpha}^{--}
	\eol & \quad
	+ 2 \,T_{mn}{}^c e_c{}^n  B^{a\,--} e_a{}^m
	- 2 e^{-1} \pa_m (e \,B^{a\,--} e_a{}^m)~.
\end{align}
Note that the last line contains an actual total derivative (which can be discarded)
and a torsion term that arises from the gravitino dependence of the
spin connection.

Before addressing this, let us recall the guiding principle discussed at
the beginning of this section. The superfield $\G_I^+$ is arctic to
order $\q^2$, so any expressions involving at most two
spinor derivatives can be treated more simply than those involving three or more.
Therefore, we will separate the first line of $T_{0. \rm TD}$,
which involves gravitinos and up to three spinor derivatives, and denote it by
$T_{0.\rm TD|Q}$.
We will analyze it shortly, but first let us discuss what happens to the
remaining terms. Denote these by $T_{0.\rm TD|rest}$. Assuming that $\G_I^+$
is an arctic superfield to this order, $B_{\alpha\dalpha}^{--}$ can be written as
\begin{align}\label{eq:V0simple}
B_{\alpha \dalpha}^{--}
	&= \frac{1}{4} z_1^{--} \nabla_\alpha^- \bar\nabla_\dalpha^- (\Gamma_I^+ \U^{I+})
	+ \frac{i}{4} (z_1^{--})^2 \nabla_{\alpha \dalpha} (\Gamma_I^+ \U^{I+})
	- \AntT
	\eol & \quad
	+ D^{--} \Big(
	\frac{1}{8} \frac{1}{v^{\1+} v^{\2+}} \, \nabla_\alpha^+ \bar\nabla_\dalpha^+ K
	\Big)
\end{align}
We emphasize that this equation holds \emph{only} if we do not apply any
further spinor derivatives.
Now when we insert this expression into $T_{0.\rm TD|rest}$, the last
term of $B_{\alpha\dalpha}^{--}$
will vanish as it leads to a total contour derivative
of a holomorphic quantity. This means that the contribution of
$T_{0. \rm TD|rest}$ to the contour integral can be taken as
\begin{align}\label{eq:TDrest}
T_{0. \rm TD|rest} &=
	- \cV^{\dalpha \alpha \, --} D^{++} \hat B_{\alpha\dalpha}^{--}
	+ D^{--} (\cV^{\dalpha \alpha\, ++} \hat B_{\alpha \dalpha}^{--})
	- f^{\dalpha \alpha}{}^b K_b \hat B_{\alpha\dalpha}^{--}
	\eol & \quad
	- \frac{1}{2} \phi^{\dalpha \alpha\, \beta j} S_{\beta j} \hat B_{\alpha\dalpha}^{--}
	- \frac{1}{2} \phi^{\dalpha \alpha}{}_{\dbeta j} \bar S^{\dbeta j} \hat B_{\alpha\dalpha}^{--}
	+ 2\, e_a{}^m T_{mn}{}^c e_c{}^n \hat B^{a\,--} ~, \\
\hat B_{\alpha \dalpha}^{--}
	&= \frac{1}{4} z_1^{--} \nabla_\alpha^- \bar\nabla_\dalpha^- (\Gamma_I^+ \U^{I+})
	+ \frac{i}{4} (z_1^{--})^2 \nabla_{\alpha \dalpha} (\Gamma_I^+ \U^{I+})
	- \AntT~.
\end{align}
The important feature of $T_{0.\rm TD|rest}$ is that it depends only on
$\G_I^+ \U^{I+}$ and its antarctic conjugate.
For now we set these aside and focus on the terms first order in the
gravitino.

\subsection{Gravitino terms}
Now we must address terms involving a single explicit gravitino field.
There are two: $T_{0.\rm TD|Q}$ arises from the total
covariant derivative $\nabla^{\dalpha\alpha} B_{\alpha\dalpha}^{--}$
and $T_1$ arises from the original action.
First we will need several formulae for the spinor derivatives of
$B_{\alpha\dalpha}^{--}$. There are a number of ways to
potentially organize the resulting expression. We wish
to construct as much as possible expressions involving the
arctic combination $\G_I^+ \U^{I+}$, the
hyperk\"ahler potential $K$ and its derivatives, and various
pullbacks of the hyperk\"ahler two-form $\Omega^{++}$.
To accomplish this, it is convenient to rewrite
$B_{\alpha\dalpha}^{--}$ into the equivalent form
\begin{align}
B_{\alpha\dalpha}^{--} &= 
		\frac{1}{4} z_1^{--} \nabla_\alpha^- \bar\nabla_\dalpha^- (\Gamma_I^+ \U^{I+})
		+ \frac{i}{4} (z_1^{--})^2 \nabla_{\alpha \dalpha} (\Gamma_I^+ \U^{I+})
		\eol & \quad
		+ \frac{1}{4} z_1^{--} \Big(\Gamma_I^+ \nabla_{\alpha}^- \bar\nabla_\dalpha^- \U^{I+}
			- \U^{I+} \nabla_{\alpha}^- \bar\nabla_\dalpha^- \Gamma_I^+\Big)
		+ \frac{i}{4} (z_1^{--})^2 (\Gamma_I^+ \nabla_{\alpha \dalpha} \U^{I+}
			- \U^{I+} \nabla_{\alpha\dalpha} \G_I^+)
		\eol & \quad
		- \AntT
		+ \frac{1}{4} \frac{i}{v^{\1+} v^{\2+}} \,\nabla_{\alpha\dalpha} K~.
\end{align}
The first line involves only the arctic combination
$\G_I^+ \U^{I+}$ while the second line is manifestly
antisymmetric in $\G_I^+$ and $\U^{I+}$ and leads to pullbacks of $\Omega^{++}$.
Using relations such as
\begin{align*}
\Gamma_I^+ \nabla_\beta^- \U^{I+} 
	&= -\frac{1}{2} \nabla_\beta^+ K
	+ \frac{1}{2} \nabla_\beta^- (\Gamma_I^+ \U^{I+})~, \\
\nabla_\beta^- (\Gamma_I^+ \nabla_{\alpha \dalpha} \U^{I+})
- \nabla_{\alpha \dalpha} (\Gamma_I^+ \nabla_\beta^- \U^{I+})
	&= - \Omega_\beta^+{}_{\alpha\dalpha}
	+ \Gamma_I^+ [\nabla_\beta^-, \nabla_{\alpha \dalpha}] \U^{I+}
\end{align*}
one can show that
\begin{align}
\nabla_\beta^+ B_{\alpha \dalpha}^{--}
	&= 
	\frac{i}{2} z_1^{--} \nabla_{\beta\dalpha} \nabla_\alpha^- (\Gamma_I^+ \U^{I+})
	- 2 z_1^{--} \eps_{\beta \alpha} \bm{\bar\lambda}_\dalpha^{-r} \,D_r^{++}
	+ (z_1^{--})^2 \eps_{\beta \alpha} \bm{\bar\lambda}_\dalpha^{+r} \, D_r^{++}
	- \antt
	\eol & \quad
	+ \frac{1}{v^{\1+} v^{\2+}} \Big(
	\frac{i}{4} \nabla_{\alpha \dalpha} \nabla_\beta^+ K
	- \frac{i}{2} \nabla_{\beta \dalpha} \nabla_\alpha^+ K
	- \frac{i}{2} \homega_{\alpha\, \beta \dalpha}^+
	\eol & \qquad \qquad
	- \frac{3}{4} \eps_{\beta \alpha}  \bar \chi^{+}_\dalpha K
	+ \frac{1}{4} \eps_{\beta \alpha} \bar W_{\dalpha \dbeta} \bar\nabla^{\dbeta +} K
	+ \frac{1}{2} \eps_{\beta \alpha} \bar W^s \, \homega_{s \dalpha}^+ \Big)~,
\end{align}
and
\begin{align}
\nabla_\beta^- B_{\alpha \dalpha}^{--}
	&= \frac{1}{4} z_1^{--} \eps_{\beta \alpha} (\nabla^-)^2 (\Gamma_I^+ \bar\nabla_\dalpha^- \U^{I+})
	+ \frac{i}{4} (z_1^{--})^2 \nabla_{\alpha \dalpha} \nabla_\beta^- (\Gamma_I^+ \U^{I+})
	- \frac{i}{4} (z_1^{--})^2 \nabla_{\alpha \dalpha} \nabla_\beta^+ K
	\eol & \quad
	- \frac{i}{2} (z_1^{--})^2 \homega_{\beta\, \alpha\dalpha}^+
	+ \frac{3}{4} (z_1^{--})^2 \eps_{\beta \alpha} \bar\chi_\dalpha^+ D^{--}(\Gamma_I^+ \U^{I+})
	+ \frac{3}{4} (z_1^{--})^2 \eps_{\beta \alpha} \bar\chi_\dalpha^+ K
	\eol & \quad
	- \frac{3}{2} (z_1^{--})^2 \eps_{\beta \alpha} \bar\chi_\dalpha^- \Gamma_I^+\U^{I+}
	+ \frac{1}{4} (z_1^{--})^2 \eps_{\beta \alpha} \bar W_{\dalpha \dbeta} \bar \nabla^{\dbeta -} (\Gamma_I^+ \U^{I+})
	\eol & \quad
	- \frac{1}{4} (z_1^{--})^2 \eps_{\beta \alpha} \bar W_{\dalpha \dbeta} \bar\nabla^{\dbeta +} K
	+ (z_1^{--})^2 \eps_{\beta \alpha} \bm{\bar\lambda}_\dalpha^{- r} D_r^{++}
	- \AntT
	\eol & \quad
	+ \frac{1}{v^{\1+} v^{\2+}} \Big(
	\frac{i}{4} \nabla_{\alpha \dalpha} \nabla_\beta^- K
	+ \frac{i}{2} \homega_{\alpha\, \beta \dalpha}^-
	- \frac{3}{4} \eps_{\beta\alpha} \bar \chi_\dalpha^- \, K
	\eol & \qquad \qquad\qquad
	+ \frac{1}{4} \eps_{\beta\alpha} \bar W_{\dalpha \dbeta} \bar\nabla^{\dbeta -} K
	+ \frac{1}{2} \eps_{\beta\alpha} \bar W^s \homega_{s\dalpha}^-
	\Big)~.
\end{align}

It helps to rewrite $T_1 + T_{0.\rm TD|Q}$ as
\begin{align}\label{eq:T1+}
T_1 + T_{0.\rm TD|Q} &=
	\frac{1}{8} \psi^{\alpha \dalpha}{}_\alpha^- \,(\nabla^-)^2 \Big(\Gamma_I^+ \bar\nabla_\dalpha^- \U^{I+}\Big)
	- \frac{1}{4} \psi^{\dalpha \alpha}{}_\alpha^+ \,\nabla^{\beta -} B_{\beta \dalpha}^{--}
	+ \frac{1}{4} \psi^{\dalpha \alpha}{}_\alpha^- \,\nabla^{\beta +} B_{\beta \dalpha}^{--}
	\eol & \quad
	- \frac{1}{2} \psi^{\dalpha \alpha}{}^{\beta -} \nabla_{(\beta}^+ B_{\alpha) \dalpha}^{--}
	+ \frac{1}{2} \psi^{\dalpha \alpha}{}^{\beta +} \nabla_{(\beta}^- B_{\alpha) \dalpha}^{--}
	+ \HC
\end{align}
The first and second terms involve a common expression
$(\nabla^-)^2 (\Gamma_I^+ \bar\nabla_\dalpha^- \U^{I+})$ given by
\begin{align}\label{eq:DDDarctic}
\frac{1}{8} (\nabla^-)^2 (\G_I^+ \bar\nabla^{\dalpha-} \U^{I+})
	&= \G_I^+ \Xi^{I\dalpha-}
	- \frac{1}{2} M_I^- \Psi^{I \dalpha }
	- \frac{i}{4} \Psi^I_\alpha A^{I\, \alpha \dalpha -}
	\eol & \quad
	+ z_1^{--} \Big(
	\frac{i}{4} \nabla^{\dalpha \alpha} \nabla_\alpha^- (\G_I^+ \U^{I+})
	- \frac{1}{4} \bar W^\dalpha{}_\dbeta \bar\nabla^{\dbeta -} (\G_I^+ \U^{I+})
	+ \frac{3}{2} \bar\chi^{\dalpha -} \G_I^+ \U^{I+}
	\eol & \qquad\quad
	- \frac{3}{4} \bar\chi^{\dalpha +} D^{--} (\G_I^+ \U^{I+})
	- \frac{i}{4} \nabla^{\dalpha \alpha} \nabla_\alpha^+ K
	- \frac{i}{2} \Omega_\alpha^+{}^{\dalpha \alpha}
	\eol & \qquad\quad
	+ \frac{1}{4} \bar W^\dalpha{}_\dbeta \bar\nabla^{\dbeta +} K
	- \frac{3}{4} \bar\chi^{\dalpha +} K
	- 2 \bm{\bar\l}^{\dalpha r -} D_r^{++}
	\Big)
	\eol & \quad
	+ (z_1^{--})^2 \Big(
		\bm{\bar\l}^{\dalpha r +} D_r^{++}
	\Big)~.
\end{align}
The coefficient of this term in \eqref{eq:T1+} involves
$\psi^{\alpha \dalpha}{}_\alpha^- 
	- z_1^{--} \psi^{\dalpha \alpha}{}_\alpha^+ 
= \psi^{\alpha \dalpha}{}_\alpha^\1 / v^{\1+}$
and so the first line of \eqref{eq:DDDarctic} contributes
a purely arctic expression to \eqref{eq:T1+} that can be discarded under the contour integral.
This had to be the case as otherwise we
would need to impose the explicit form for $\Xi^{I \dalpha -}$,
which involves the field equation for the physical fermion.
Now we can analyze the remaining terms in \eqref{eq:T1+}. Discarding a total
contour derivative,
\begin{align}
T_1 + T_{0.\rm TD|Q}
	&\sim
	-\frac{1}{4} z_1^{--} \psi^{\alpha \dalpha}{}_\alpha^- \bar W_{\dalpha\dbeta}
			\bar\nabla^{\dbeta -} (\G_I^+ \U^{I+})
	+ \frac{1}{8} (z_1^{--})^2 \psi^{\alpha \dalpha}{}_\alpha^+
		\bar W_{\dalpha\dbeta} \bar\nabla^{\dbeta -} (\G_I^+ \U^{I+})
	\eol & \quad
	+ \frac{3}{2} z_1^{--} \psi^{\alpha \dalpha}{}_\alpha^- \bar\chi^{\dalpha -} \G_I^+ \U^{I+}
	- \frac{3}{4} z_1^{--} \psi^{\alpha \dalpha}{}_\alpha^- \bar\chi^{\dalpha +} D^{--} (\G_I^+ \U^{I+})
	\eol & \quad
	- \frac{3}{4} (z_1^{--})^2 \psi^{\alpha \dalpha}{}_\alpha^+ \bar\chi^{\dalpha -} \G_I^+ \U^{I+}
	+ \frac{3}{8} (z_1^{--})^2 \psi^{\alpha \dalpha}{}_\alpha^+ \bar\chi^{\dalpha +} D^{--} (\G_I^+ \U^{I+})
	\eol & \quad
	- \frac{3i}{8} z_1^{--} \psi^{\alpha \dalpha}{}_\alpha^- \nabla_{\beta\dalpha } \nabla^{\beta-} (\G_I^+ \U^{I+})
	+ \frac{3i}{16} (z_1^{--})^2 \psi^{\alpha \dalpha}{}_\alpha^+ \nabla_{\beta\dalpha } \nabla^{\beta-} (\G_I^+ \U^{I+})
	\eol & \quad
	+ \frac{i}{4} \Big(\frac{1}{2} (z_1^{--})^2 \psi^{\dalpha (\alpha}{}^{\beta) +} 
	- z_1^{--} \psi^{\dalpha (\alpha}{}^{\beta) -} \Big)\nabla_{\alpha\dalpha} \nabla_\beta^- (\Gamma_I^+ \U^{I+})
	- \antt
	\eol & \quad
	+ \frac{1}{4} \frac{i}{v^{\1+} v^{\2+}} \psi^{\alpha \dalpha}{}_{\alpha j} \Big(
		\nabla_{\beta\dalpha} \nabla^{\beta j} K 
		+ \Omega^{\beta j}{}_{\beta\dalpha}
		- i \bar W_{\dalpha \dbeta} \bar\nabla^{\dbeta j} K
		+ i \bar\chi_\dalpha^j K 
		\eol & \qquad \qquad
		- i \bar W^s \homega_{s\dalpha}^j
		+ 4 i \bm{\bar\lambda}_\dalpha^{r}{}_k D_r^{jk}
		\Big)
\end{align}

Combining all of the above results, we find
\begin{align}\label{eq:LmmFinal}
T_0 + T_1 &=
	\frac{i}{v^{\1+} v^{\2+}} \bigg[
	\frac{1}{2} K_\mu \widehat {\Box} \phi^\mu
	- \frac{i}{4} \, g_{\rb \bar \rb}\,
		(\z^{\alpha \rb} \overleftrightarrow{{\widehat\nabla}}_{\alpha\dalpha} \bar\zeta^{\dalpha \bar \rb})
	+ \frac{1}{16} \z^\ra \z^\rb \bar \z^{\bar \ra} \bar\z^{\bar \rb}\, R_{\ra \bar \ra\, \rb \bar \rb}
	\eol & \qquad\qquad
	+ Y^r_{ij} \, D_r^{ij}
	+ (\bm\lambda^\alpha_i{}^r \homega_{r \,\alpha}^i + \HC)
	- \frac{1}{2} W^r \bar W^s J_r^\mu J_s^\nu g_{\mu\nu}
	\eol & \qquad\qquad
	- \frac{1}{4} W^r \z^\ra \z^\rb \nabla_\ra (\omega_{\rb \rc} J_r^\rc)
	- \frac{1}{4} \bar W^r \bar\z^{\bar\ra} \bar\z^{\bar\rb} \nabla_\ra (\omega_{\bar\rb \bar\rc} J_r^{\bar\rc})
	\eol & \qquad\qquad
	-\frac{3}{2} D\, K
	- \frac{1}{4} (W^{\alpha\beta} \z_\beta^\rb \z_\alpha^\ra \,\omega_{\rb \ra} + \HC)
	- \frac{3}{4} (\bar \chi_\dalpha^j \,\bar\nabla^{\dalpha}_j K  + \HC)
	\eol & \qquad\qquad
	+ \frac{1}{4} \psi^{\alpha \dalpha}{}_{\alpha j} K_\mu \widehat \nabla_{\beta\dalpha} \z^{\beta j \mu} 
	+ \frac{3i}{4} \psi^{\alpha \dalpha}{}_{\alpha j} \bar\chi_\dalpha^j K
	\eol & \qquad\qquad
	+ \frac{i}{4} \psi^{\dalpha \alpha}{}_\alpha^j \bar W_{\dalpha \dbeta} \bar\nabla^{\dbeta}_j K
	- \frac{i}{4} \psi^{\dalpha \alpha}{}_\alpha{}_j \bar W^s \homega_{s\dalpha}^j
	+ i \psi^{\alpha \dalpha}{}_\alpha{}_j \bm{\bar\lambda}_\dalpha^{r}{}_k D_r^{jk}
	\bigg]
	+ \mathscr{R}^{--}~.
\end{align}
The terms within square braces are independent of $v^{i+}$,
so their contour integral can be done immediately. This actually gives
the final Lagrangian $\cL$. The component action is
\begin{align}\label{eq:CompActionRemainders}
S = \int \rd^4x\, e\, \cL
	-\frac{1}{2\pi} \oint_\cC v_{i}^+ \rd v^{i+} \int \rd^4x\, e\, \mathscr{R}^{--}
	+ S_{\rm rest}~.
\end{align}
In addition to the many terms $S_{\rm rest}$ that we separated out at the 
start of the calculation, there are still a large number of terms left within
the remainder term $\mathscr{R}^{--}$,
\begin{align}\label{eq:Rmm}
\mathscr{R}^{--} &= 
	\frac{3i}{2} z_1^{--} \bar \chi_\dalpha^- \bar\nabla^{\dalpha -} (\Gamma_I^+ \U^{I+})
	- \frac{3i}{4} (z_1^{--})^2 \bar \chi_\dalpha^+ \bar\nabla^{\dalpha -} (\Gamma_I^+ \U^{I+})
	\eol & \quad
	+ i z_1^{--} \psi_a{}^{\alpha -} (\sigma^{ab})_\alpha{}^\beta \nabla_b \nabla_\beta^{-} (\G_I^+ \U^{I+})
	- \frac{i}{2} (z_1^{--})^2 \psi_a{}^{\alpha +} (\sigma^{ab})_\alpha{}^\beta \nabla_b \nabla_\beta^{-} (\G_I^+ \U^{I+})
	\eol & \quad
	+ \frac{3}{2} z_1^{--} \psi^{\alpha \dalpha}{}_\alpha^- \bar\chi_\dalpha^{-} (\G_I^+ \U^{I+})
	- \frac{3}{4} z_1^{--} \psi^{\alpha \dalpha}{}_\alpha^- \bar\chi_\dalpha^{+} D^{--} (\G_I^+ \U^{I+})
	\eol & \quad
	- \frac{3}{4} (z_1^{--})^2 \psi^{\alpha \dalpha}{}_\alpha^+ \bar\chi_\dalpha^{-} (\G_I^+ \U^{I+})
	+ \frac{3}{8} (z_1^{--})^2 \psi^{\alpha \dalpha}{}_\alpha^+ \bar\chi_\dalpha^{+} D^{--} (\G_I^+ \U^{I+})
	\eol & \quad
	-\frac{1}{4} z_1^{--} \psi^{\alpha \dalpha}{}_\alpha^- \bar W_{\dalpha\dbeta}
			\bar\nabla^{\dbeta -} (\G_I^+ \U^{I+})
	+ \frac{1}{8} (z_1^{--})^2 \psi^{\alpha \dalpha}{}_\alpha^+
		\bar W_{\dalpha\dbeta} \bar\nabla^{\dbeta -} (\G_I^+ \U^{I+})
	\eol & \quad
	- \AntT+ \HC
	+ D^{--} B
	+ T_{0.\rm TD|rest}~,
\end{align}
where $B$ 
and $T_{0.\rm TD|rest}$ are given by \eqref{eq:DefB} and \eqref{eq:TDrest}.
All of these are written in terms of the arctic combination
$\Gamma_I^+ \U^{I+}$ and its antarctic conjugate, using the relation
\eqref{eq:Fhomo}. In fact, it turns out that all of these terms
actually vanish up to a total derivative -- that is,
\begin{align}
0 = -\frac{1}{2\pi} \oint_\cC v_{i}^+ \rd v^{i+} \int \rd^4x\, e\, \mathscr{R}^{--}
	+ S_{\rm rest}~.
\end{align}
The proof is somewhat indirect,
so we postpone it until appendix \ref{app:ArcticAction}.

\subsection{Final result}
Our final component Lagrangian is
\begin{align}
\cL &=
	\frac{1}{2} K_\mu \widehat {\Box} \phi^\mu
	- \frac{i}{4} \, g_{\rb \bar \rb}\,
		(\z^{\alpha \rb} \overleftrightarrow{{\widehat\nabla}}_{\alpha\dalpha} \bar\zeta^{\dalpha \bar \rb})
	+ \frac{1}{16} \z^\ra \z^\rb \bar \z^{\bar \ra} \bar\z^{\bar \rb}\, R_{\ra \bar \ra\, \rb \bar \rb}
	\eol & \quad
	+ Y^r_{ij} \, D_r^{ij}
	+ (\bm\lambda^\alpha_i{}^r \homega_{r \,\alpha}^i + \bm{\bar\lambda}_\dalpha^{i\,r} \homega_r{}^\dalpha_i)
	- \frac{1}{2} W^r \bar W^s J_r^\mu J_s^\nu g_{\mu\nu}
	\eol & \quad
	- \frac{1}{4} W^r \z^\ra \z^\rb \nabla_\ra (\omega_{\rb \rc} J_r^\rc)
	- \frac{1}{4} \bar W^r \bar\z^{\bar\ra} \bar\z^{\bar\rb} \nabla_{\bar\ra} (\omega_{\bar\rb \bar\rc} J_r^{\bar\rc})
	\eol & \quad
	-\frac{3}{2} D\, K
	- \frac{1}{4} (W^{\alpha\beta} \z_\beta^\rb \z_\alpha^\ra \,\omega_{\rb \ra}
		+ \bar W^{\dalpha\dbeta} \bar\z_\dbeta^{\bar\rb} \bar\z_\dalpha^{\bar\ra} \,\omega_{\bar\rb \bar\ra})
	- \frac{3}{4} K_\mu (\chi^\alpha_j \,\z_\alpha^{j\mu} +\bar \chi_\dalpha^j \,\bar\z^{\dalpha}_j{}^\mu)
	\eol & \quad
	+ \psi^{\alpha \dalpha}{}_{\alpha j} \Big(
	\frac{1}{4} K_\mu \widehat \nabla_{\beta\dalpha} \z^{\beta j \mu} 
	- \frac{i}{4} K_\mu \bar W_{\dalpha \dbeta} \bar\z^{\dbeta j \mu}
	+ \frac{3i}{4} \bar\chi_\dalpha^j K
	- \frac{i}{4} \bar W^r \homega_{r\dalpha}^j
	+ i \bm{\bar\lambda}_\dalpha^{r}{}_k D_r^{jk}\Big)
	\eol & \quad
	+ \psi_{\alpha \dalpha}{}^{\dalpha j} \Big(
	\frac{1}{4} K_\mu \widehat \nabla^{\dbeta\alpha} \bar\z_{\dbeta}{}_j^\mu
	+ \frac{i}{4} K_\mu W^{\alpha \beta} \z_{\beta j}{}^\mu
	+ \frac{3i}{4} \chi^\alpha_j K
	+ \frac{i}{4} W^r \homega_{r\alpha j}
	- i \bm\lambda^{r \alpha k} D_{r\,jk}\Big)~.
\end{align}
It is convenient to relabel the tensor auxiliary field as
$T_{ab}^- = 4 (\sigma_{ab})_\alpha{}^\beta W_\beta{}^\alpha$,
in accordance with tensor calculus conventions, where
$T_{ab}^- \equiv T_{ab}{}^{ij} \veps_{ij}$.
(Note that this changes the definition of self-duality used
in \cite{Butter:CSG4d_2} so as to agree with component conventions.)
Also, a number of terms can be rewritten to involve the fields $A_i{}^\ra$, with the result
\begin{align}\label{eq:CompActionFinal}
\cL &=
	\frac{1}{2} K_\mu \widehat {\Box} \phi^\mu
	-\frac{3}{2} D\, K
	- \frac{1}{2} W^r \bar W^s J_r^\mu J_s^\nu g_{\mu\nu}
	+ Y^r_{ij} \, D_r^{ij}
	+ \frac{1}{16} \z^\ra \z^\rb \bar \z^{\bar \ra} \bar\z^{\bar \rb}\, R_{\ra \bar \ra\, \rb \bar \rb}
	\eol & \quad
	- \frac{1}{4} (\bar \z_\dalpha^{\bar \ra} - i (\psi_{m j} \sigma^m)_\dalpha A^{j \bar \ra}) \times
	\eol & \qquad\quad
		\Big(
			i g_{\bar \ra \rb} \widehat\nabla^{\dalpha \alpha} \z_\alpha^\rb
			+ \frac{1}{8} \omega_{\bar \ra \bar\rb} (\bsigma^{bc} \bar\z^{\bar \rb})^\dalpha T_{bc}^+
			+ 3 g_{\bar \ra \rb} A_k{}^\rb \bar\chi^{\dalpha k}
			+ \bar\z^{\dalpha \bar \rb} \bar W^r  \nabla_{\bar \ra} (\omega_{\bar \rb \bar\rc} J_r^{\bar \rc})
			- 2 \bm{\bar\l}^{\dalpha r k} J_{r \bar \ra k}
		\Big)
	\eol & \quad
	- \frac{1}{4} (\z^{\alpha\ra} - i (\bpsi_{m}^j \bsigma^m)^\alpha A_j{}^{\ra}) \times
	\eol & \qquad\quad
		\Big(
			i g_{\ra \bar\rb} \widehat\nabla_{\alpha\dalpha } \bar\z^{\dalpha \bar \rb}
			+ \frac{1}{8} \omega_{\ra \rb} (\sigma^{bc} \z^{\rb})_\alpha T_{bc}^-
			+ 3 g_{\ra \bar \rb} A^{k \bar \rb} \chi_{\alpha k}
			+ \z_\alpha^{\rb} W^r  \nabla_{\ra} (\omega_{\rb \rc} J_r^{\rc})
			- 2 \bm\l^r_{\alpha k} J_{r \ra}{}^k
		\Big)
	\eol & \quad
	+ \frac{1}{2} \z^{\alpha \ra} \bm\l^r_{\alpha i} J_{r \ra}{}^i
	+ \frac{1}{2} \bar\z_\dalpha^{\bar \ra} \bm{\bar\l}^{r \dalpha i} J_{r \bar \ra i}
\end{align}
Two useful checks can be made. First, in the rigid
supersymmetric limit, this matches the component Lagrangian constructed from
\eqref{eq:N1version}. Second, the Lagrangian must be $S$-supersymmetric. To check this,
it helps to note that the parenthetical terms in the third and fifth lines,
which multiply explicit gravitinos, must vanish under $S$-supersymmetry.
It is straightforward to check that the remaining terms all cancel against each
other.

This result can be compared with eq. (5.4) of \cite{dWKV}, where superconformal
tensor calculus conventions (see e.g. \cite{dWLP:Lag}) were used. The relation between those conventions
and ours for conformal supergravity are spelled out in \cite{Butter:CSG4d_2}; for example,
one must swap the locations of $\SU{2}$ indices, taking care to observe that
tensor calculus conventions employ $\veps^{\1\2} = \veps_{\1\2} = 1$ while we
use $\eps^{\1\2} = \eps_{\2\1} = 1$. This amounts to the exchange of
$\veps^{ij} \rightarrow -\eps_{ij}$ and $\veps_{ij} \rightarrow \eps^{ij}$.
The target space conventions of \cite{dWKV} also differ in several ways from ours.
For example, their \Sp{n} indices $(\bar \alpha, \alpha)$ correspond to our $(\ra, \bar \ra)$,
and we use a different normalization for the fermions.
For the sigma model component fields, one must exchange
\begin{align}
\phi^A \rightarrow \phi^\mu~, \qquad
A_i{}^\alpha \rightarrow A^{i \bar \ra}~, \qquad
A^{i\bar\alpha} \rightarrow A_i{}^{\ra}~, \qquad
\z^{\bar\beta} \rightarrow \frac{1}{2} \z_\alpha^{\rb}~, \qquad
\z^{\beta} \rightarrow \frac{1}{2} \bar \z^{\dalpha \bar \rb}~.
\end{align}
The target space geometric quantities are related as
\begin{gather}
\Omega_{\bar\alpha \bar\beta} \rightarrow \omega_{\ra\rb}~, \qquad
G_{\bar\alpha \beta} \rightarrow g_{\ra \bar\rb}~, \qquad
\gamma^A_{i \bar\alpha} \rightarrow f_\ra{}^{i \mu}~, \qquad
V_A^{i \bar \alpha} \rightarrow f_\mu{}_i{}^\ra~, \eol
(J^{ij})^A{}_B \rightarrow -(\cJ_{ij})^\mu{}_\nu~, \qquad
(k_I)^\alpha{}_\beta \rightarrow \nabla_{\bar \rb} J_r^{\bar \ra}~, \qquad
P_I^{ij} \rightarrow D_{r\, ij}~.
\end{gather}
The components of the vector multiplet are also defined slightly differently:
\begin{align}
X^I \rightarrow -\frac{1}{2} W^r~, \qquad
\Omega_i^{I} \rightarrow -\,\eps^{ij} \bm\l^{r}_{\alpha j}~, \qquad
\Omega^{I i} \rightarrow \eps_{ij} \bm{\bar \l}^{r \dalpha j}~, \qquad
Y^I_{ij} \rightarrow Y^{r\,ij}~.
\end{align}
Other quantities can be derived from these relations and the equations
given in \cite{dWKV}.

\section{Conclusion}

Our goal in this paper was quite specific: to recover the component action of
a hyperk\"ahler cone coupled to conformal supergravity given in
\cite{dWKV} using projective
superspace methods. Our approach was based on the formal solution of the
equations \eqref{eq:N2Eoms}, corresponding to the requirement that
the dual multiplets $\Gamma_I^+$ and $\breve \Gamma_{\bar I}^+$
were respectively arctic and antarctic, as advocated in \cite{GK:CNM,LR:Prop}.
We have not attempted to actually solve these equations for any
specific models -- the formal solutions were sufficient to yield the
component action --
but it should be mentioned that many specific cases do admit
solutions.
A major class involve $\cO(2n)$ hypermultiplets and the generalized
Legendre transform construction \cite{LR88, HiKLR, KLR, IvRo:AH}:
here the number of auxiliary fields is finite from the outset.
Another major class involve target spaces that are cotangent bundles
of Hermitian symmetric spaces \cite{GK:CNM, GK2, AN, AKL1, AKL2, KuNo:CTBundle, AraiBlaschke:E7/E6}
(see also \cite{Kuzenko:Lectures} for a pedagogical discussion).

As discussed in \cite{dWKV, dWRV:QK}, the component
action \eqref{eq:CompActionFinal} does not directly yield the general
action of canonically normalized supergravity coupled to matter.
One must fix the Weyl gauge and the $\SU{2}_R$ gauge:
this effectively eliminates four degrees of
freedom from the hypermultiplet manifold, reducing the $4n$-dimensional
hyperk\"ahler cone to a $4 (n-1)$-dimensional quaternion-K\"ahler
manifold. In addition, one must add another matter sector
(e.g. vector multiplets parametrizing a special K\"ahler manifold)
to yield a consistent equation of motion for the auxiliary $D$.
This can all be understood at the component level, but what about in
superspace?

Here again comparing with $\cN=1$ superspace is helpful.
Recall the geometry of an $\cN=1$ superconformal sigma model is a K\"ahler
cone. If the chiral superfields are reorganized into the set
$\{\phi_0, \phi^i\}$ with $\chi^\ra = (\phi_0, 0)$, then without
loss of generality
the K\"ahler cone potential can be written as\footnote{The
overall sign must be negative to
yield the correct sign for the Einstein-Hilbert term.}
\begin{align}\label{eq:N1ConformalK}
K &= -3 \,\phi_0 \bar\phi_0 \,e^{-\cK / 3}~, 
\end{align}
for a real function $\cK(\phi^i, \bar\phi^{\bar \imath})$, subject to the K\"ahler transformations,
\begin{align}
\phi_0 &\rightarrow \phi_0 \,e^{F/3}~, \qquad \cK \rightarrow \cK + F + \bar F~, \qquad F = F(\phi^i)~.
\end{align}
Imposing the dilatation+$\gU{1}_R$ gauge $\phi_0 = e^{\cK / 6}$
leads to the standard formulation of an $\cN=1$ supergravity-matter system
(including a superpotential is straightforward)
in the so-called K\"ahler superspace \cite{BGG} with a simple Lagrangian
\begin{align}
-3 \int \rd^4x\, \rd^2\q\, \rd^2\bar\q\, E~.
\end{align}
The function $\cK$ is absorbed within the superspace structure and
the original K\"ahler transformations become associated with
an effective $\gU{1}_R$ symmetry of the component fields.
This K\"ahler superspace is extremely useful: for example,
higher derivative interactions can easily be incorporated
in a K\"ahler-covariant way. It would be extremely interesting
to construct the $\cN=2$ analogue of the above superspace geometry:
a proposal along the lines discussed above has already been made in
\cite{KLRT-M1}. It would be especially intriguing if a direct link
could be found to the harmonic description of quaternionic sigma models
\cite{BaGIO, GIO:QK}, as was done for the hyperk\"ahler case \cite{Butter:HarmProj}.
It would also be nice to understand the quaternionic examples in harmonic superspace
(see e.g. the quaternionic Taub-NUT \cite{IV:TN} and
the general two-center metrics \cite{CIV}) in this context.

We should note that the projective superspace description of quaternionic sigma models
takes a particularly elegant form if the hyperk\"ahler cone possesses
an additional $\gU{1}$ isometry that separately rotates the arctics
and antarctics. Then it is always possible to perform a duality
transformation, exchanging one polar multiplet for a tensor multiplet $\cG^{++}$,
giving the supergravity-matter action \cite{Kuzenko:DualForm, KLvU}
\begin{align}\label{eq:GKaction}
-\frac{1}{2\pi} \oint_\cC \rd\tau \int \rd^4x\, \rd^4\q^+ \, \cE^{--}
	\Big(
	\cG^{++} \log (\cG^{++} / \ri \U_0^+ \breve \U_0^+)
	+ \cG^{++} K(\U, \breve \U) \Big)
\end{align}
where $K(\U, \breve \U)$ is a real function of weight-zero arctics
$\U$ and antarctics $\breve \U$. (The arctic $\U_0^+$ drops out of the
component action but is necessary to keep the argument of the logarithm
dimensionless.)
This is the natural
$\cN=2$ generalization of the $\cN=1$ new minimal supergravity action
coupling the tensor multiplet compensator to a K\"ahler potential.
The component Lagrangian of \eqref{eq:GKaction} gives the general
supergravity-matter system (after gauge-fixing) involving a
quaternion-K\"ahler target space arising from a hyperk\"ahler cone
with a triholomorphic $\gU{1}$ isometry.

An interesting open question would be to address hypermultiplet couplings
in the presence of various higher derivative terms. Higher derivative
tensor multiplet actions were discussed at the component level
in \cite{dWS:Tensor}; their lift to projective superspace (and generalization
to include $\cO(2n)$ multiplets) was addressed in \cite{BuKu:HigherDerv}.
In principle, one should be able to apply the latter construction to
include higher derivative actions for off-shell polar multiplets.
However, the elimination of the auxiliaries in the presence of such
terms becomes (even formally) a formidable process. Nevertheless,
superspace would seem to be the natural mechanism for constructing such
terms in a systematic way, and the calculation undertaken here is an
important first step.
It would be intriguing to explore these issues further.

\section*{Acknowledgements}
I am grateful to Sergei Kuzenko, Ulf Lindstr\"om, Martin
Ro\v{c}ek, and Gabriele Tartaglino-Mazzucchelli
for valuable comments and suggestions.
This work was supported in part by the ERC Advanced Grant no. 246974,
{\it ``Supersymmetry: a window to non-perturbative physics''}
and by the European Commission
Marie Curie International Incoming Fellowship grant no.
PIIF-GA-2012-627976.	

\appendix

\section{Vector multiplet conventions and supersymmetry transformations}\label{app:VectorMult}
The introduction of a gauge connection to the curved projective
superspace \cite{Butter:CSG4d.Proj} is
straightforward and can be found in \cite{KLRT-M2,Kuzenko:DualForm}.
We introduce a superspace connection
$\cA_{\ul M} = \cA_{\ul M}{}^r X_r$,
with $X_r$ the formal gauge generator which acts on a field $\Psi$ in 
a given representation as $X_r \Psi = \mathbf T_r \Psi$ for matrices $\mathbf T_r$.
Note that the algebra of the operators possesses a different sign from the
algebra of the matrices,
\begin{align}
[X_r, X_s] = -f_{rs}{}^t X_t \quad \implies\quad 
[\mathbf T_r, \mathbf T_s] = +f_{rs}{}^t \mathbf T_t ~.
\end{align}
For a compact gauge group, $\mathbf T_r$ are anti-Hermitian matrices, and
so are frequently defined with an additional factor of $i$ to make them Hermitian.
For a non-linear sigma model with gauged isometries, the scalars
$\phi^\mu = (\phi^\ra, \bar \phi^{\bar \ra})$
transform as $X_r \phi^\mu = J_r{}^\mu$ into a Killing vector $J_r{}^\mu$.
The fermions $\z_\alpha^\rb$ and $\bar\z_\dalpha^{\bar \rb}$
transform as
\begin{align}
X_r \z_\alpha^\rb = \z_\alpha^\rc \,\pa_\rc J_r{}^\rb ~,\qquad
X_r \bar\z_\dalpha^{\bar\rb} = \bar\z_\dalpha^{\bar\rc} \,\pa_{\bar\rc} J_r{}^{\bar\rb} ~.
\end{align}

The gauge covariant derivative $\nabla_{\ul A}$ is defined implicitly by
\begin{align}
\pa_{\ul M} &= E_{\ul M}{}^{\ul A} \nabla_{\ul A}
	+ \frac{1}{2} \Omega_{\ul M}{}^{ab} M_{ba}
	+ A_{\ul M} \bbA
	+ B_{\ul M} \bbD
	+ F_{\ul M}{}^A K_A
	+ \cA_{\ul M}{}^r X_r~.
\end{align}
The algebra of the $\SU{2}$ covariant derivatives $\nabla^{\pm\pm}$
and $\nabla^0$ with themselves and with the other operators remains
unchanged. Similarly, the algebra of spinor covariant derivatives
obeys the integrability conditions
$\{\nabla_{\ul \alpha}^\pm, \nabla_{\ul \beta}^\pm\} = 0$,
which implies the dimension-1 curvatures
\begin{align}
\{\nabla_\alpha^\pm, \bnabla_{\dbeta}^\mp\} = \mp 2i \nabla_{\alpha \dbeta}~, \quad
\{\nabla_\alpha^\pm, \nabla_\beta^\mp\} = \pm 2 \eps_{\alpha \beta} \bar \cW~, \quad
\{\bnabla^{\dalpha \pm}, \bnabla^{\dbeta \mp}\} = \pm 2 \, \eps^{\dalpha \dbeta} \cW~.
\end{align}
Now the curvature operator $\cW$ receives a new contribution $W^r X_r$,
\begin{align}
\cW &= \frac{1}{2} W^{\alpha \beta} M_{\beta \alpha}
	+ \frac{1}{4} \nabla^{\beta +} W_{\beta}{}^\alpha S_{\alpha}^-
	- \frac{1}{4} \nabla^{\beta -} W_{\beta}{}^\alpha S_{\alpha}^+
     + \frac{1}{4} \nabla^{\dalpha \beta} W_\beta{}^\alpha K_{\alpha \dalpha}
	+ W^r X_r
\end{align}
and similarly for its complex conjugate. The new superfield $W^r$ is the
covariant non-abelian vector multiplet. It is chiral, inert under the $\SU{2}$
covariant derivatives, and obeys the Bianchi identity
$(\nabla^+)^2 W^r = (\bar\nabla^+)^2 \bar W^r$.
The dimension-3/2 curvatures
\begin{align}
[\nabla_{\beta}^\pm, \nabla_{\alpha \dalpha}] = -2 \eps_{\beta \alpha} \bar\cW_\dalpha^\pm~, \qquad
[\bar\nabla_{\dbeta}^\pm, \nabla_{\alpha \dalpha}] = -2 \eps_{\dbeta \dalpha} \cW_\alpha^\pm~.
\end{align}
receive a new contribution from the gauge sector,
\begin{align}
\cW_\alpha^\pm \ni -\frac{i}{2} \nabla_\alpha^\pm W^r X_r~, \qquad
\bar\cW_\dalpha^\pm \ni -\frac{i}{2} \bar\nabla_\dalpha^\pm \bar W^r X_r~.
\end{align}
Finally, the dimension-2 curvatures
$[\nabla_b, \nabla_a] = -\cF_{ba}$
include a contribution $\cF_{ba}{}^r X_r$, with
\begin{align}
\cF_{ba}{}^r = -\frac{1}{4} (\sigma_{ba})^{\alpha\beta}
	\Big(\nabla_{(\alpha}^+ \nabla_{\beta)}^- W^r - 2 W_{\alpha\beta} \bar W^r\Big)
	- \frac{1}{4} (\bsigma_{ba})^{\dalpha\dbeta}
	\Big( \bar\nabla_{(\dalpha}^+ \bar\nabla_{\dbeta)}^- \bar W^r
		- 2 \bar W_{\dalpha\dbeta} W^r \Big)~.
\end{align}

We also use the symbol $W^r$ to denote the lowest component of the vector multiplet.
The other covariant components of the vector multiplet are given by
\begin{align}
\bm\l^r_{\alpha i} = \frac{1}{2} \eps_{ij} \nabla_\alpha^j W^r~, \qquad
\bm{\bar\l}^{r \dalpha i} = -\frac{1}{2} \eps^{ij} \bar\nabla^\dalpha_j \bar W^r~, \qquad
Y^{r\, ij} = \frac{1}{4} \nabla^{ij} W^r~,
\end{align}
while the component one-form $A_m{}^r$ and two-form $F_{mn}{}^r$ are given by
the projections of the corresponding superspace quantities
$\cA_m{}^r$ and $\cF_{mn}{}^r$. The supersymmetry and $S$-supersymmetry transformations are given by
\begin{align}
\delta A_m{}^r &= i (\xi_i \sigma^m \bm{\bar \l}^{ir}) - \eps^{ij} (\xi_i \psi_m{}_j) \bar W^r + \HC~, \eol
\delta W^r &= 2 \,\xi_i \bm\l^{i r}~, \eol
\delta \bm\l^r_{\alpha i} &= (\sigma^{ab} \xi_i)_\alpha \Big(\cF_{ab}{}^r + \frac{1}{8} T_{ab}^- \bar W^r\Big)
	- Y_i{}^j \xi_{\alpha j}
	+ \frac{1}{2} \xi_{\alpha i} \bar W^s W^t f_{ts}{}^r
	+ i \nabla_{\alpha \dbeta} W^r \bar \xi^{\dbeta j}
	- 2 \,\eta_{\alpha i} W^r ~, \eol
\delta Y^{r\,ij} &= 2 i \xi^{\alpha i} \nabla_{\alpha \dalpha} \bm{\bar \l}^\dalpha_j{}^r
	-  \xi^{\alpha i} \bm\l_\alpha^{j}{}^s \,\bar W^t f_{ts}{}^r
	- 2 i \bar \xi_\dalpha{}^i \nabla^{\dalpha \alpha} \bm\l_\alpha^j{}^r
	- \bar \xi_\dalpha{}^i \bm{\bar\l}^{\dalpha j}{}^s \,W^t f_{ts}{}^r~.
\end{align}
The transformation law of the gaugino involves the supercovariant
curvature tensor $\cF_{ab}{}^r$, whose component form is given by
\begin{align}
e_m{}^a e_n{}^b \cF_{ab}{}^r &:= F_{mn}{}^r
	- \frac{i}{2} (\psi_{m j} \sigma_n \bm{\bar\l}^{j r} - \psi_{n j} \sigma_m \bm{\bar\l}^{j r})
	- \frac{i}{2} (\bpsi_{m}{}^j \bsigma_n \bm\l_j^{r} - \bpsi_{n}{}^j \bsigma_m \bm\l_j^{r})
	\eol & \quad
	+ \frac{1}{2} \eps^{ij} (\psi_{m}{}_i \psi_{n}{}_j) \,\bar W^r
	- \frac{1}{2} \eps_{ij} (\bpsi_m{}^i \bpsi_{n}{}^j) \, W^r~.
\end{align}

\section{Vanishing of a pure arctic action and the remainder terms in eq. \eqref{eq:CompActionRemainders}}\label{app:ArcticAction}

We begin this appendix by reviewing an important lemma: the projective superspace action
\begin{align}\label{eq:PSAction}
-\frac{1}{2\pi} \oint_\cC \rd\tau \int \rd^4x\, \rd^4\q^+ \, \cE^{--} \L^{++}
\end{align}
vanishes if $\L^{++}$ is a purely arctic (or antarctic) superfield.
A nearly identical statement was established in \cite{KT-M:DiffReps}
for the choice $\L^{++} = \cG^{++} \L$ for $\cG^{++}$ a tensor multiplet
and $\L$ an arctic (or antarctic) superfield; the proof is exactly the same
so let's briefly review it.

As discussed in \cite{Butter:CSG4d.Proj}, it is possible to complexify the auxiliary
manifold $\SU{2}$ to $\rm SL(2, \mathbb C)$,
taking $v^{i+} = v^i$ and $v_i^- = u_i / (v, u)$. This modifies the
component Lagrangian only by a total derivative.
In particular, we may choose $u_i$ to be a fixed coordinate
so long as $(v,u) \neq 0$ along the contour $\cC$.
This has the benefit of eliminating the second integral in
\eqref{eq:ProjCompAction}, and so significantly simplifies the
evaluation of component actions. This approach is exactly the formulation of
projective superspace presented in \cite{KLRT-M1, KLRT-M2, KT-M:DiffReps, Kuzenko:DualForm}.

Let us suppose now that $\L^{++}$ is arctic. Following \cite{KT-M:DiffReps},
we may choose
$u_i = (1,0)$, so that $v_i^- = (1, 0) / v^{\1+}$ while $v^{i+} = v^{\1+} (1, \z)$.
Examining the component Lagrangian \eqref{eq:Lmm} with $\cF^{++} = \L^{++}$,
it is immediately apparent that $\cL^{--}$ is a purely arctic expression.
That is, after factoring out a common factor of $1 / (v^{\1+})^2$,
the remaining terms are all free of singularities at $\z=0$.
Such a Lagrangian exactly vanishes under the contour integral.

This is the most direct proof, but not the only one.
One could also remain with the real $\SU{2}$ manifold and explicitly
analyze the component reduction of \eqref{eq:PSAction}. This would require
defining the component fields of $\L^{++}$ as we did for $\U^{I+}$ in
section \ref{sec:Blocks.Comps} and then proceeding to massage each
term of the component action meticulously until everything vanished.
This approach is clearly much more laborious.

It turns out that this more difficult approach actually has an excellent use. If we
recall the situation in section \ref{sec:CompActonCurved}, we argued that
the projective superspace Lagrangian
$\cF^{++} = \frac{i}{2} \G_I^+ \U^{I+} + \HC$
is the sum of an arctic and antarctic piece up to the $\q^2$ level
without imposing dynamical equations of motion. Then in analyzing the component
action of $\cF^{++}$, we stopped at \eqref{eq:CompActionRemainders} once everything had been reduced to the
physical component fields or to expressions involving $\G_I^+ \U^{I+}$ and
its complex conjugate. These we isolated into a remainder term $\mathscr{R}^{--}$, given in
\eqref{eq:Rmm}, as well as the terms $S_{\rm rest}$ neglected in \eqref{eq:CleverCompAction}.
All of these residual terms involved $\G_I^+ \U^{I+}$ (and its conjugate)
with at most two spinor derivatives.
We claimed that the sum of all these terms vanished.
Now we will provide a proof. We will start by analyzing the component action
of \eqref{eq:PSAction} using the real $\SU{2}$ manifold, proceeding
systematically order by order in the number of gravitinos and other
connection fields. We will stop after eliminating the leading terms:
what remains will be those terms left over
in \eqref{eq:CompActionRemainders} after choosing
$\L^{++} = \frac{i}{2} \G_I^+ \U^{I+}$, which is indeed arctic up to
the $\q^2$ level. These must vanish, of course, because the
original action \eqref{eq:PSAction} vanishes.

We begin anew with the action \eqref{eq:PSAction} and evaluate the component Lagrangian
\eqref{eq:Lmm} with $\cF^{++} = \L^{++}$. As before, we will organize the calculation as
in \eqref{eq:CleverCompAction}, taking the leading terms $T_0$ and $T_1$ and placing
the others into $S_{\rm rest}$. The leading term $T_0$,
involves the highest component of $\L^{++}$. This can be decomposed as
\begin{align}
(\nabla^-)^4 \L^{++}
	&= P^{--}
	- \frac{i}{2} z_1^{--} \nabla^{\dalpha \alpha} \nabla_\alpha^- \bar\nabla_\dalpha^- \L^{++}
	+ \frac{1}{2} (z_1^{--})^2 \nabla^{\dalpha \alpha} \nabla_{\alpha \dalpha} \L^{++}
	\eol & \quad
	+ \frac{3}{2} z_1^{--} \chi^{\alpha +} D^{--} \nabla_\alpha^- \L^{++}
	- \frac{3}{2} z_1^{--} \bar\chi_\dalpha^{+} D^{--} \bar\nabla_\dalpha^- \L^{++}
	\eol & \quad
	- \frac{3}{2} z_1^{--} \chi^{\alpha -} \nabla_\alpha^- \L^{++}
	+ \frac{3}{2} z_1^{--} \bar\chi_\dalpha^{-} \bar\nabla_\dalpha^- \L^{++}
	\eol & \quad
	- 3 z_1^{--} \,D \,D^{--} \L^{++} + 3 (z_1^{--})^2 \,D \,\L^{++}
\end{align}
where $P^{--}$ is a purely arctic expansion. This can be discarded under
the contour and so the relevant contributions to $T_0$ are
\begin{align}
T_0 &=
	- 3 z_1^{--} \chi^{\alpha -} \nabla_\alpha^- \L^{++}
	+ \frac{3}{2} (z_1^{--})^2 \chi^{\alpha +} \nabla_\alpha^- \L^{++}
	+ \HC
	\eol & \quad
	+ \nabla^{\dalpha \alpha} B_{\alpha \dalpha}^{--}
	+ D^{--} \Big(\frac{3}{2} z_1^{--} \chi^{\alpha +} \nabla_\alpha^- \L^{++} + \HC
	- 3 z_1^{--} \,D \,\L^{++} \Big)
\end{align}
with
\begin{align}
B_{\alpha\dalpha}^{--} = -\frac{i}{2} z_1^{--} \nabla_\alpha^- \bar\nabla_\dalpha^- \L^{++}
	+ \frac{1}{2} (z_1^{--})^2 \nabla_{\alpha\dalpha} \L^{++}~.
\end{align}
Denoting the contribution of the $B_{\alpha\dalpha}^{--}$ to 
$T_0$ as $T_{0. \rm TD}$, we evaluate it as in section \ref{sec:CompActionCurved.Simp},
\begin{align}
T_1 + T_{0.\rm TD|Q}
	&= -\frac{i}{8} (\psi^{\dalpha \alpha}{}_\alpha^- - z_1^{--} \psi^{\dalpha \alpha}{}_\alpha^+)
			(\nabla^-)^2 \bar\nabla_\dalpha^- \L^{++}
		\eol & \quad
		- \frac{1}{4} \Big(z_1^{--} \psi^{\dalpha \alpha}{}_\alpha^- + \frac{1}{2} (z_1^{--})^2 \psi^{\dalpha\alpha}{}_\alpha^+\Big)
				\nabla_{\beta \dalpha} \nabla^{\beta -} \L^{++}
		\eol & \quad
		- \frac{3i}{2} (z_1^{--})^2 \psi^{\dalpha \alpha}{}_\alpha^+ \Big(\bar \chi_\dalpha^- \L^{++}
			 -\frac{1}{2} \bar\chi_\dalpha^+ D^{--}\L^{++}\Big)
		\eol & \quad
		+ \frac{i}{4} (z_1^{--})^2 \psi^{\dalpha \alpha}{}_\alpha^+ \bar W_{\dalpha\dbeta} \bar\nabla^{\dbeta -} \L^{++}
	\eol & \quad
	- \frac{1}{2} z_1^{--} \psi^{\dalpha (\alpha}{}^{\beta) -} \nabla_{\beta \dalpha} \nabla_\alpha^- \L^{++}
	+ \frac{1}{4} (z_1^{--})^2 \psi^{\dalpha (\alpha}{}^{\beta) +} \nabla_{\beta\dalpha} \nabla_{\alpha}^- \L^{++}
	+ \HC
\end{align}
In the first line, we require the expression
\begin{align}
\frac{1}{8} (\nabla^-)^2 \bar\nabla^{\dalpha -} \L^{++}
	&= \Xi^{\dalpha-} + z_1^{--} \Big(
		\frac{i}{2} \nabla^{\dalpha\beta} \nabla_\beta^{-} \L^{++}
		- \frac{1}{2} \bar W^\dalpha{}_{\dbeta} \bar\nabla^{\dbeta -} \L^{++}
		\eol & \qquad\qquad
		+ 3 \bar\chi^{\dalpha-} \L^{++}
		- \frac{3}{2} \bar\chi^{\dalpha +} D^{--} \L^{++} \Big)~.
\end{align}
The contribution of $\Xi^{\dalpha -}$ to the action is purely arctic so we can discard it,
leaving
\begin{align}
T_1 + T_{0.\rm TD|Q}
	&= 
	\Big(z_1^{--} \psi^{\dalpha \alpha}{}_\alpha^- - \frac{1}{2} (z_1^{--})^2 \psi^{\dalpha \alpha}{}_\alpha^+\Big) \times
	\eol & \qquad \qquad
	\Big(
		\frac{i}{2} \bar W_{\dalpha\dbeta} \bar \nabla^{\dbeta -} \L^{++}
		-3i \bar \chi_\dalpha^- \L^{++} + \frac{3i}{2} \bar \chi_\dalpha^+ D^{--} \L^{++}\Big)
	\eol & \quad
	+ (2 z_1^{--} \psi_a^- \sigma^{ab} - (z_1^{--})^2 \psi_a^+ \sigma^{ab})^\alpha \,\nabla_b \nabla_\alpha^- \L^{++}
	+ \HC
\end{align}
for the terms involving a single gravitino.
Now we define $\mathscr{R}^{--}$ as everything that has not yet canceled out from
$T_0$ and $T_1$:
\begin{align}
\mathscr{R}^{--}
	&=
	- 3 z_1^{--} \chi^{\alpha -} \nabla_\alpha^- \L^{++}
	+ \frac{3}{2} (z_1^{--})^2 \chi^{\alpha +} \nabla_\alpha^- \L^{++}
	\eol & \quad
	+ \Big(z_1^{--} \psi^{\dalpha \alpha}{}_\alpha^- - \frac{1}{2} (z_1^{--})^2 \psi^{\dalpha \alpha}{}_\alpha^+\Big)
	\Big(
		\frac{i}{2} \bar W_{\dalpha\dbeta} \bar \nabla^{\dbeta -} \L^{++}
		-3i \bar \chi_\dalpha^- \L^{++} + \frac{3i}{2} \bar \chi_\dalpha^+ D^{--} \L^{++}\Big)
	\eol & \quad
	+ \Big(2 z_1^{--} (\psi_a^- \sigma^{ab})^\alpha
		- (z_1^{--})^2 (\psi_a^+ \sigma^{ab})^\alpha \Big) \,\nabla_b \nabla_\alpha^- \L^{++}
	+ \HC
	+ T_{0.\rm TD|rest} 
	\eol & \quad
	+ D^{--} \Big(\frac{3}{2} z_1^{--} \chi^{\alpha +} \nabla_\alpha^- \L^{++} 
	- \frac{3}{2} z_1^{--} \bar\chi_\dalpha^{+} \bar\nabla^{\dalpha-} \L^{++} 
	- 3 z_1^{--} \,D \,\L^{++} \Big)~.
\end{align}
When combined with the terms in $S_{\rm rest}$, this must vanish up to a total derivative.

Note that $\L^{++}$ now has at most two spinor derivatives acting on it,
so we can choose $\L^{++} = \frac{i}{2} \Gamma_I^+ \U^{I+}$ to be arctic
without any difficulty.
This exactly matches the arctic part of \eqref{eq:Rmm}.
The part of the action $S_{\rm rest}$ that did not involve $T_0$ and $T_1$ is
also identical. But we know that all these contributions must vanish,
and so it follows that the unevaluated terms in \eqref{eq:CompActionRemainders}
involving $\G_I^+ \U^{I+}$ all cancel out. The antarctic ones also vanish in
like fashion.

Lest the reader find this proof too indirect, it should be added
that we have \emph{explicitly} checked that the remaining
terms do indeed vanish. To do so is fairly involved and requires
the contributions from both integrals in \eqref{eq:ProjCompAction}.
To demonstrate some of the manipulations that occur, we will show
here how the cancellation occurs when all fermionic terms are turned off.
The relevant bosonic terms remaining in $\cL^{--}$ are
\begin{align}
\cL^{--} &\sim 
	-3 D^{--} \Big(z_1^{--} \,D \,\L^{++} \Big)
	- \frac{i}{4} \cV_m{}^{--} (\bsigma^m)_{\alpha \dalpha} [\nabla^{\alpha -}, \bnabla^{\dalpha -}] \L^{++}
	\eol & \quad
	- \cV^{\dalpha \alpha\, --} D^{++} B_{\alpha\dalpha}^{--}
	+ D^{--} (\cV^{\dalpha \alpha ++} B_{\alpha\dalpha}^{--})
	- f^{\dalpha \alpha b} K_b B_{\alpha\dalpha}^{--} \eol
	&= - D^{--} \Big(
	 \frac{i}{2} z_1^{--} \cV^{\dalpha \alpha ++} \nabla_\alpha^- \bnabla_\dalpha^- \L^{++}
	 - \frac{1}{2} (z_1^{--})^2 \cV^{\dalpha \alpha ++} \nabla_{\alpha\dalpha} \L^{++}
	 \eol & \qquad \qquad
	+ 3 z_1^{--} \,D \,\L^{++}
	+ 4 z_1^{--} f_a{}^a \L^{++}
	\Big)~.
\end{align}
Those contributing to $\cL^{++}$ are given by
\begin{align}
\cL^{++}
	&= -\Big(3 D + 4 f_a{}^a \Big) \L^{++}
	- \frac{i}{2} \cV^{\dalpha \alpha ++} \nabla_\alpha^- \bar\nabla_\dalpha^- \L^{++}~,
\end{align}
which can be rewritten as
\begin{align}
\cL^{++} &= - D^{++} \Big(
	 \frac{i}{2} z_1^{--} \cV^{\dalpha \alpha ++} \nabla_\alpha^- \bnabla_\dalpha^- \L^{++}
	 - \frac{1}{2} (z_1^{--})^2 \cV^{\dalpha \alpha ++} \nabla_{\alpha\dalpha} \L^{++}
	 \eol & \qquad \qquad
	+ 3 z_1^{--} \,D \,\L^{++}
	 + 4 z_1^{--} f_a{}^a \L^{++}
	\Big)~.
\end{align}
The combination of $\cL^{--}$ and $\cL^{++}$ is a total contour derivative
and can be discarded. Similar cancellations occur with the fermionic terms,
but these require much more work.

\end{document}

%% file: sugracom_v2.tex
\newcommand {\cA}{{\cal A}}

\newcommand {\cC}{{\cal C}}
\newcommand {\cD}{{\cal D}}
\newcommand {\cE}{{\cal E}}
\newcommand {\cF}{{\cal F}}
\newcommand {\cG}{{\cal G}}

\newcommand {\cJ}{{\cal J}}
\newcommand {\cK}{{\cal K}}
\newcommand {\cL}{{\cal L}}

\newcommand {\cN}{{\cal N}}
\newcommand {\cO}{{\cal O}}

\newcommand {\cQ}{{\cal Q}}
\newcommand {\cR}{{\cal R}}
\newcommand {\cS}{{\cal S}}

\newcommand {\cV}{{\cal V}}
\newcommand {\cW}{{\cal W}}

\def\G{\Gamma}

\def\l{\lambda}

\def\q{\theta}

\def\z{\zeta}

\def\L{\Lambda}

\def\U{\Upsilon}

\newcommand{\ra}{{\mathrm a}}
\newcommand{\rb}{{\mathrm b}}
\newcommand{\rc}{{\mathrm c}}
\newcommand{\rd}{{\mathrm d}}

\newcommand{\ri}{{\mathrm i}}

\newcommand{\dalpha}{{\dot{\alpha}}}
\newcommand{\dbeta}{{\dot{\beta}}}

\newcommand{\1}{{\underline{1}}}
\newcommand{\2}{{\underline{2}}}

\newcommand{\bm}[1]{\mbox{\boldmath$#1$}}
\newcommand{\veps}{\varepsilon}
\newcommand{\eps}{{\epsilon}}
\newcommand{\eol}{\notag \\}

\newcommand{\bsigma}{\bar{\sigma}}
\newcommand{\ul}{\underline}
\newcommand{\pa}{\partial}
\newcommand{\bphi}{{\bar\phi}}
\newcommand{\bpsi}{{\bar\psi}}
\newcommand{\HC}{{\mathrm{h.c.}}}